\documentclass[preprint,12pt]{elsarticle}
\usepackage[top=1.0in,left=0.80in,bottom=1.0in,right=0.80in]{geometry}
\usepackage[]{algorithm}
\usepackage{algorithmicx}

\usepackage{algpseudocode}
\usepackage{enumitem}

\usepackage{amssymb}
\usepackage{mathtools}
\usepackage{placeins}  

\usepackage{comment}
\usepackage{url} 
 \usepackage{hyperref}

 \usepackage{color}
\usepackage[displaymath, mathlines]{lineno}

\newcommand{\cG}{\mathcal G}

\journal{Energy and Buildings}
\begin{document}
\begin{frontmatter}


 \title{Quantifying uncertainty in thermophysical properties of walls by means of Bayesian inversion}
 \author[label2]{Lia De Simon}
\author[label1]{Marco Iglesias}
\author[label2]{Benjamin Jones}
\author[label2]{Christopher Wood}



\address[label2]{Department of Architecture and Built Environment, University of Nottingham, University Park, Nottingham, NG7 2RD, UK}
\address[label1]{School of Mathematical Sciences, University of Nottingham, University Park, Nottingham, NG7 2RD, UK}

\begin{abstract}

We introduce a computational framework to statistically infer thermophysical properties of any given wall from in-situ measurements of air temperature and surface heat fluxes. The proposed framework uses these measurements, within a Bayesian calibration approach, to sequentially infer input parameters of a one-dimensional heat diffusion model that describes the thermal performance of the wall. These inputs include spatially-variable functions that characterise the thermal conductivity and the volumetric heat capacity of the wall. We encode our computational framework in an algorithm that sequentially updates our probabilistic knowledge of the thermophysical properties as new measurements become available, and thus enables an on-the-fly uncertainty quantification of these properties. In addition, the proposed algorithm enables us to investigate the effect of the discretisation of the underlying heat diffusion model on the accuracy of estimates of thermophysical properties and the corresponding predictive distributions of heat flux. By means of virtual/synthetic and real experiments we show the capabilities of the proposed approach to (i) characterise heterogenous thermophysical properties associated with, for example, unknown cavities and insulators;  (ii) obtain rapid and accurate uncertainty estimates of effective thermal properties (e.g. thermal transmittance); and (iii) accurately compute an statistical description of the thermal performance of the wall which is, in turn, crucial in evaluating possible retrofit measures. 

\end{abstract}

\end{frontmatter}


\section{Introduction}\label{Intro}

There is continuous global evidence of a discrepancy between the predicted energy performance of buildings and the measured performance \cite{DEWILDE201440,fabric1}. This, often called \textit{energy performance gap}, has been attributed to occupancy, poor construction quality \cite{azar2012comprehensive,gupta2016deep,hong2006impact,hong2004impact} and uncertainties in the thermophysical composition of the building fabric \cite{hong2006impact,li2014solid,cesaratto2013measuring,doran2000detr,asdrubali2014evaluating}. In the context of the existing housing stock, the performance gap hinders predictive capabilities which are essential to inform optimal retrofitting strategies, cost-effective energy-saving measures and ultimately, to produce sound and robust carbon saving policies towards international decarbonisation targets \cite{climate2008change}.

Using measured data to calibrate simulations of the energy performance of a building is perhaps the most obvious pathway to minimise discrepancies between measured and predicted performance \cite{DEWILDE201440,BROUNS2016160}. However extensive reviews of current approaches for the calibration of Building Energy Performance Simulation (BEPS) tools/software reveal that there is no generic framework suitable for a wide class of dwellings (see \cite{COAKLEY2014123} and references therein). One substantial challenge faced by existing calibration workflows is that thousands of BEPS input variables/parameters need to be optimised/inferred given only very limited amount of displayed measurements. A particularly relevant subset of inputs are those that characterise the thermophysical properties of the building fabric. It has been shown that incorrect assumptions on the these properties, for example, based on tabulated values and/or visual inspections, may lead to inaccurate predictions of energy performance \cite{li2014solid,cesaratto2013measuring} and hence, unreliable estimates of carbon reductions from retrofit interventions \cite{doi:10.1177/0143624416647758}.  More accurate and robust characterisations of the building fabric can be achieved by inferring thermophysical properties from in-situ monitored data of the thermal behaviour of individual fabric elements \cite{fabric2,biddulph2014inferring,gori2017inferring}. These inferred estimates can be used to better inform inputs of BEPS tools prior to calibration workflows.

Since the thermal transmission through external walls accounts for the largest share of heat losses in a typical dwelling, using in-situ measurement to infer the thermal transmittance (or U-value [W/m$^2$K]) of external walls is a primary goal when assessing the thermal performance of an existing building. Although theoretical estimates of the U-value  can be obtained from (steady-state) calculations, these assume that a wall comprises a number of clearly defined layers with known geometry and homogeneous thermophysical properties, such as conductivity and heat capacity \cite{yunus2011heat}. In practice, the number and properties of layers may be unknown and heterogenous due to the presence of thermal bridges including those arising from material defects, moisture penetration and residual cavities. Consequently, theoretical computations of U-values based on visual inspections and the use of tabulated values often provide an inaccurate characterisation of a walls thermal performance. In contrast, inferred U-values from in-situ measurements can better capture the thermal properties of walls thereby providing a more accurate description of their thermal performance \cite{li2014solid,biddulph2014inferring,gori2017inferring}.

The current practice to infer U-value from in-situ measurement is defined by the ISO9869:2014  \cite{ISO9869:2014} as the average method, equation (1), where the U-value is derived from $M$ measurements of the internal surface heat flux, $q_{I,m}$ ($m=1,\dots M$), and internal and external air temperature measurements, $T_{I,m}$ and $T_{E,m}$:
\begin{equation}\label{average_method}
\mathcal{U}_{AV}=\frac{\sum_{m}^{M} q_{I,m}}{\sum_{m=1}^{M}(T_{I,m} - T_{E,m})}
\end{equation}
The monitoring time (i.e. number of measurements $M$) necessary to obtain a reliable measurement indicated by the ISO:9869:2014 \cite{ISO9869:2014} is at least 72h but, in practice, monitoring periods of above 10 days are  common   \cite{baker2011u}.  These lengthy testing periods lead to the practice of in-situ measurements being seen as a rare exception rather than the norm. Additionally, the uncertainty associated to this practice can be as high as 25\% as indicated by the ISO:9869:2014 \cite{ISO9869:2014}. This uncertainty in the U-value follows through into energy saving predictions, which in turn, affects estimated pay back periods and makes investment decisions challenging \cite{booth2013decision}.

It is the limitations associated with the average method calculation, both in terms of test duration and the need for a better understanding of the associated uncertainty within the result, that have led to novel approaches for estimating thermophysical properties within statistical frameworks that enable uncertainty quantification. In particular, \cite{biddulph2014inferring,gori2017inferring} has recently proposed to infer thermal properties (U-value and thermal mass) of solid walls given in-situ measurements via the Bayesian calibration of \textit{lumped thermal mass models}. By means of an electrical network analogy, lumped thermal mass models provide a simplified description of the heat transfer process through the wall. The input parameters of these models include the nominal values of a prescribed number of resistances and capacitances which, in turn, characterise the thermophysical properties of the wall under investigation. The approach proposed in \cite{biddulph2014inferring,gori2017inferring} infers these parameters given air temperature and heat flux measurements. 

The small number of input parameters of lumped thermal mass models constitute their main advantage over fully descriptive heat transfer models. The calibration of those parameters from in-situ measured data is often computationally tractable either via standard optimisation \cite{gutschker2008parameter} or statistical approaches such as those used in \cite{biddulph2014inferring,gori2017inferring}. The Bayesian approach, in particular, provides the tools necessary to characterise uncertainties in the thermophysical composition of a wall via the probability distributions of inferred model parameters. The variance arising from these distributions can be used, in turn, to quantify the uncertainty on the predictions of thermal performance of the building element under investigation and thus inform decision-making workflows during retrofit interventions. However, the simplicity of the underlying simplified lumped thermal mass models  calibrated with existing approaches \cite{biddulph2014inferring,gori2017inferring} can reduce both their accuracy and their portability, as the model topology (number of resistances/capacitors) requires to be adjusted for different wall types. For example, the single thermal mass model successfully calibrated to fit the experimental data reported in \cite{biddulph2014inferring}, was unsuitable to characterise the thermal performance of the walls investigated in \cite{gori2017inferring} and which was, in turn, properly characterised via the Bayesian calibration of a more complex lumped thermal mass model with two capacitors. Some further limitations of lumped thermal mass models for walls that are thick, well insulated, or have large indoor convection coefficients have been discussed in \cite{kircher2015lumped,xu2007optimal}.

\subsection{Contribution of this work}\label{contribution}

With the development of recent Bayesian methodologies that enable the calibration of large sets of parameters \cite{Kantas} and the wide availability of more computer power, it is timely to consider more physically realistic models of the building fabric in order to understand their applicability to larger class of dwellings and to provide a more accurate quantification of the uncertainties that arise from inhomogeneities within the fabric. In this paper we propose to use in-situ measured data to calibrate, within a sequential Bayesian approach, a high-dimensional heat transfer model capable of describing the thermal performance of an arbitrary wall regardless of its (possibly heterogenous) thermophysical composition. In contrast to existing Bayesian approaches \cite{biddulph2014inferring,gori2017inferring} where simple lumped thermal mass models have been used for the inference of effective properties, our heat transfer model of the wall is based on the 1D heat equation. Within this model, the wall's thermophysical properties are unknown/unobservable input parameters characterised by the internal and external surface resistance, $R_{I}$ and $R_{E}$, as well as two heterogenous spatially-varying functions of the thickness of the wall:  thermal conductivity of the wall $\kappa(x)$ and the volumetric heat capacity $c(x)$.  For a given wall under investigation, the proposed sequential Bayesian methodology approximates the posterior probability distribution of $R_{I}$, $R_{E}$, $\kappa(x)$ and $c(x)$ conditioned to in-situ measurements of near-air and surface heat fluxes. Upon discretisation, the estimates of $\kappa(x)$ and $c(x)$ provide a statistical characterisation of the wall's thermophysical properties (e.g. number of multiple layers and thermophysical properties on each layer). Moreover, our computational approach enables us to convert Bayesian posteriors of these thermal properties into probability distributions of (i) (averaged) thermal properties such as the wall's U-value and C-value (heat capacitance per unit of area) and (ii) predictions of observable quantities (e.g. surface heat fluxes) which are essential to compute risks of retrofit options and thus to inform optimal energy saving measures.

The proposed Bayesian approach is embedded in a computational algorithm that uses an ensemble Kalman methodology \cite{SMC_REnKA} to merge in-situ measurements with computer simulations of heat fluxes, and generate an ensemble of realisations of the thermal properties that approximate the posterior distribution in a sequential fashion. The Kalman-based methodology at the core of the proposed algorithm is derived from a Sequential Monte Carlo (SMC) approach which, in contrast to the standard all-at-once existing Bayesian approaches \cite{biddulph2014inferring,gori2017inferring}, enables us to update our probabilistic knowledge of the thermal properties as new in-situ measurements are collected. The capabilities of the proposed approach are demonstrated by means of numerical experiments with both synthetic and real data. We show that by incorporating a high-dimensional characterisation of the spatial variability in thermal properties (via the inference of  $\kappa(x)$ and $c(x)$), the proposed computational framework can reveal internal inhomogeneities of the wall (e.g. residual cavities) that are unknown a priori. We further demonstrate that the proposed framework can provide more accurate uncertainty estimates of the effective thermophysical properties (e.g. U-value) and higher degree of confidence in the predictions of the wall's surface heat fluxes, compared to those obtained via low-dimensional (coarse-grid) heat transfer models which are the basis for lumped thermal mass models used in existing Bayesian approaches for in-situ characterisation. We additionally show that the proposed sequential Bayesian approach can be used to monitor the stability of the uncertainty estimates of the thermophysical properties, thus providing us with a tool to determine, on the fly, the duration of the measurement campaign needed to achieve a desired level of accuracy. 

The rest of the paper is organised as follows. The mathematical framework for the proposed approach to infer thermal properties is introduced in Section \ref{maths}. Both the synthetic and experimental data used for the validation of the proposed inferential approach are described in Section \ref{data}. In Section \ref{synthetic} and Section \ref{BSRIA} we report and discuss the numerical results obtained from the application of the proposed framework. Some conclusions and final remarks are presented in Section \ref{conclusions}.

\begin{figure}[h]
	\centering
	\includegraphics[trim = 0cm 0cm 0cm 0cm, clip=true, totalheight=0.3\textheight]{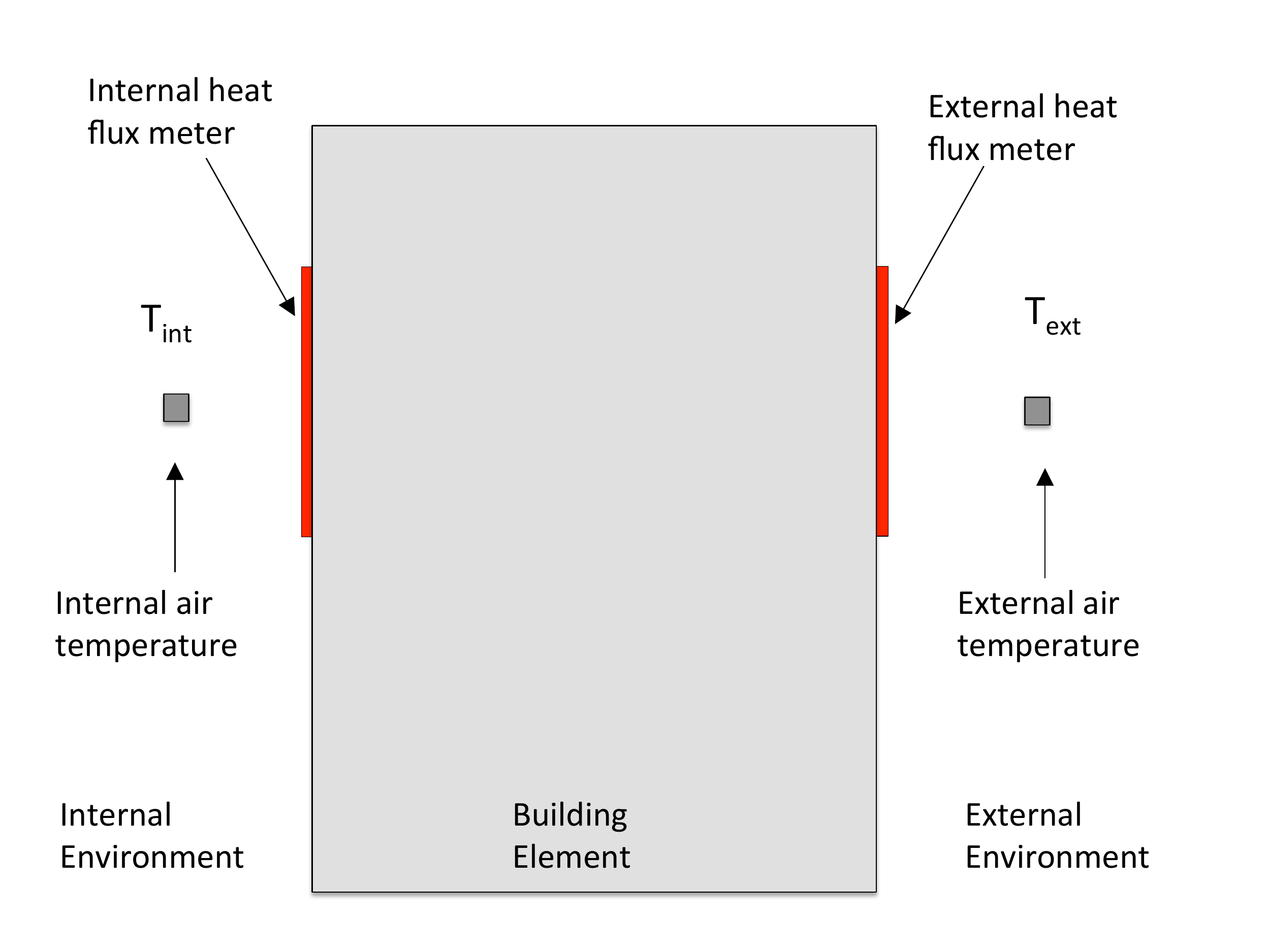}
	\caption{Measurement configuration.}
	\label{RC}
\end{figure}

\section{Mathematical background}\label{maths}

In this section we describe a Bayesian approach to infer thermal properties of any given wall given in-situ measurements of the wall's thermal performance. Following the preliminary definitions and notation introduced in subsection \ref{notation}, in subsection \ref{HDM} we introduce a heat diffusion model (hereon known as the HDM) aimed at describing the thermal performance of any wall. The wall's thermal properties are characterised by unobservable inputs of this HDM. A Bayesian computational framework is developed in subsection \ref{bayesian_theo} in order to infer these inputs from in-situ measurements of surface heat flux and near-air temperatures. Computational aspects and the discretisation of the algorithm are discussed in subsection \ref{bayesian_compu} and subsection \ref{discretisation}, respectively.

\subsection{Preliminaries and notation.}\label{notation}

We study a simple one-dimensional model of a generic wall with thickness $L$ (m); see  Figure \ref{RC}. Any location along the thickness of the wall is denoted by $x$ (m). The location of the internal and external surfaces correspond to $x=0$ and $x=L$, respectively, thus $x\in [0,L]$. We are interested in monitoring the thermal performance of the wall over an interval of time $[0,t_{f}]$ (seconds). In particular, we consider a collection of $M$ observation times denoted by $\{\tau_{m}\}_{m=1}^{M}$, with $0<\tau_{1}<\dots<\tau_{M}=t_{f}$. At each observation time $\tau_{m}$, we collect  in-situ measurements of internal and external wall's surface heat flux (Wm$^{-2}$); these measurements are denoted by $q_{I,m}$ and $q_{E,m}$, respectively.  In addition, internal and external near-air temperature measurements denoted by $T_{I,m}^{\dagger}$ and $T_{E,m}^{\dagger}$ (K) are collected at time $\tau_{m}$. For each $m=1,\dots, M$, we define the variable
\begin{eqnarray}\label{eq:1}
q_{m}=(q_{I,m},q_{E,m}),
\end{eqnarray}
with measurements of both internal and external heat fluxes observed at each observation time $\tau_{m}$. Furthermore, we define
\begin{eqnarray}\label{eq:2}
q_{1:m}=(q_{1},\dots,q_{m}),
\end{eqnarray}
the vector with all measurements of heat fluxes collected at all observation times up to $\tau_{m}$. Similarly, we use the notation $T_{I,1:m}^{\dagger}$ (resp. $T_{E,1:m}^{\dagger}$) for the vector that contains internal (resp. external) near-air temperature measurements collected at all observation times within the time frame $[0,\tau_{m}]$.

As discussed in subsection \ref{contribution}, we characterise the thermal conductivity of the wall and the heat capacity via spatially-varying functions $\kappa(x)$ and $c(x)$, respectively. These are unobservable properties of the wall that we aim at inferring given in-situ measurements of surface heat flux and near-air temperatures denoted as above. Thermal properties are inferred with a Bayesian approach (subsection \ref{bayesian_theo}) that captures the uncertainty of $\kappa(x)$ and $c(x)$, and will, in turn, enable us to infer and quantify the uncertainty of the wall's effective/averaged thermal properties such as the U-value and the heat capacity per unit area (hereon known as C-value). In terms of $\kappa(x)$ and $c(x)$, these effective properties can be defined by
\begin{eqnarray}\label{eq:3}
\mathcal{U}=\Bigg[R_{I}+R_{E}+\int_{0}^{L}\frac{dx}{\kappa(x)}\Bigg]^{-1}\qquad  \textrm{and} \qquad \mathcal{C}=\int_{0}^{L}c(x)dx,
\end{eqnarray}
respectively, where $R_{I}$ and $R_{E}$ are the internal and external surface resistances.  Although book values for these variables are typically used \cite{cibse2015environmental,biddulph2014inferring}, these values are valid under conditions which may not necessarily apply for the specific wall under investigation. In our approach we incorporate the uncertainty in these variables and infer them within the proposed Bayesian methodology.

In the following subsection we develop a heat diffusion model (HDM) to describe the thermal performance of a generic wall.  This model has a set of input parameters that includes the wall's thermal properties $\kappa(x)$ and $c(x)$, surface resistances $R_{I}$ and $R_{E}$, as well as the internal and external near-air temperatures. The outputs of the HDM are aimed to predict, at each observation time $\tau_{m}$, heat flux at both the internal $(x=0)$ and external $(x=L)$ surface of the wall. The objective of this HDM is to establish a relationship between observable variables and outputs that we measured in situ, and the unobservable inputs, $R_{I}$, $R_{E}$, $\kappa(x)$ and $c(x)$, that we wish to infer via the Bayesian inference approach developed in subsection \ref{bayesian_theo}.

\subsection{The heat diffusion model}\label{HDM}

Let us assume that there are no internal heat sources and that the temperature inside the wall varies only with $x$. Then, at any location $x\in [0,L]$, and any moment in time $t\ge 0$ the wall's heat flux $Q(x,t)$ (Wm$^{-2}$) is defined by
\begin{eqnarray} \label{eq:4}
Q(x,t)=-\kappa(x)\frac{\partial}{\partial x}T(x,t),
\end{eqnarray}
where $T(x,t)$ (K) is the temperature distribution within the wall at location $x$ and time $t$. Denoting by $T_{0}(x)$ the temperature distribution within the wall at the initial simulation time $t=0$,
the internal temperature of the wall $T(x,t)$ (at subsequent times $t>0$) is then given by the solution of the heat diffusion equation \cite{yunus2011heat}:
\begin{eqnarray}\label{eq:6}
c(x)\frac{\partial T}{\partial t} =\frac{\partial }{\partial x}\Bigg [\kappa(x) \frac{\partial T}{\partial x}\Bigg],\qquad (x,t)\in (0,L),\quad t>0
\end{eqnarray}
with initial condition 
 \begin{eqnarray} \label{eq:6B}
T(x,0)&=&T_{0}(x), ~~~~~~~x\in [0,L]
\end{eqnarray}
and with the following convective boundary conditions that describe the heat transfer between the wall and local air:
\begin{eqnarray} \label{eq:7}
Q(0,t)&=&-R_{I}^{-1}[T(0,t)- T_{I}(t)] ~~~~~~~t\ge 0\\
Q(L,t)&=&-R_{E}^{-1}[T_{E}(t)-T(L,t)], ~~~~~t\ge 0. \label{eq:8}
\end{eqnarray}
In the previous expressions, $T_{I}(t)$ and  $T_{E}(t)$ denote the internal and external near-air temperatures. The set of input parameters for the heat transfer problem defined by (\ref{eq:6})-(\ref{eq:8})  then comprises 
$$( \kappa(x), c(x), T_{0}(x),R_{E},R_{I},T_{I}(t), T_{E}(t)).$$ 
Once these parameters are prescribed, equations (\ref{eq:6})-(\ref{eq:8}) can be uniquely solved for the temperature $T(x,t)$ (with $x\in [0,L]$ and $t\ge 0$) and predictions of heat flux $Q(x,t)$ can be obtained via (\ref{eq:4}). In particular, we may compute heat flux at the internal and external wall's surface at the observation time of interest $\tau_{m}$. These are comprised in the following variable:
 \begin{eqnarray}\label{eq:9}
Q_{m}=(Q(0,\tau_{m}),Q(L,\tau_{m})), \qquad m=1,\dots,M.
\end{eqnarray}
Note that, for each observation time $\tau_{m}$, the HDM model defined by equations (\ref{eq:6})-(\ref{eq:9}) induces the following parameter-to-output map $\cG_{m}$:
\begin{eqnarray}\label{eq:10}
(u(x),T_{I}(t), T_{E}(t))\xrightarrow[]{~~~\cG_{m}~~~}Q_{m},\qquad m=1,\dots,M,
\end{eqnarray}
where, for ease in the notation, we combined the unobservable variables in
\begin{eqnarray}\label{eq:10B}
u(x)\equiv (\kappa(x), c(x), T_{0}(x),R_{E},R_{I}).
\end{eqnarray}
Given the set of input parameters, the map (\ref{eq:10}) produces HDM predictions of internal and external surface heat fluxes at the observation time $\tau_{m}$. We define this map as the relation:
\begin{eqnarray}\label{eq:11}
\cG_{m}( u,T_{I}, T_{E})=Q_{m},
\end{eqnarray}
where we have dropped the dependence of $u$, $T_{I}$ and $T_{E}$, on the independent variables $x$ and $t$ in order to emphasise that the parameter-to-out map $\cG_{m}$ is a relation defined for functions rather than only for the values of these functions. 

In general, each parameter-to-output map $\cG_{m}$ ($m=1,\dots, M$) cannot be expressed analytically and so the HDM outputs (heat fluxes) need to be computed by means of a numerical solver (see subsection \ref{discretisation}).

\subsection{Development of the Bayesian inversion framework.}\label{bayesian_theo}

In this subsection we develop a computational Bayesian approach that, for any type of wall under investigation, infers the unobservable inputs ($\kappa(x)$, $c(x)$, $T_{0}(x)$, $R_I$ and $R_E$) of the parameter-to-output maps $\{\cG_{m}\}_{m=1}^{M}$ defined in subsection \ref{HDM} from in-situ measurements of internal and external near-air temperatures ($T_{I,1:M}^{\dagger}$,$T_{E,1:M}^{\dagger}$) and measurement of surface heat fluxes $q_{1:M}$. The inference of the unobservable parameters is an inverse problem which involves ``inverting'' the parameter-to-output maps in order to learn their input parameters from observed outputs. Once these parameters have been inferred, or more precisely, their probability distributions computed, we may then proceed to compute the corresponding probability distributions of model predictions which are, in turn, crucial to inform decision-making workflows involved during retrofit interventions.

\subsubsection{Assumptions on measurements.}

For simplicity we assume that temperature measurements errors are negligible, and propose to use in-situ measurements $T_{I,1:m}^{\dagger}$ and $T_{E,1:m}^{\dagger}$ to construct deterministic approximations (e.g. by smooth interpolation) of the observable parameter $T_{I}(t)$ and $T_{E}(t)$ over the time frame $[0,\tau_{m}]$. Although measurements are often contaminated with errors, our assumption is reasonable for experimental settings where standard high precision sensors of temperature are available (e.g. error less that $1\%$). In contrast, heat flux meters are typically prone to larger measurement errors (at least $5\%$) and so the uncertainty in the measurements of heat flux, $q_{1:M}$, must be taken into account within the inference process. To this end, we assume that for each observation time $\tau_{m}$, unobservable input parameters $u=(\kappa(x),c(x),T_{0}(x),R_{I},R_{E})$ are related to heat flux measurements $q_{m}$, and the measurement-based approximations of near-air temperatures $(T_{I,1:m}^{\dagger},T_{E,1:m}^{\dagger})$, via
\begin{eqnarray}\label{eq:12}
q_{m}=\cG_{m}(u, T_{I,1:m}^{\dagger}, T_{E,1:m}^{\dagger})+\eta_{m},
\end{eqnarray}
where $\eta_{m}$ is a two-dimensional vector of random noise, independent and identically distributed according a prescribed distribution denoted by ${\mathbb P}_{\eta}(\eta_{m})$. Note that equation (\ref{eq:12}) simply states that the empirical measurements of heat flux $q_{m}$ (at time $\tau_m$) can be obtained, from the corresponding HDM prediction $\cG_{m}(u, T_{I,1:m}^{\dagger}, T_{E,1:m}^{\dagger})=Q_{m}$, by accounting for and additive random error in the heat flux measurements. In this work, we use the standard assumption that ${\mathbb P}_{\eta}(\eta_{m})$  ($m=1,\dots, M$) is a Gaussian distribution with zero mean and covariance denoted by $\Gamma_{m}$. However, the Bayesian approach introduced below can be applied to more general cases for which measurements errors are characterised by non-Gaussian distributions.

\subsubsection{The Bayesian approach}

In order to infer the unobservable/unknown variable $u(x)=(\kappa(x),c(x),T_{0}(x),R_{I},R_{E})$ within the context of the parameter-to-output map (\ref{eq:11}), we adopt the Bayesian framework \cite{Andrew} in which these unknown parameters are random functions/variables with a specified (joint) prior probability distribution. The prior encapsulates our probabilistic prior knowledge of the wall before in-situ measurements are collected, and it may incorporate information of the thermal properties obtained from the wall's design and/or visual inspection. We denote the prior by $\mathbb{P}(u)$ and refer the reader to \ref{prior} where we construct this distribution and discuss algorithms to produce the corresponding  samples.

Starting with the prior ${\mathbb P}(u)$ we use a sequential Bayesian approach \cite{Kantas} to update our probabilistic knowledge of $u(x)=(\kappa(x),c(x),T_{0}(x),R_{I},R_{E})$ comprised in the conditional (posterior) distribution of $u(x)$ given all measurements of heat fluxes $q_{1:m-1}=(q_{1},\dots, q_{m-1})$ collected up to a given observation time $\tau_{m-1}$. Once new heat flux measurements $q_{m}$ are collected at time $\tau_{m}$, the posterior, denoted by ${\mathbb P}(u|q_{1:m-1})$, is then updated via the following recursive version of Bayes' rule \cite{Kantas}:
\begin{eqnarray}\label{eq:13}
{\mathbb P}(u |q_{1:m}) = \frac{ {\mathbb P}(q_{1:m-1})}{ {\mathbb P}(q_{1:m})} {\mathbb P}(q_{m}|u){\mathbb P}(u |q_{1:m-1}),
\end{eqnarray}
where $ {\mathbb P}(q_{m}|u)$ is the likelihood, namely, the probability of the observed measurements of heat flux $q_{m}$ given a particular realisation of the unknown parameter $u(x)=(\kappa(x),c(x),T_{0}(x),R_{I},R_{E})$. In expression (\ref{eq:13}), the terms ${\mathbb P}(q_{1:m-1})$ and ${\mathbb P}(q_{1:m})$ denote the probabilities of $q_{1:m-1}$ and $q_{1:m}$, respectively. These are normalisation constants (with respect to $u(x)$) defined by
\begin{equation}\label{eq:14}
{\mathbb P}(q_{1:m}) =\int  {\mathbb P}(q_{1}|\kappa, c,T_{0})\cdots   {\mathbb P}(q_{m}|u){\mathbb P}(du).
\end{equation}
Once the updated posterior ${\mathbb P}(u|q_{1:m})$ has been determined, the marginal distributions of $\kappa(x)$, $c(x)$, $R_{E}$ and $R_{I}$ can be used, via equation (\ref{eq:3}), to compute the posterior distributions of the U-value and the C-value; these distributions are denoted by ${\mathbb P}(\mathcal{U}\vert q_{1:m})$ and ${\mathbb P}(\mathcal{C}\vert q_{1:m})$, respectively. Similarly, the uncertainty quantified by the posterior distribution of the unknown parameters can be used to predict uncertainty in heat flux predictions (via the parameter-to-output map \eqref{eq:11}). In other words, we may consider the (predictive) distribution, $\mathbb{P}(Q_{m+1}, \dots, Q_{m+p}\vert q_{1:m})$, of internal and external surface heat flux predictions over an interval $(\tau_{m},\tau_{m+p}]$ given all measurements, $q_{1:m}$, collected up to time $\tau_{m}$. These predictive distributions will be crucial in assessing the degree of confidence of our uncertainty estimates of the thermal performance of the wall.

It is important to mention that both (\ref{eq:13}) and (\ref{eq:14}) are valid under the assumption of independence of the heat flux measurement error $\eta_{m}$. Furthermore, since $\eta_{m}$ is distributed according to ${\mathbb P}_{\eta}(\eta_{m})$, it follows from (\ref{eq:12}) that the likelihood is given by
\begin{eqnarray}\label{eq:15}
{\mathbb P}(q_{m}|u)={\mathbb P}_{\eta}(q_{m}-\cG_{m}(u, T_{I}^{\dagger}, T_{E}^{\dagger})),
\end{eqnarray}
which, in turn, allows us to rewrite (\ref{eq:13}) as 
\begin{equation}\label{eq:16}
{\mathbb P}(u|q_{1:m})= \frac{ {\mathbb P}(q_{1:m-1})}{ {\mathbb P}(q_{1:m})}  {\mathbb P}_{\eta}(q_{m}-\cG_{m}(u, T_{I,1:m}^{\dagger}, T_{1:mE}^{\dagger})){\mathbb P}(u |q_{1:m-1}).
\end{equation}

For the sake of clarity in the previous exposition of the sequential Bayesian framework, we have assumed that only one measurement of internal/external heat flux (i.e. $q_{m}=(q_{m,I},q_{m,E})$) is assimilated at the observation time $\tau_{m}$. For computational efficiency, the numerical implementation of the proposed Bayesian approach (see subsection below), we consider a batch-sequential version whereby not only one but a set of heat flux measurements collected within the time frame $(\tau_{m-1},\tau_{m}]$ are assimilated via (\ref{eq:16}) at time $\tau_{m}$. This batch-sequential approach follows a very similar formulation to the one presented above and so we omit it. Nevertheless, hereon we refer to $\tau_{m}$'s as ``assimilation times'' to emphasise that these are observation times at which the distribution of the unknown parameters is updated in the Bayesian setting given by (\ref{eq:16}). 

\subsection{The computational approach to the Bayesian inference framework}\label{bayesian_compu}

Given the posterior ${\mathbb P}(u |q_{1:m-1})$ computed at the assimilation time $\tau_{m-1}$, expression (\ref{eq:16}) provides us with a formula to compute the updated posterior ${\mathbb P}(u |q_{1:m})$ defined at time $\tau_{m}$. This formula involves the normalisation constant $ {\mathbb P}(q_{1:m-1})$ and ${\mathbb P}(q_{1:m})$ (see (\ref{eq:14})), which from \eqref{eq:15}, can be written as
\begin{equation}\label{eq:27}
 {\mathbb P}(q_{1:m}) =\int \prod_{j=1}^{m}{\mathbb P}_{\eta}(q_{j}-\cG_{j}(u, T_{I}^{\dagger}, T_{E}^{\dagger})){\mathbb P}(du).
\end{equation}
Due to the nonlinearity of the parameter-to-output maps $\cG_{m}$ which appears in \eqref{eq:27}, these normalisation constants, in general, cannot be computed analytically, and so the resulting posterior distribution ${\mathbb P}(u |q_{1:m})$ cannot be expressed in closed form. Sampling/particle methods then need to be applied for the sequential approximation of the Bayesian posterior \cite{Kantas,Doucet}. A generic particle-based approach, applied to the present problem, is displayed in Algorithm \ref{IF}. This approach is initialised with an ensemble $\{u_{0}^{(j))}\}_{j=1}^{J}=\{\kappa_{0}^{(j)}(x),c_{0}^{(j)}(x),T_{0,0}^{(j)}(x),R_{I,0}^{(j)},R_{E,0}^{(j)}\}_{j=1}^{J}$ of $J$ realisations (often called particles) from the prior ${\mathbb P}(u)$. Suppose that at the assimilation time $\tau_{m-1}$ the algorithm produces an ensemble $\{u_{m-1}^{(j)}(x)\}_{j=1}^{J}$ that approximates ${\mathbb P}(u |q_{1:m-1})$, where $q_{1:m-1}$ contains all the measurements in the interval $[0,\tau_{m-1}]$. Once new measurements in the subinterval $(\tau_{m-1},\tau_{m}]$ are collected, the aim of the particle-based Bayesian approach is to use a framework stemming from (\ref{eq:16}) to update these particles so that the new ensemble $\{u_{m}^{(j)}(x)\}_{j=1}^{J}=\{\kappa_{m}^{(j)}(x),c_{m}^{(j)}(x),T_{0,m}^{(j)}(x),R_{I,m}^{(j)},R_{E,m}^{(j)}\}_{j=1}^{J}$ approximates the posterior ${\mathbb P}(u|q_{1:m})$.

The ensemble of particles obtained via Algorithm \ref{IF} can be used to compute approximation to the posterior expectations of thermal properties such as the posterior mean and posterior variance. For example, the posterior mean of the thermal conductivity $\kappa(x)$, at time $\tau_{m}$, can be approximated in terms of the ensemble mean $\overline{\kappa}_{m}(x)$ defined by
\begin{eqnarray}\label{eq:17A}
\overline{\kappa}_{m}(x)=  \frac{1}{J}\sum_{j=1}^{J}\kappa_{m}^{(j)}(x).
\end{eqnarray}
We emphasize, by means of the dependence on $x$ in (\ref{eq:17A}), that statistical measures of the unknown parameter $\kappa(x)$ are, in general, functions that vary across the thickness of the wall. Similar definitions can be used to define posterior means $\overline{c}_{m}(x)$ and $\overline{T}_{0,m}(x)$ of $\kappa(x)$ and $T_{0}(x)$, respectively. In the present work we also consider equal tail $(1-\alpha)100\%$ (pointwise) credible intervals \cite{gelmanbda04}. At each location $x$, these intervals contain, with a $(1-\alpha)100\%$ (posterior) probability, the unobserved parameters $\kappa(x)$, $c(x)$ and $T_{0}(x)$, that we aim at inferring with the Bayesian approach. Credible intervals thus provide us with a measure of the uncertainty in our estimates of inferred parameters at each location within the wall. We use the ensemble $\{\kappa_{m}^{(j)}(x)\}_{j=1}^{J}$, $\{c_{m}^{(j)}(x)\}_{j=1}^{J}$ and $\{T_{0,m}^{(j)}(x)\}_{j=1}^{J}$ to compute particle approximations of the aforementioned credible intervals. 

Expectations of the marginal posteriors for the scalars $R_{E}$ and $R_{I}$ can be approximated directly from the ensembles $\{R_{I,m}^{(j)}\}_{j=1}^{J}$ and $\{R_{E,m}^{(j)}\}_{j=1}^{J}$, respectively. These marginals are denoted by  ${\mathbb P}(R_{I}\vert q_{1:m})$ and ${\mathbb P}(R_{E}\vert q_{1:m})$. Similarly, we may use the posterior ensemble to compute, via (\ref{eq:3}), samples of the U-value and C-value:
\begin{eqnarray}\label{eq:19B}
\mathcal{U}_{m}^{(j)}=\Bigg[R_{I,m}^{(j)}+R_{E,m}^{(j)}+\int_{0}^{L}\frac{dx}{\kappa_{m}^{(j)}(x)}\Bigg]^{-1},\qquad 
\mathcal{C}_{m}^{(j)}=\int_{0}^{L}c_{m}^{(j)}(x)dx, \qquad j=1,\dots,J
\end{eqnarray}
These ensembles can be used to approximate statistical measures of the posterior distributions ${\mathbb P}(\mathcal{U}\vert q_{1:m})$ and ${\mathbb P}(\mathcal{C}\vert q_{1:m})$, respectively. 

We use the proposed ensemble approach to approximate the predictive distribution of internal and external surface heat flux $\mathbb{P}(Q_{m+1},\dots, Q_{m+p}\vert q_{1:m})$ introduced earlier. This predictive distribution is characterised by the ensemble of HDM predictions of heat fluxes:
\begin{eqnarray}\label{eq:19C}
(Q_{m+1}^{(j)},\dots, Q_{m+p}^{(j)})\equiv \{\cG_{m+1}(u_{m}^{(j)}, T_{I,1:p+1}^{\dagger}, T_{E,1:m+1}^{\dagger}),\dots,\cG_{m+p}(u_{m}^{(j)}, T_{I,1:m+p}^{\dagger}, T_{E,1:m+p}^{\dagger})\}_{j=1}^{J}.
\end{eqnarray}
Computationally, (\ref{eq:19C}) involves solving the HDM (\ref{eq:6})-(\ref{eq:9}), on the interval $[0,\tau_{m+p})$, 
 for each ensemble member $u_{m}^{(j)}$, inferred at the assimilation time $\tau_{m}$. Statistical measures including credible intervals can be approximated from the ensemble of model unknown $\{(Q_{m+1}^{(j)},\dots, Q_{m+p}^{(j)}\}_{j=1}^{J}$ in a similar fashion to the ones discussed above. These measures will enable us to assess the accuracy of our posterior model predictions together with their degree of confidence.

For the present work, we propose to conduct the Bayesian updating step of Algorithm \ref{IF} via the Regularising ensemble Kalman Algorithm (REnKA) that has been recently proposed in \cite{SMC_REnKA} as a Gaussian approximation from the (fully-Bayesian) adaptive-tempering Sequential Monte Carlo (SMC) of \cite{Kantas}. The work of \cite{SMC_REnKA} has shown that REnKA provides, at a reasonable computational cost, accurate approximations of the Bayesian posterior that arises from similar PDE-constrained inference problems such as the one defined by the HDM discussed earlier. In \ref{REnKA} we dicuss how we adapt REnKA to the present application. The algorithm (see Algorithm \ref{REnKA_al}), can be used in a black-box fashion; for further details of this numerical scheme the reader is referred to \cite{SMC_REnKA}.

\begin{algorithm}
\caption{Framework to infer $u(x)=(\kappa(x),c(x),T_{0}(x),R_{I},R_{E})$}\label{IF}{~}
\begin{algorithmic}
\Statex  Construct an initial ensemble $\{(\kappa_{0}^{(j)}(x),c_{0}^{(j)}(x),T_{0,0}^{(j)}(x),R_{I,0}^{(j)},R_{E,0}^{(j)})\}_{j=1}^{J}$of $J$ samples from the prior distributions $\mathbb{P}(u)$ (see Algorithm \ref{prior_al}). 
\For{$m=1,\dots,M$}
\begin{enumerate}
\item[(1)] Collect internal and external near-air temperature $T_{I,m}^{\dagger}$ and $T_{E,m}^{\dagger}$ in the interval $(\tau_{m-1},\tau_{m}]$. 
\item[(2)] Collect internal and external surface heat flux measurements $q_m$ within  $(\tau_{m-1},\tau_{m}]$
\item[(3)] Use $T_{I,1:m}^{\dagger}(t)$ and $T_{E,1:m}^{\dagger}(t)$ to define the parameter-to-put map in the interval $[0,\tau_{m})$:
$$\cG_{m}( u,T_{I,1:m}^{\dagger}, T_{E,1:m}^{\dagger})=Q_{m},$$
\item[(4)] Use a Bayesian updating algorithm (e.g. Algorithm \ref{REnKA}) (with $\cG_{m}$, $q_m$ and $\Gamma_{m}$ to update the posterior ensemble (for $j=1,\dots J$):  
\footnotesize
$$u_{m}^{(j)}=(\kappa_{m}^{(j)}(x),c_{m}^{(j)}(x),T_{0,m}^{(j)}(x),R_{I,m}^{(j)},R_{E,m}^{(j)}) \gets u_{m-1}^{(j)}=(\kappa_{m-1}^{(j)}(x),c_{m-1}^{(j)}(x),T_{0,m-1}^{(j)}(x),R_{I,m-1}^{(j)},R_{E,m-1}^{(j)}),$$
\normalsize
so that $\{u_{m}^{(j)}\}_{j=1}^{J} $ approximates the Bayesian posterior  $\mathbb{P}(u\vert q_{1:m})$.
\item[(5)] Use the updated posterior ensemble to compute an ensemble of quantities of interest such as the U-value of the C-value (see expressions (\ref{eq:3})) and the predictive distributions of internal and external heat flux $\mathbb{P}(Q_{m+1}\vert q_{1:m})$.  

\end{enumerate}
\EndFor
\end{algorithmic}
\end{algorithm}

We emphasize that our work is based on the assumptions that measurements errors in near-air temperature are sufficiently small so that these can be used to approximate the corresponding terms that arise from convective boundary conditions in the HDM. This assumption is the basis for most existing Bayesian work \cite{biddulph2014inferring,gori2017inferring} that infers thermal properties of walls. However, we recognise that, failing to account the uncertainty in these errors can introduce bias on the uncertainty estimates of the thermal properties. Incorporating those uncertainties is beyond the scope of this manuscript. Nevertheless, the proposed Bayesian approach is flexible enough and can be further extended to include those uncertainties via marginalisation techniques such as those recently proposed in \cite{marginalised}.

\subsection{Discretisation, implementation and computational cost of the Algorithm.}\label{discretisation}

It is important to emphasize that the proposed methodology encoded in Algorithm \ref{IF} includes the inference of function parameters $(\kappa(x),c(x),T_{0}(x))$ of the continuous HDM of the wall introduced subsection \ref{HDM}. In terms of Algorithm \ref{IF}, the HDM needs to be run (see Step 3 of Algorithm \ref{IF}) in order to evaluate the parameter-to-output maps $\cG_{m}$. In practice, however, the HDM must  be discretised and the solution approximated on a finite dimensional mesh/grid. This discretisation involves also the discretisation of the function parameters $(\kappa(x),c(x),T_{0}(x))$, thus the Bayesian inference of these functions reduces to the inference of the corresponding approximations at the nodes/elements of the finite dimensional domain. For the validation of the proposed methodology (see Section \ref{synthetic} and Section \ref{BSRIA}), the spatial discretisation of the HDM model is conducted by means of a standard Finite Element with linear basis functions \cite{becker1981finite}. The thermal properties $\kappa(x)$ and $c(x)$ are approximated with piece wise constant functions defined on each element of the discretisation scheme. The resulting semi-discrete time dependent problem is solved with at Backward Euler scheme. This numerical scheme of the HDM is implemented in MATLAB. The corresponding code is used within a MATLAB implementation of Algorithm \ref{IF} that uses the sequential updating approach provided by REnKA (Algorithm \ref{REnKA_al}). These codes are available on GitHub at \url{https://github.com/Marco-Iglesias-Nottingham/REnKA_Walls}.

While the practical implementation of Algorithm \ref{IF} involves the inference of the nodal values of some of the unknown parameters of the discretised HDM (recall $R_I$ and $R_E$ are scalars), the formulation in terms of the continuous HDM is crucial to our goal of characterising thermal properties of walls. Indeed, we wish to take advantage of the ability to use a fine mesh in the HDM in order to potentially resolve, via the inference of thermal properties, the fine structure within the wall which may reveal different constituents of the wall unknown a priori. The benefits that we obtain from approximating the HDM are lost if we subsequently embed it into a calibration method which degenerates when the mesh is refined. This is the case of conventional Bayesian methodologies which have been shown to collapse as the size of input space increases \cite{MCMC,Failure}. In contrast, the Bayesian methodology at the core of Algorithm \ref{IF}, based on REnKA \cite{SMC_REnKA}, is not only independent of the solver for the HDM, but it is also robust/stable with respect to increasing dimension of the underlying unknown (i.e. robust with respect to mesh refinement). This will allow us to study (subsection \ref{mesh_impact}) the effect of mesh refinement in the accuracy and degree of confidence in our posterior uncertainty estimates of (i) effective properties of the wall such as the U-value and C-value and (ii) model predictions of internal and external surface heat flux. 

At each assimilation time $\tau_{m}$, each iteration of REnKA (Algorithm \ref{REnKA_al}), requires to solve  the HDM, over the time window $[0,\tau_{m}]$, for each ensemble member of unknown parameters. Additional costs for updating the ensemble via REnKA are negligible. Therefore, at each $\tau_{m}$, the computational cost of the algorithm is $J$ (ensemble size) x cost of running the HDM x total number of iterations required for convergence. For the experiments reported in Section \ref{synthetic}, the average number of iterations is 2.7 while the cost of running the HDM (with a discretisation of $2^5$ elements) over the whole assimilation window $(0,\tau_{M}]$ is  0.0143 seconds of CPU time (on a Macbook pro 2-core 3.1 GHz Intel Core i7). A selection of a large ensemble of size $J=1000$, thus yields a maximum computational cost of 38.2 seconds. If measurements are assimilated at every five minutes, the proposed algorithm can be easily executed to compute posterior thermophysical properties as soon as these new measurements become available.

\section{Synthetic data generation and experimental data collection}\label{data}

In order to demonstrate the capabilities of the proposed Bayesian approach for inferring thermal properties of walls we apply Algorithm \ref{IF} to a set of synthetic/virtual data that we generate as discussed in subsection \ref{synthetic_data}. Further validations are carried out with the real data collected in-situ as described in subsection \ref{BSRIA_data}. In both cases, data sets comprise near-air temperature measurements of the internal and/or external environment and surface heat flux over several days.

\subsection{Synthetic/virtual data}\label{synthetic_data}

In this subsection we describe the procedure to generate synthetic data that we use to validate the proposed Bayesian approach under an idealised scenario, for which (i) we have perfect knowledge of the wall's thermal properties, initial internal temperature and near-air temperature measurements; and (ii) assumptions on the HDM and the measurement errors are perfectly satisfied. By means of virtual measurements of heat flux generated by using the HDM with these thermal properties, our aim in Section \ref{synthetic} is to recover those thermal properties within the posterior uncertainty band that we generate from the ensemble approach encoded in Algorithm \ref{IF}. Furthermore, we aim at computing predictions of internal and external surface heat flux that capture the corresponding measurements within a credible interval of high confidence. In real experiments with existing walls, thermal properties and the initial temperature profile would be unknown and/or difficult to characterise a priori. Moreover, the aforementioned assumptions are often difficult to verify. Therefore, by means of synthetic experiments, we reduce the potential lack of inconsistency between real conditions and modelling assumptions, and thus we focus entirely on assessing the capabilities of the proposed approach to characterise the thermal performance of a virtual wall under investigation.

\begin{figure}[htbp]
\begin{center}

\vspace{2.5mm}
\includegraphics[scale=0.4]{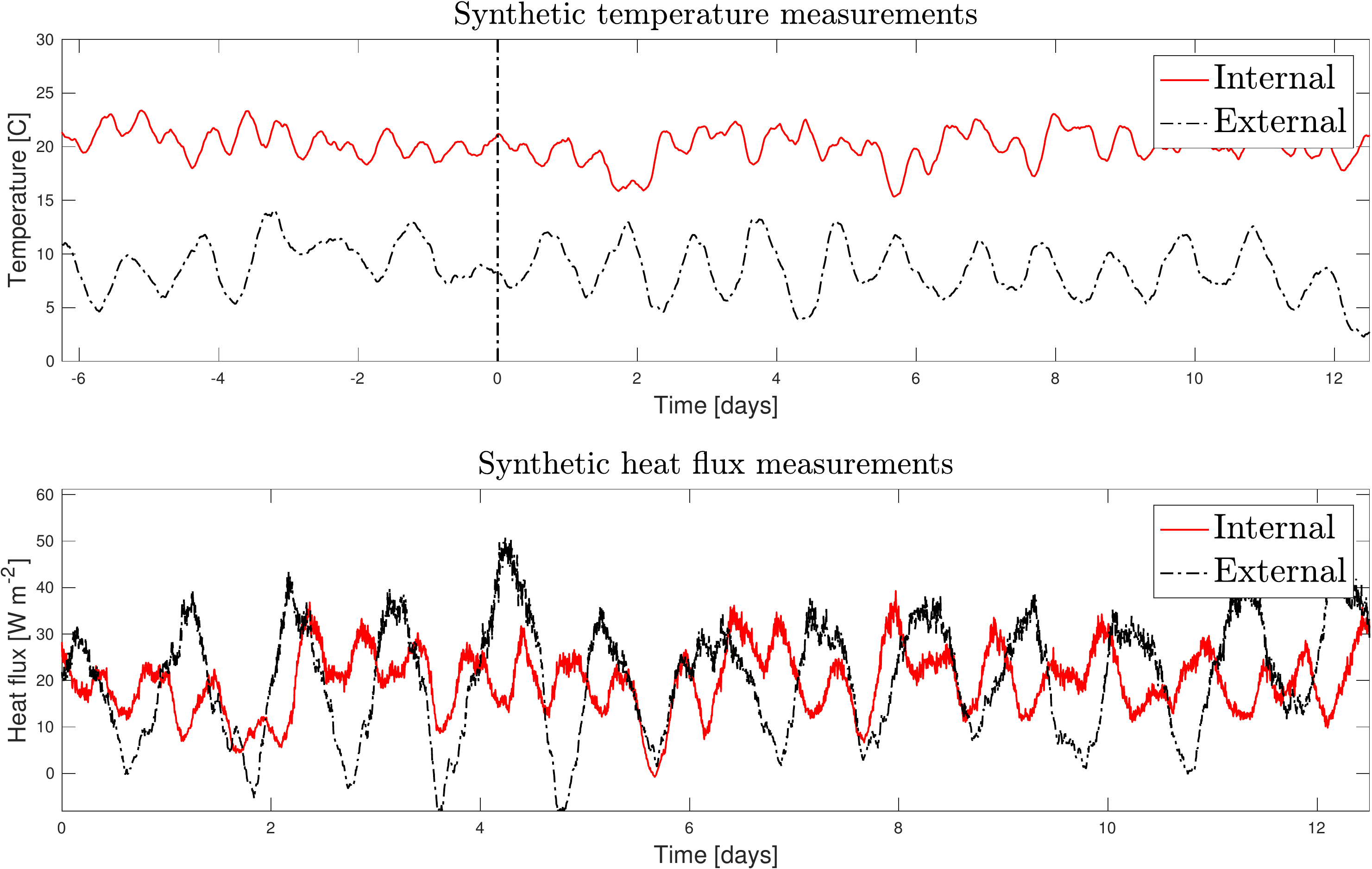}\\
\includegraphics[scale=0.4]{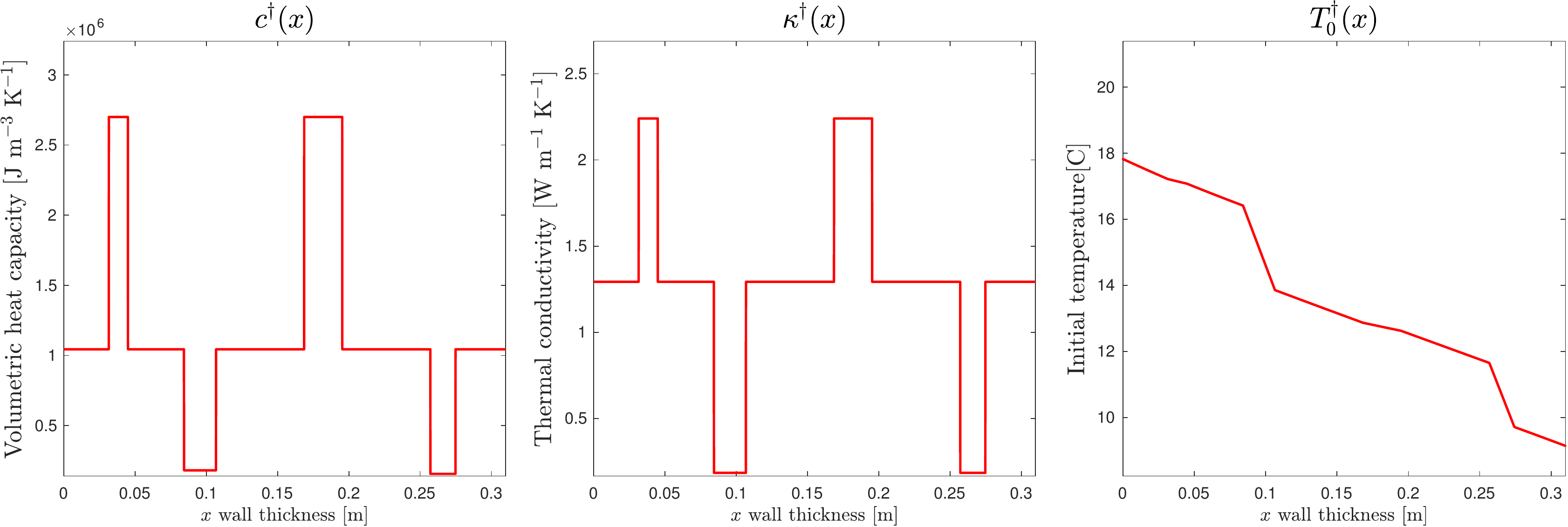}
 \caption{Synthetic experiment. Top: Synthetic internal (red) and external (black) near-air temperatures. Measurements to the left of the vertical line are used to generate $T_{0}^{\dagger}$; subsequent temperatures are used for the generation of synthetic heat fluxes. Middle: Synthetic internal (red) and external (black) surface heat flux measurements with $5\%$ of measurement noise. Bottom: True parameters $c^{\dagger}(x)$ (left), $\kappa^{\dagger}(x)$ (middle) and $T_{0}^{\dagger}(x)$ (right) }\label{Fig1}
\end{center}
\end{figure}

For the synthetic experiments that we conduct, we define a virtual wall of thickness $L=0.31$m, that we wish to monitor over a time interval [0,13.5 days]. In order to have perfect knowledge of the thermal properties of our hypothetical wall, we specify the ``true''  thermal conductivity and volumetric heat capacity, denoted by  $\kappa^{\dagger}(x)$ and $c^{\dagger}(x)$, respectively. More specifically, we define these as piece-wise constant functions with graphs displayed in Figure \ref{Fig1} (bottom-middle and bottom-left). With these definitions we aim at characterising a hypothetical wall with different constituents of various conductivities and capacitances, such as a cavity wall or a structure comprising internal insulation layers that may be overlooked during visual inspection. Following \cite{biddulph2014inferring,cibse2015environmental} we assume true surface resistances given by
\begin{eqnarray}\label{eq:19D}
R_{I}^{\dagger}= 0.13~W^{-1}m^2 K\qquad R_{E}^{\dagger}=0.04~W^{-1}m^2 K
\end{eqnarray}
For these definitions of $\kappa^{\dagger}(x)$, $c^{\dagger}(x)$, $R_{I}^{\dagger}$ and $R_{E}^{\dagger}$, the corresponding ``true" U-value and C-value calculated via equations \eqref{eq:3} are 
\begin{eqnarray}\label{eq:20B}
\mathcal{U}^{\dagger}= 1.715~W/m^2K\qquad \textrm{and}\qquad \mathcal{C}^{\dagger}= 3.55\times 10^5~J/m^2K,
\end{eqnarray}
respectively.

Internal and external near-air (virtual) temperatures are generated by adding a stochastic zero-mean Gaussian process to sine and cosine functions with prescribed amplitudes and frequencies. This process is generated with similar approaches to the ones discussed in \ref{prior}. The plots of these surface temperatures are displayed in Figure \ref{Fig1} (top). These synthetic/virtual measurements are generated over an interval of 18.75 days and discretised uniformly at every 5 minutes. The interval [-6.25 days, 0] will be used for the generation of the true initial temperature of the wall, $T_{0}^{\dagger}(x)$, as described below. Measurements defined on the subsequent interval [0, 13.5 days] will be used as the surface temperature measurements, $T_{I,1:M}^{\dagger}$ and $T_{E,1:M}^{\dagger}$, to apply and validate the proposed scheme.

\begin{figure}[htbp]
\begin{center}
\includegraphics[scale=0.45]{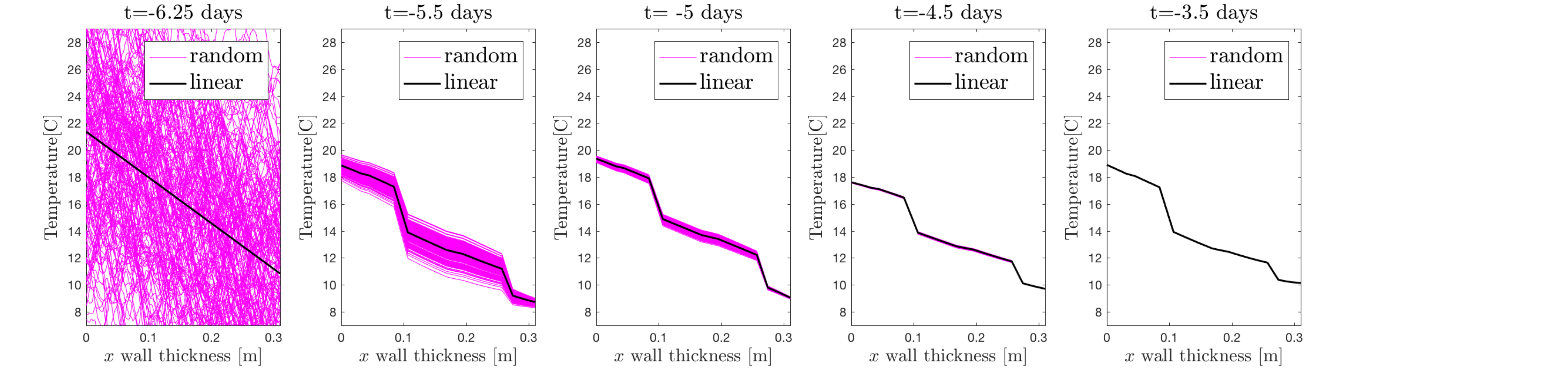}

 \caption{Synthetic experiment. Wall's internal temperature profile at different simulation times computed via the HDM for different (linear and random) initial conditions specified at $t=-6.25$ (left panel).}\label{Fig1A}
 
\end{center}
\end{figure}

In order to generate synthetic heat flux data $q_{1:m}$ on the time frame of interest [0,13.5 days], we first prescribe the ``true'' temperature profile across the wall, $T_{0}^{\dagger}(x)$ at the initial time $t=0$. This initial temperature must be consistent with the proposed virtual setting introduced above. To this end, we select a linear temperature profile that interpolates, at time $t=-6.25$ days, the surface temperature measurements introduced above (see Figure \ref{Fig1} (top)). This linear profile is then used as the initial condition for the 1D heat equation, which we run with the true parameters $\kappa^{\dagger}$, $c^{\dagger}$, $R_{I}^{\dagger}$ and $R_{E}^{\dagger}$, and the surface temperatures on the interval [-6.25 days, 0] defined earlier. The resulting internal temperature profile after 6.25 days (i.e. at $t=0$), is used as the true initial temperature profile, $T_{0}^{\dagger}(x)$, that we, in turn, use to simulate synthetic heat flux measurements over the monitoring interval [0, 13.5 days]. This parameter will be subsequently inferred together, with the parameters $\kappa^{\dagger}(x)$, $c^{\dagger}(x)$, $R_{I}^{\dagger}$ and $R_{E}^{\dagger}$, via the proposed Bayesian approach. The plot of $T_{0}^{\dagger}(x)$ is displayed in Figure \ref{Fig1} (bottom-right). It is important to mention that the selection of $T_{0}^{\dagger}(x)$ does not depend on the linear profile prescribed as the initial condition at time $t=-6.25$ days. Indeed, in Figure \ref{Fig1A} we show the internal temperature profile of the wall simulated with different initial conditions generated stochastically. Note that after less than 3 days of simulation, the temperature profile no longer depends on the initial temperature that we used to initialised the HDM. This comes as no surprise since the effect of initial condition on the HDM outputs decays exponentially fast as a consequence of the fundamental properties of the heat equation.

The ``true" internal and external surface heat fluxes on the time frame of interest [0,13.5 days] are now generated by solving the HDM defined by equations \eqref{eq:6}-\eqref{eq:9}, using the ``true" thermal properties  $\kappa^{\dagger}(x)$, $c^{\dagger}(x)$, $R_{I}^{\dagger}$ and $R_{E}^{\dagger}$, the initial temperature $T_{0}^{\dagger}(x)$, and the near-air temperatures $T_{I,1:M}^{\dagger}$ and $T_{E,1:M}^{\dagger}$ generated as described earlier. For the spatial discretisation of the HDM we use a grid of $2^9$ elements (see subsection \ref{discretisation} for details of the numerical solver that we use). Heat fluxes are discretised at a 5 minutes time interval. Measurements are grouped into 120 subintervals $[\tau_{m-1},\tau_{m}]$, each of them comprising $\beta=30$ observations times. In order to simulate a realistic scenario in which measurements are contaminated with errors, we add Gaussian noise to the true heat flux measurements with standard deviation proportional to the average relative error over each assimilation interval. More specifically, for each set of internal heat flux measurements $q_{I,m}=(q_{I,m}^{1},\dots q_{I,m}^{\beta})$ within the assimilation interval $[\tau_{m-1},\tau_{m}]$, we define the standard deviation via
\begin{eqnarray}\label{std}
\sigma_{I,m}\equiv \frac{1}{\beta}\sum_{i=1}^{\beta} \epsilon \vert q_{I,m}\vert,
\end{eqnarray}
where $\epsilon$ is the relative error of measurements. We use an analogous definition for the standard deviation of the noise added to the external heat flux measurements. A similar approach to specify the variance of measurement errors has been proposed in  \cite{gori2017inferring}. In Figure \ref{Fig1} (middle) we display the synthetic measurements of heat flux that we generate with measurement error of $5\%$ (i.e. $\epsilon=0.05$).

\subsection{Experimental data: BSRIA data collection}\label{BSRIA_data}

\begin{figure}[htbp]
\begin{center}

\includegraphics[scale=0.42]{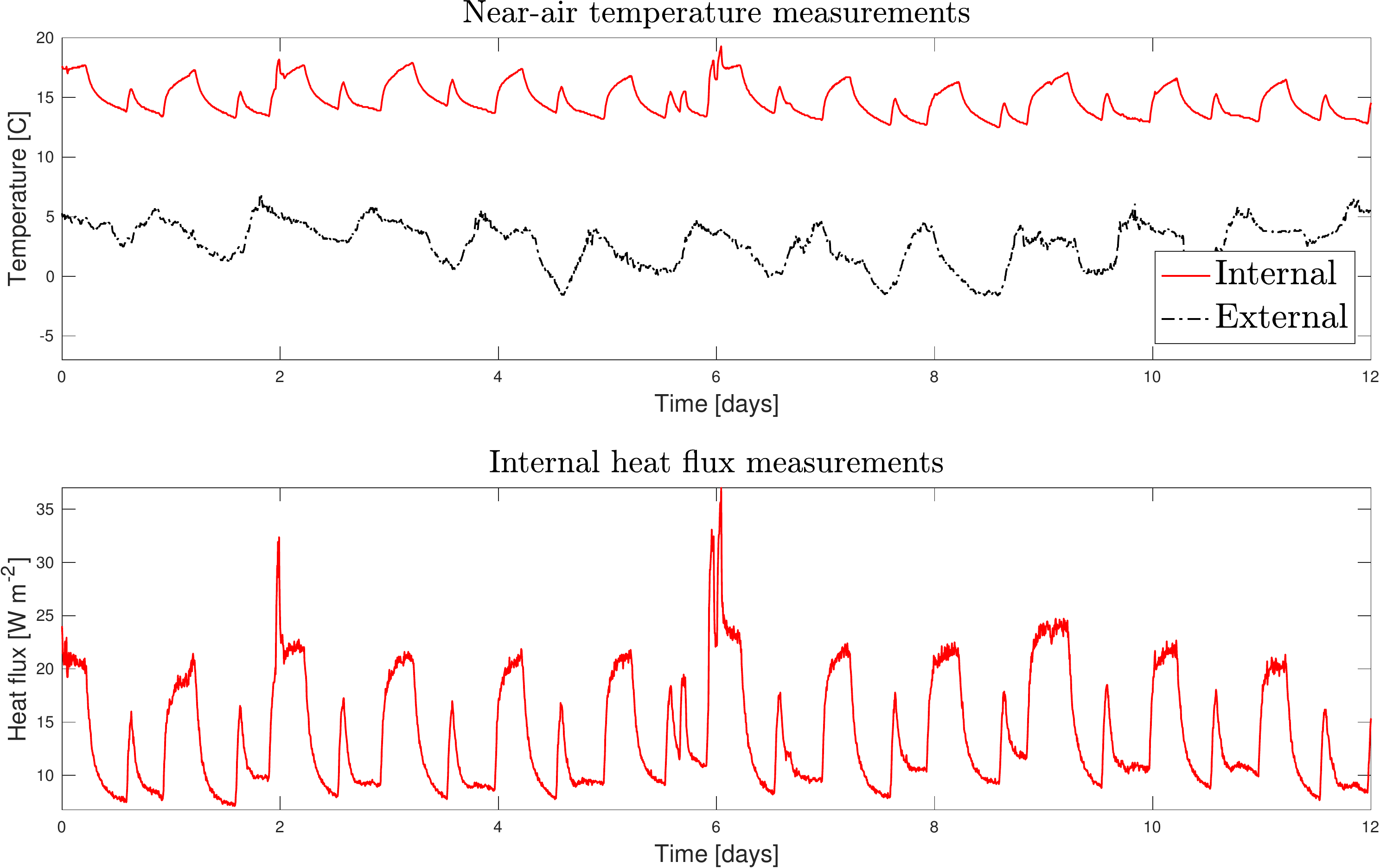}

 \caption{BSRIA experiment. Top: Internal (red) and external (black) near-air temperatures. Bottom: Interior surface heat flux measurements. }\label{Fig13}

\end{center}
\end{figure}

The proposed Bayesian approach is also applied with measurements collected by the Building Services Research and Information Association (BSRIA) as part of an investigation into the U-values of solid walls of occupied UK dwellings. These measurements were collected at 5 minutes intervals over a 14 day period during the winter of 2010 and include observations of near-air internal and external temperatures and (only) internal surface heat flux.  In Figure \ref{Fig13} we display the plots of the BSRIA data used in this section. For full details of the measurement methods and outputs, see Biddulph et al. \cite{biddulph2014inferring}. Measurements of external heat flux are not available in this case and so the approach was modified to include only surface flux from the wall's internal surface. The estimated thickness of the wall is $L=0.310\textrm{m}$, which includes a $0.01\textrm{m}$ layer of plaster on its internal surface. The first row of Table \ref{literature_data}  displays the U-values of a standard solid wall construction reported by the CIBSE guide A \cite{cibse2015environmental}. The C-value was calculated from the density and specific heat capacity values for bricks reported in the same guide. These values provide a reference that we use for the analysis of experimental data via the proposed Bayesian approach from Algorithm \ref{IF}.
\begin{table}[h!]
\centering
\caption{Internal and external surface resistances ($R_I$, $R_E$), U-value and C-value}
\label{literature_data}
\small
	\begin{tabular}{|lcccc|}
	\hline
	
	            &    $R_{I}$ ($W^{-1}m^2 K$)      &   $R_{E}$ ($W^{-1}m^2 K$) &      $\mathcal{C}$ ($\times 10^5 ~J/m^2K$)&    $\mathcal{U}$ ($W/m^2 K$)        \\
	\hline
CIBSE Guide A      &0.13                                  & 0.04  &                                         $[2.77,4.4]$                        &  [1.41,2.09]  \\
	\hline
	Syn  CI (prior)&       $[0.029,0.347]$ &	$[0.017, 0.254]$&$[1.890, 4.991]$&	$[0.917,2.182]$\\
		Syn.	CI (pos )	&    $[0.096,0.137]$&$[0.033, 0.046]$&	$[3.371,3.730]$ &  $[1.670,1.761]$\\
	Syn.		CI (pos, only internal HF)	&$[0.083,0.127]$&$[0.035,0.134]$&	$[3.066,3.576]$&  $[1.689,1.748]$ \\
	\hline
	\hline
	BSRIA	CI (prior)	&$[0.031,0.493]$&$[0.011,0.153]$&$[2.939,4.969]$&   $[0.994,2.283]$\\
	BSRIA	CI (pos)	&$[0.188,0.208]$&$[0.064,0.144]$&	$[4.162,5.129]$ &  $[1.144,1.160]$\\
	\hline
	\end{tabular}
\end{table}

\section{Validation with synthetic data}\label{synthetic}

In this Section we report the results obtained by applying the Bayesian approach encoded in Algorithm \ref{IF} to the synthetic data described in subsection \ref{synthetic_data}. For the Bayesian updating we use the REnKA scheme from Algorithm \ref{REnKA_al} with $J=10^3$ particles and tunable parameter $J_{tresh}=J/3$. Although smaller samples can be used with this algorithm, this relatively large selection of samples have been chosen in order to reduce the dependence of the algorithm with respect to the selection of the initial ensemble \cite{SMC_REnKA}. The covariance matrix of measurement errors, $\Gamma_{m}$, is constructed from the standard deviations that we use in subsection \ref{synthetic_data} to generate synthetic measurements.

\subsection{The prior uncertainty}\label{prior_num}

\begin{figure}[htbp]
\begin{center}

\includegraphics[scale=0.545]{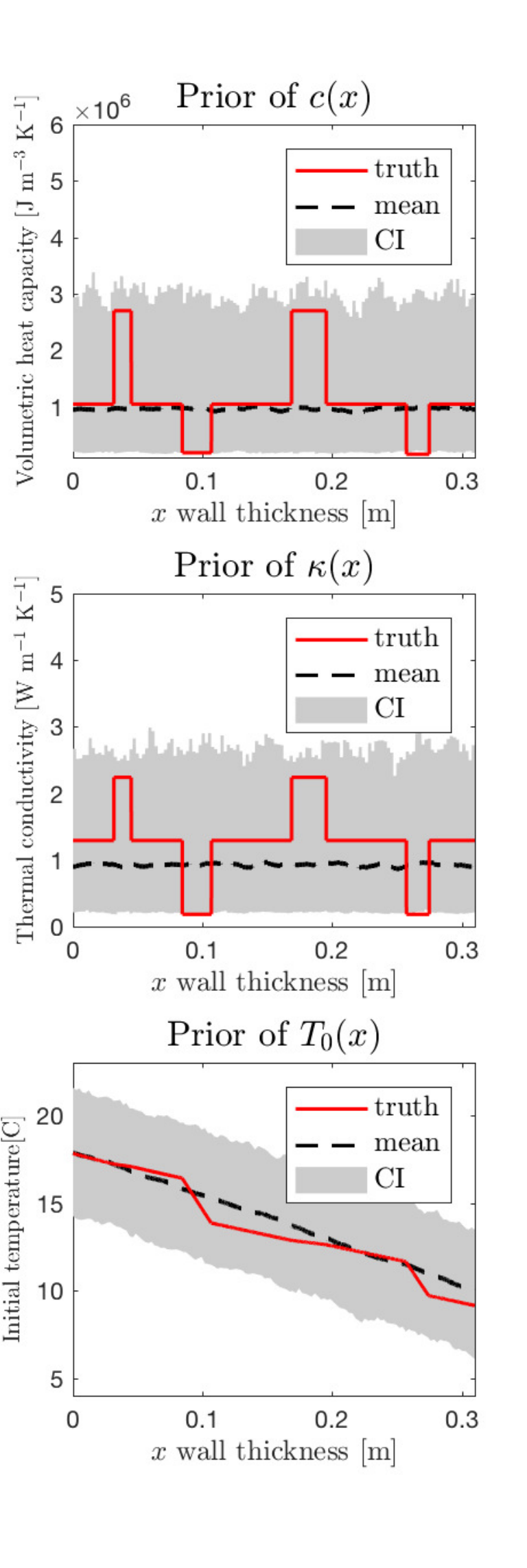}
\includegraphics[scale=0.545]{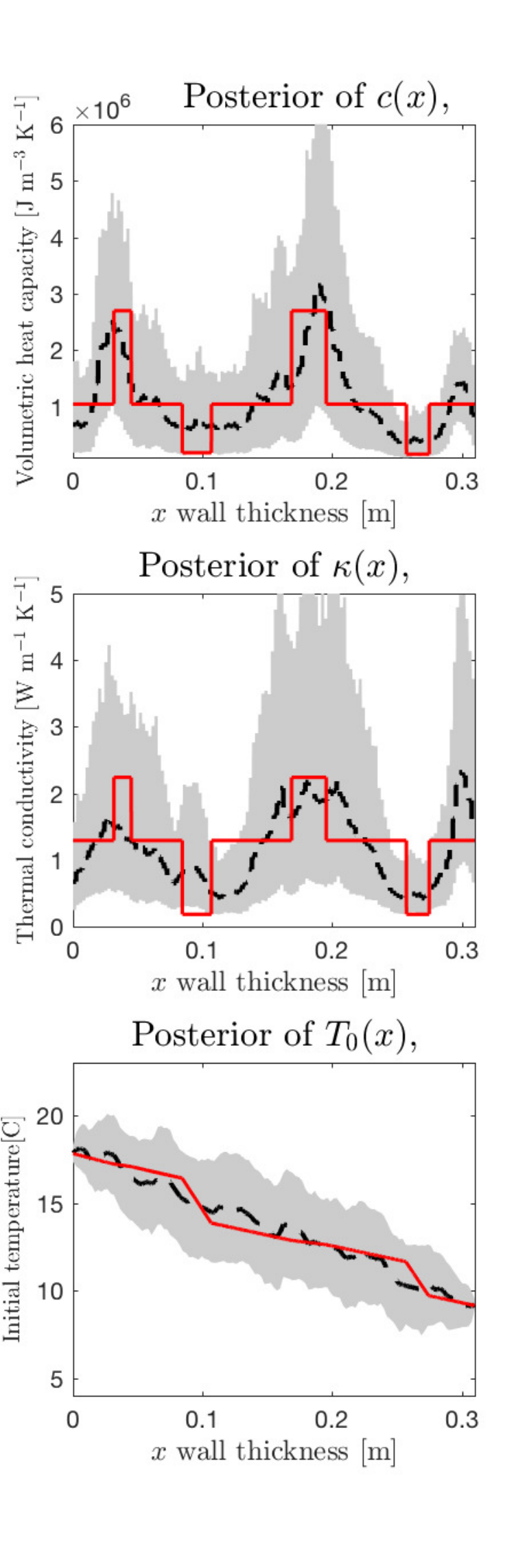}
\includegraphics[scale=0.545]{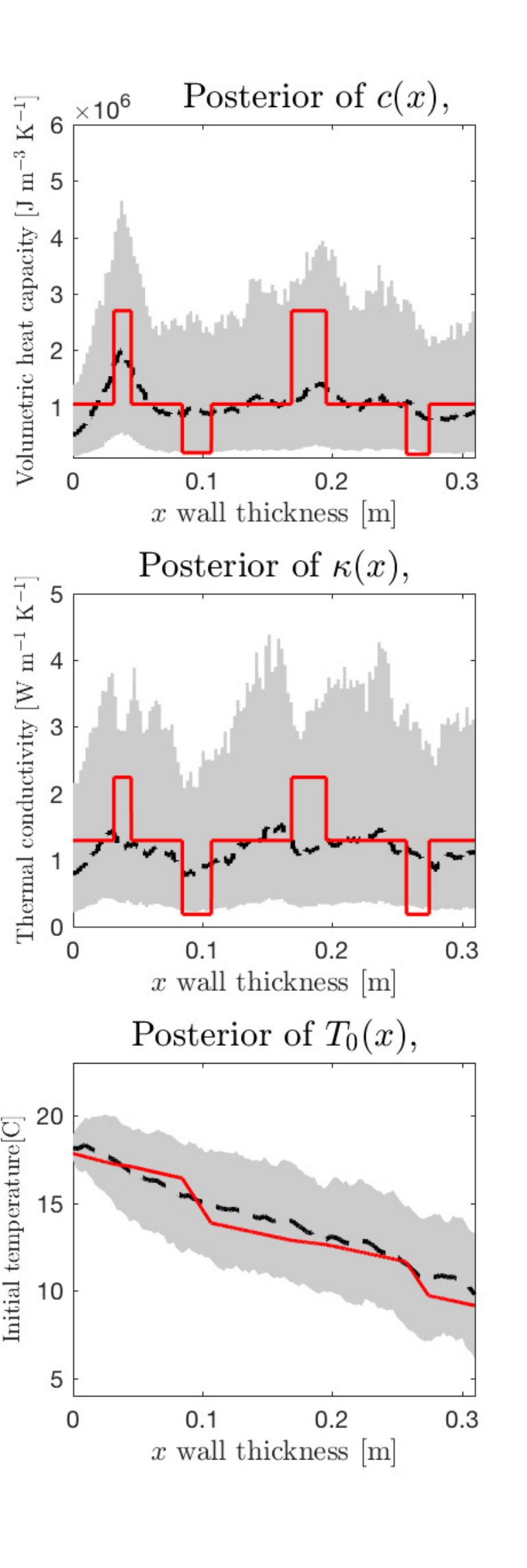}
\vspace{-3.5mm}

 \caption{Synthetic experiment. Left column: Prior mean and $95\%$ prior credible intervals of $\mathbb{P}(c)$ (top), $\mathbb{P}(\kappa)$ (middle) and $\mathbb{P}(T_{0}) $ (bottom). Middle column: Final-time (i.e. computed at $\tau_{M}=6.25$) posterior mean and posterior $95\%$ credible intervals of $\mathbb{P}(c\vert q_{1:M})$ (top), $\mathbb{P}(\kappa\vert q_{1:M})$ (middle) and $\mathbb{P}(T_{0}\vert q_{1:M}) $ (bottom), computed via Algorithm \ref{IF} using both internal and external measurements of heat flux. Right column: same as middle but computed using only synthetic measurements of internal heat flux. Solid red lines are the plots of the true $c^{\dagger}(x)$ (left), $\kappa^{\dagger}(x)$ (middle), and $T_{0}^{\dagger}(x)$ (right).}\label{Fig5}
\end{center}
\end{figure}

\begin{figure}[htbp]
\begin{center}

\includegraphics[scale=0.47]{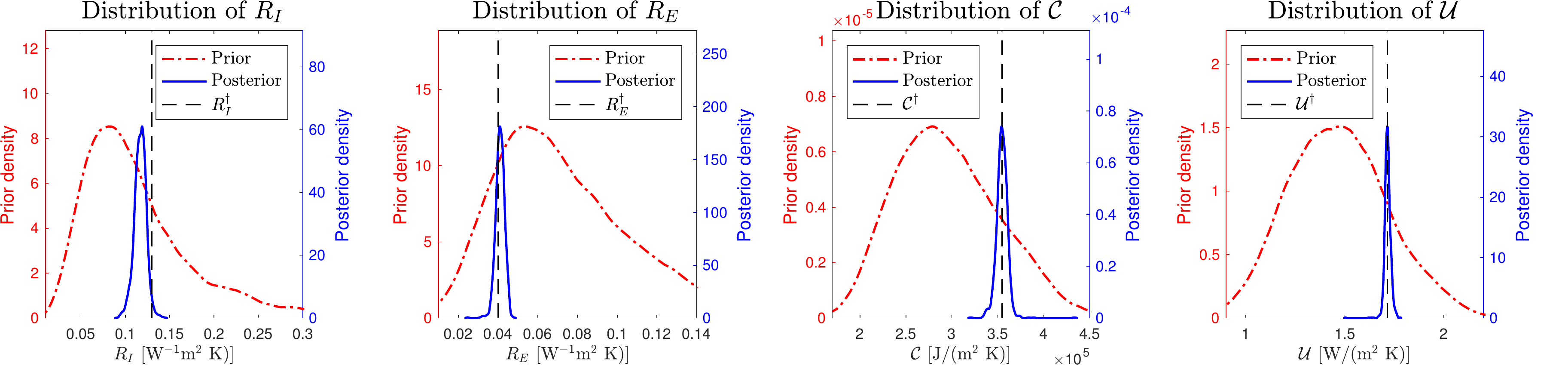}
\includegraphics[scale=0.47]{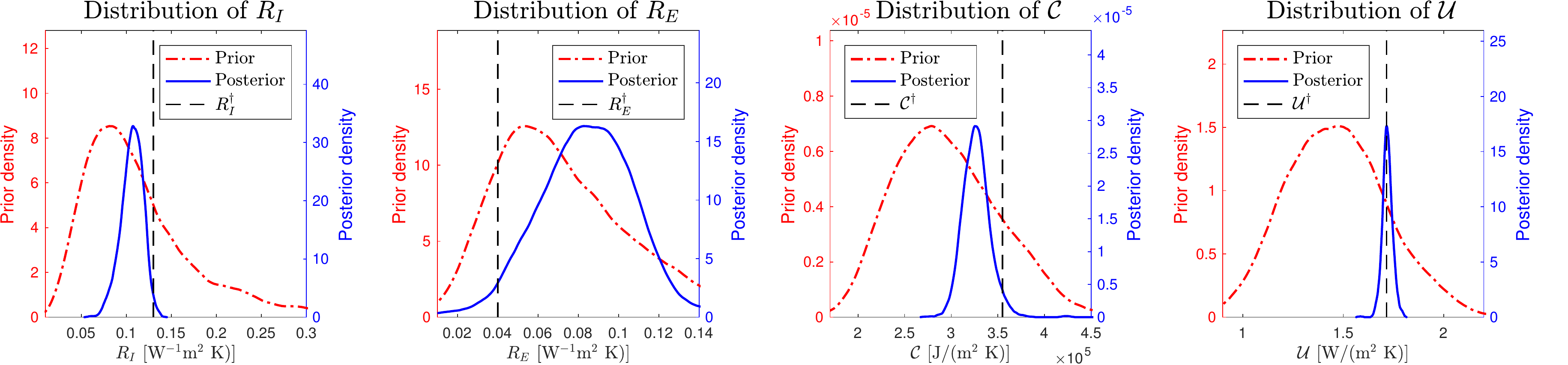}
 \caption{Synthetic experiment. Top row: Prior (dashed-red line) and final-time posterior (solid blue line) of $R_{I}$ (left), $R_{E}$ (left-middle), $\mathcal{C}$ (middle-right) and  $\mathcal{U}$ (right). The vertical line indicates the true value $R_{I}^{\dagger}$ (left), $R_{E}^{\dagger}$(left-middle) $\mathcal{C}^{\dagger}$ (middle-right) and $\mathcal{U}^{\dagger}$ (right), respectively. Bottom row: same as top row but computed using only synthetic measurements of internal heat flux. }
 \label{Fig6}
\end{center}
\end{figure}

In order to initialise Algorithm \ref{IF} we first use Algorithm \ref{prior_al} to generate $J=10^3$ samples from the prior $\mathbb{P}(u)=\mathbb{P}(\kappa)\mathbb{P}(c)\mathbb{P}(T_{0})\mathbb{P}(R_{I})\mathbb{P}(R_{E})$ that we define in \ref{prior}. Samples from $\mathbb{P}(\kappa)$, $\mathbb{P}(c)$ and $\mathbb{P}(T_{0})$ are discretised on a computational domain with $2^7$ elements. In Figure \ref{Fig5} (left column) we display the prior ensemble mean and the $95\%$ prior credible intervals computed for each point $x$ within the wall, or more precisely, at each of the nodes of the computational domain on which the samples from $\mathbb{P}(\kappa)$, $\mathbb{P}(c)$ and $\mathbb{P}(T_{0})$ are discretised. Note that our selection of the initial ensemble reflects a large variability in the internal structure of the input parameters. Furthermore, true parameters (see solid red lines in Figure \ref{Fig5}) $\kappa^{\dagger}(x)$, $c^{\dagger}(x)$, and $T^{\dagger}_{0}(x)$ are enclosed within the prior uncertainty band determined by these prior intervals. Prior (log-normal) densities of $R_{I}$ and $R_{E}$ are displayed (dotted red-line) in the left and left-middle panels of Figure \ref{Fig6} (top); vertical lines indicate the corresponding true $R_{I}^{\dagger}$ and $R_{E}^{\dagger}$ specified in equation (\ref{eq:19D}). 

With the aid of equation (\ref{eq:3}), we use samples from the priors $\mathbb{P}(\kappa)$, $\mathbb{P}(c)$, $\mathbb{P}(R_{I})$ and $\mathbb{P}(R_{E})$ to generate Monte Carlo approximations of the prior distributions, $\mathbb{P}(\mathcal{U})$ and $\mathbb{P}(\mathcal{C})$, of the U-value and the C-value, respectively. These priors are displayed (dotted red line) in the middle-right and right panels of Figure \ref{Fig6} (top). Vertical lines in these plots indicate the corresponding true values $\mathcal{U}^{\dagger}$ and $\mathcal{C}^{\dagger}$ from (\ref{eq:20B}). Equal tail $95\%$ credible intervals for the priors of $R_{I}$, $R_{E}$, $\mathcal{U}$ and $\mathcal{C}$ are reported in Table \ref{literature_data}. Note that these intervals include the range of values provided by the literature (see Table \ref{literature_data}) as well as the corresponding true values $R_{I}^{\dagger}$, $R_{E}^{\dagger}$, $\mathcal{U}^{\dagger}$ and $\mathcal{C}^{\dagger}$. 

\begin{figure}[htbp]
\begin{center}

\includegraphics[scale=0.42]{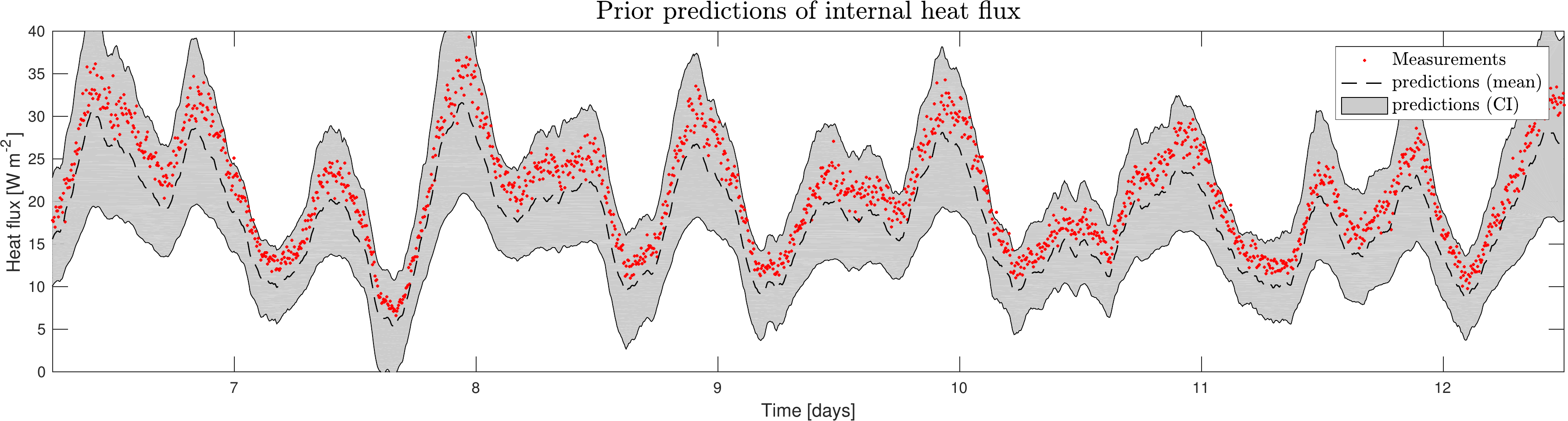}
\includegraphics[scale=0.42]{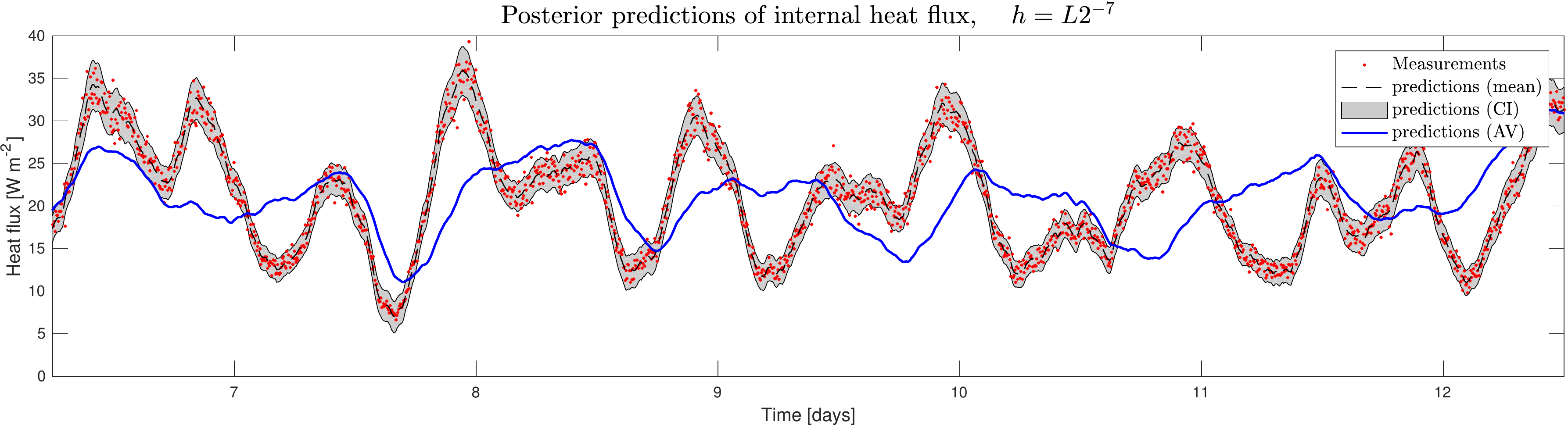}\\
\includegraphics[scale=0.42]{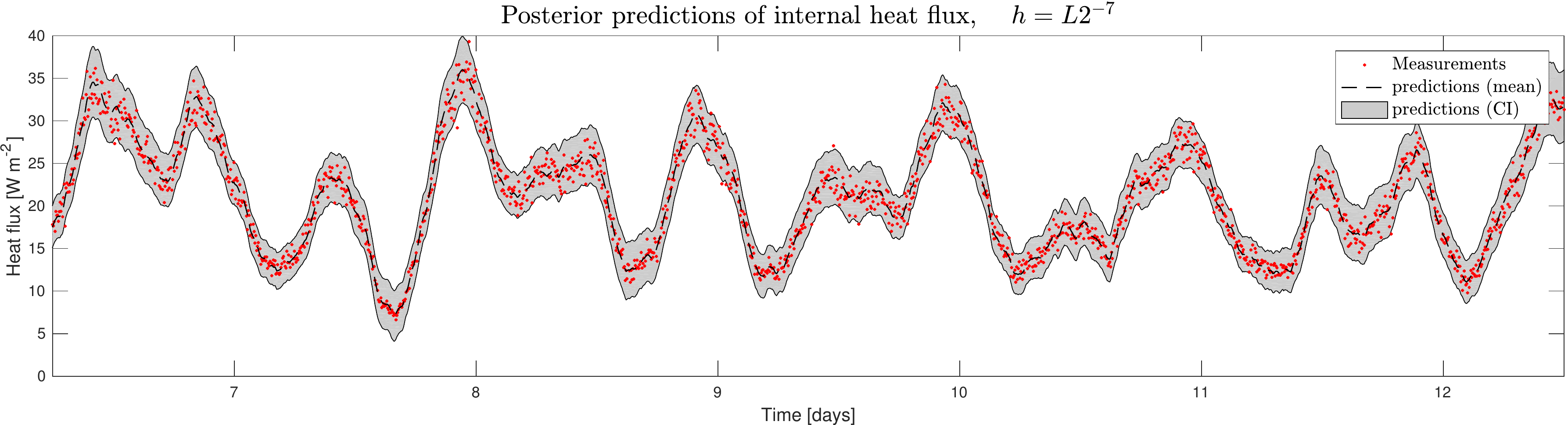} 
\caption{Synthetic experiment. Mean and $95\%$ credible intervals of internal heat flux predictions generated from the prior (top) and the posterior (middle) using the HDM with a mesh size of $h=L/2^7$; blue line is the plot of heat flux predictions computed via no-thermal-mass model (see (\ref{nomass})). Bottom: Same as middle panel but computed using only synthetic measurements of internal heat flux. Synthetic measurements are displayed with red dots. }\label{Fig3}

\end{center}
\end{figure}

\begin{figure}[htbp]
\begin{center}

\includegraphics[scale=0.42]{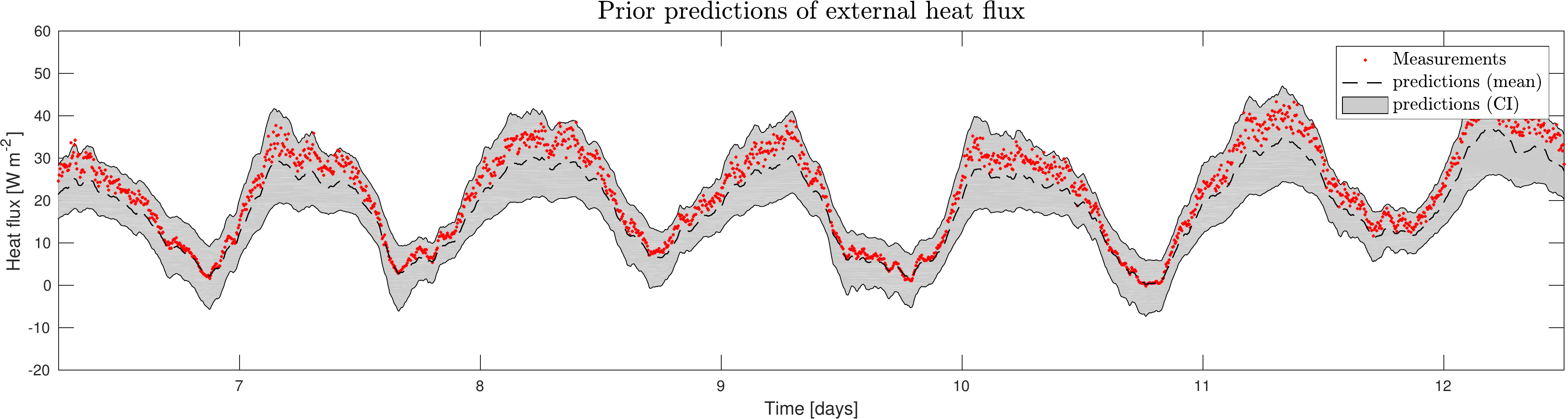}\\
\includegraphics[scale=0.42]{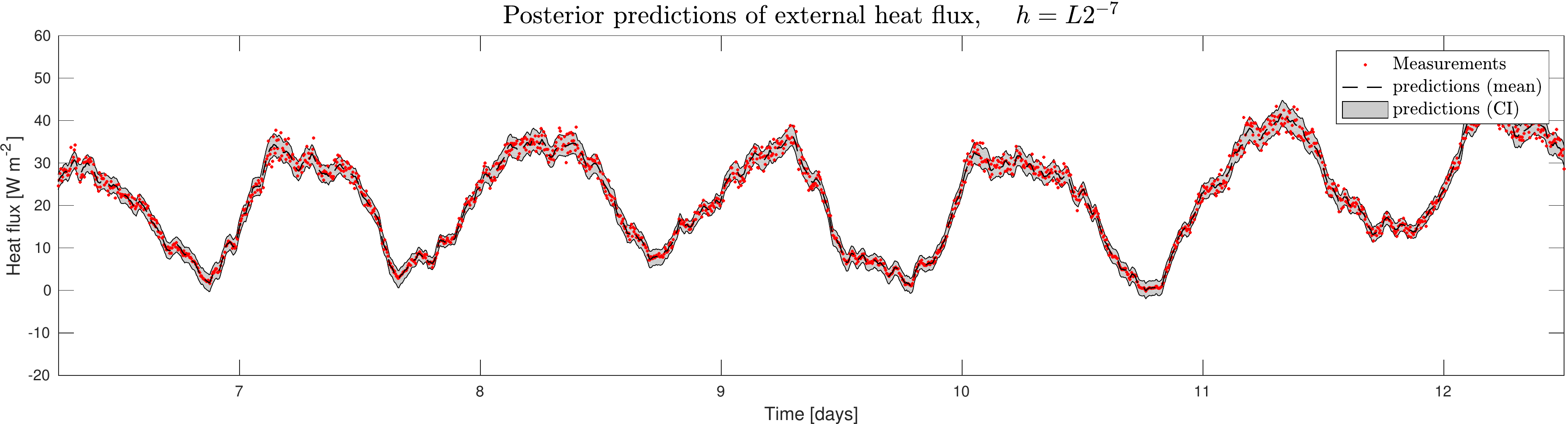}\\
\includegraphics[scale=0.42]{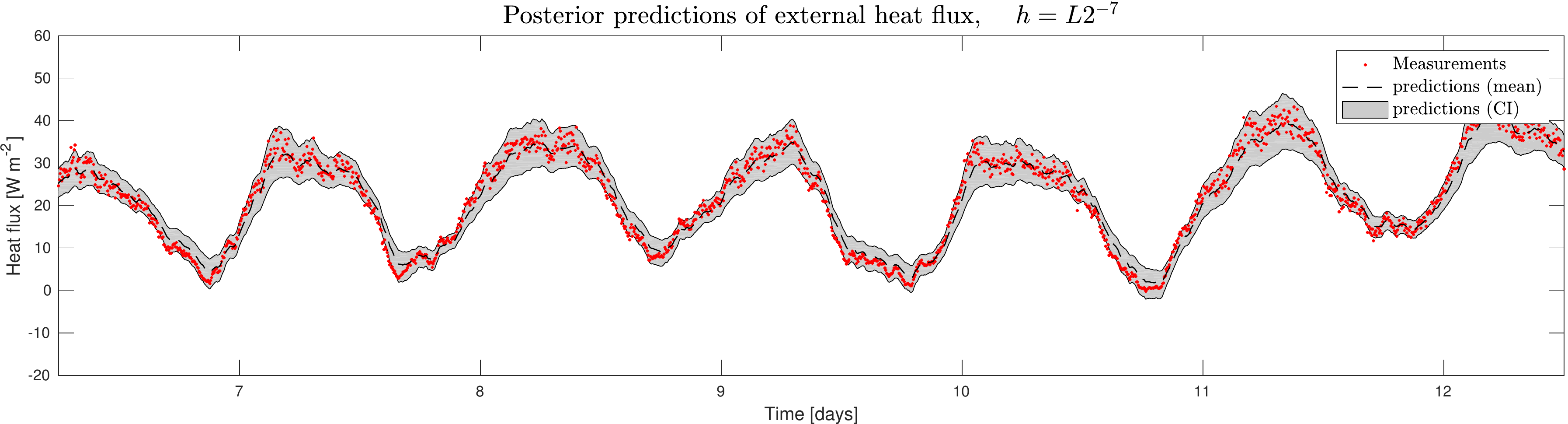} \caption{Synthetic experiment. Mean and $95\%$ credible intervals of external heat flux predictions generated from the prior (top) and the posterior (middle) using the HDM with a mesh size of $h=L/2^7$. Bottom: Same as middle panel but computed using only synthetic measurements of internal heat flux. Synthetic measurements are displayed in red dots. }\label{Fig3B}

\end{center}
\end{figure}

We also use a Monte Carlo approach to observe the effect of the prior uncertainty of the unknown parameters  $\kappa(x)$, $c(x)$, $T_{0}(x)$, $R_{I}$ and $R_{E}$ in the corresponding model predictions of surface heat fluxes. More specifically, we run the HDM (\ref{eq:6})-(\ref{eq:9}) for each of the samples from the prior and thus characterise, at each observation time within the time window $[0~\textrm{days}, 13.5~\textrm{days}]$, the prior distributions of internal and external heat flux model predictions. From the ensemble of prior predictions we compute mean and equal tail 95\% credible intervals. These statistics, visualised for the interval $[6.25~\textrm{days}, 13.5~\textrm{days}]$, can be found in the top panels of Figure \ref{Fig3} (internal heat flux) and Figure \ref{Fig3B} (external heat flux). These figures also display (red dots) the synthetic data corrupted with 5\% noise as described in subsection \ref{synthetic_data}. Note that our selection of priors gives rise to a distribution of heat flux predictions with credible intervals that capture the measurements. However, we note that (i) the mean of these distributions do not fit the data; and (ii) there is a large variability in heat flux predictions around the corresponding  mean. By means of the Bayesian approach embedded in Algorithm \ref{IF}, our objective is to reduce the prior uncertainty of the unknown parameters and, consequently, reduce the uncertainty in model predictions of surface heat fluxes.

\subsection{The posterior uncertainty}\label{posterior_num}

We apply Algorithm \ref{IF} initialised with the ensemble of draws from the prior described in the preceding subsection. We use a mesh size $h=L/2^7$ for the HDM that we discretised as discussed in subsection \ref{discretisation}. We assimilate/invert measurements within the interval $[0,6.25~\textrm{days}]$; measurements from the time window $[6.25~\textrm{days}, 13.5~\textrm{days}]$ will be used for the validation of the predictive capabilities of the proposed approach. 
At each assimilation time $\tau_{m}$, each component of the posterior ensemble $\{(\kappa_{m}^{(j)}(x),c_{m}^{(j)}(x),T_{0,m}^{(j)}(x),R_{m,I}^{(j)},R_{m,E}^{(j)})\}_{j=1}^{J}$ generated via Algorithm \ref{IF} is used to approximate the posterior mean and the posterior $95\%$ credible intervals as discussed in subsection \ref{bayesian_compu}. In particular, posterior means and credible intervals of $\mathbb{P}(c\vert q_{1:m})$, $\mathbb{P}(\kappa\vert q_{1:m})$ and $\mathbb{P}(T_{0}\vert q_{1:m})$, computed at the final assimilation time ($m=M$) $\tau_{M}= 6.25~\textrm{days}$, are shown in Figure \ref{Fig5} (middle column). We note that certain features of the spatial variability of the corresponding posterior means (see dotted black line in Figure \ref{Fig5}) are consistent with the true parameters $\kappa^{\dagger}(x)$, $c^{\dagger}(x)$, and $T^{\dagger}_{0}(x)$ (solid red line). These features reveal regions of high/low thermal conductivity and volumetric heat capacitance which are unknown a priori. More importantly, these uncertainty estimates capture the true parameters within the uncertainty measure determined by the posterior credible intervals. These results suggest that the proposed approach can be used as a non-destructive test to determine, albeit under uncertainty, regions of a wall with different thermal properties including cavities or insulators. 

Despite of the capability of the proposed approach to infer spatial variability in the thermal properties within the posterior credible intervals, we note that the posterior mean of $\kappa(x)$ and $c(x)$ were not able to accurately detect the sharp edges/discontinuities of the true thermal properties. This limitation arises from the selection of log-normal priors within the proposed methodology (see \ref{prior}). More specifically, our selection of priors for these functions enforces smoothness/regularity of the posterior estimates which do not reflect/capture the discontinuous features of the truth ($\kappa^{\dagger}(x)$ and $c^{\dagger}(x)$). Further extensions of the proposed Bayesian approach should be conducted to incorporate more realistic priors; this could be accomplished, for example, via the level-set approach of \cite{Level} to infer piece-wise constant functions.

The posterior ensembles of surface resistances $\{R_{I,M}^{(j)}\}_{j=1}^{J}$ and $\{R_{E,M}^{(j)}\}_{j=1}^{J}$, computed at the final assimilation time, are used to approximate the posterior densities $\mathbb{P}(R_{I}\vert q_{1:M})$ and $\mathbb{P}(R_{E}\vert q_{1:M})$ displayed in the left and middle-left panels of Figure \ref{Fig6} (solid blue line). As discussed in subsection \ref{bayesian_compu} the posterior ensembles $\{\mathcal{U}_{m}^{(j)}(x)\}_{j=1}^{J}$ and $\{\mathcal{C}_{m}^{(j)}(x)\}_{j=1}^{J}$ (see equation (\ref{eq:19B})) are used to approximate, at each assimilation time $\tau_{m}$, the posterior distributions of the U-value and  C-value, $\mathbb{P}(\mathcal{U}\vert q_{1:m})$ and $\mathbb{P}(\mathcal{C}\vert q_{1:m})$, respectively. The plot of these posterior distributions, computed at the final assimilation time are shown (solid blue line) in the middle-right and right panels of Figure \ref{Fig6} (top). Note that these posterior densities have a substantially smaller variance compared to the corresponding priors (see dotted red-line in the same figure) while enclosing the true values of the parameters that we aim at inferring (vertical dashed black lines). In fact, the true parameters $R_{E}^{\dagger}$, $\mathcal{C}^{\dagger}$ and $\mathcal{U}^{\dagger}$ are captured within the high probability region of these posteriors. Although the true value $R_{I}^{\dagger}$ is captured on the tail of the corresponding posterior, it is clear from these results that the proposed approach can successfully identify, under small uncertainty, these thermophysical properties of the wall. A quantitative assessment of these distributions is displayed in Table \ref{literature_data} (3rd row), where we report approximations for the 99\% credible intervals of the final time posteriors of $R_I$, $R_{E}$, $\mathcal{C}$ and $\mathcal{U}$. These intervals not only contain the true values stated in (\ref{eq:19D}) -(\ref{eq:20B}), but also have lengths which are  $ 87.08\%$,  $   94.59\%$, $   88.43\%$ and $92.78\%$ smaller than the prior credible intervals (second row of Table \ref{literature_data}). Our results indicate that, in spite of the large uncertainty in the posterior estimates of $\kappa(x)$, $c(x)$, and $T_{0}(x)$ (see left-middle column of Figure \ref{Fig5}), the posteriors of the $\mathcal{U}$ and $\mathcal{C}$ enable us to identify these effective properties under small uncertainty.

\subsubsection{Sequential posterior estimates.}

\begin{figure}[htbp]
\begin{center}

\includegraphics[scale=0.65]{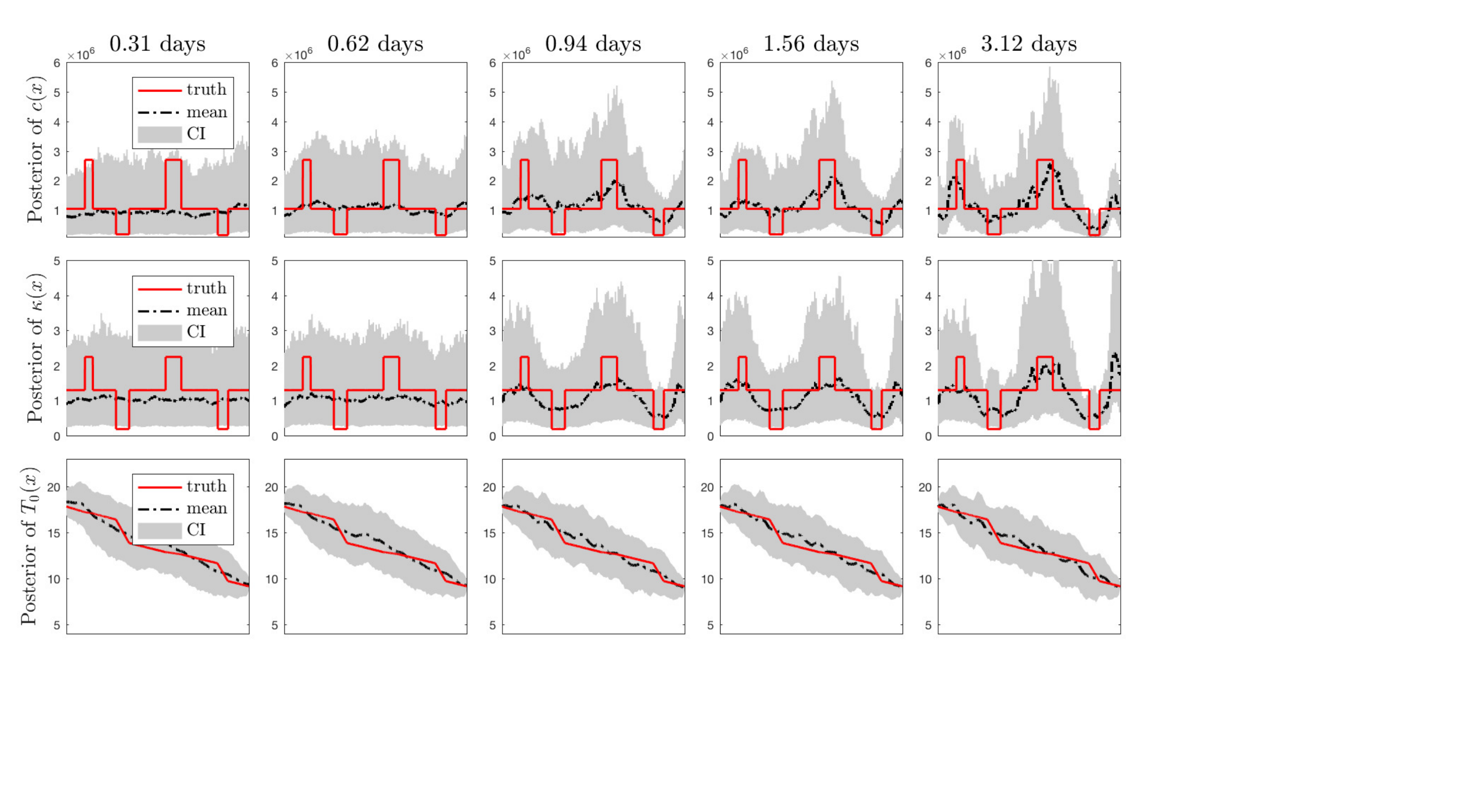}
\vspace{-3.0cm}
 \caption{Synthetic experiment. Posterior mean and $95\%$ posterior credible intervals of $\mathbb{P}(c\vert q_{1:m})$ (top), $\mathbb{P}(\kappa\vert q_{1:m})$ (middle) and $\mathbb{P}(T_{0}\vert q_{1:m}) $ (bottom), computed at different assimilation times $\tau_{m}$'s. Solid red line are the plots of the true $c^{\dagger}(x)$ (top), $\kappa^{\dagger}(x)$ (middle), and $T_{0}^{\dagger}(x)$ (bottom).}\label{Fig5B}
\end{center}
\end{figure}

While the discussion of the preceding paragraph involves posterior distributions of the unknown parameters computed at the final assimilation time $\tau_{M}$, the proposed Bayesian approach enables to monitor these posteriors at each assimilation time $\tau_{m}$ during the measurement campaign. As we now show, this information can be used to assess whether a sufficient number of measurements have been assimilated to achieve stable estimates of these properties. In Figure \ref{Fig5B} we display the posterior mean and credible intervals for $c(x)$, $\kappa(x)$ and $T_{0}(x)$ obtained at some of the intermediate assimilation times $\tau_{m}'s$. We note that, the posterior uncertainty of the initial temperature of the wall, $T_{0}(x)$, is substantially decreased during the first assimilation interval (i.e. $[0, 0.31~\textrm{days}$)) only  at the internal and external surfaces of the wall. No significant changes are observed when further measurements are assimilated. This is expected since, as stated earlier, the effect of the initial condition of the HDM outputs decays exponentially fast. In contrast, a more gradual reduction in the posterior uncertainties of the thermal properties $\kappa(x)$ and $c(x)$ is observed as more measurements are assimilated. However, from Figure \ref{Fig5B} we note that, after the initial 3 days of the measurement campaign, these posterior measures of uncertainty do not display substantial changes, suggesting that posteriors uncertainties are not further informed by subsequent measurements.

\begin{figure}[htbp]
\begin{center}
\includegraphics[scale=0.45]{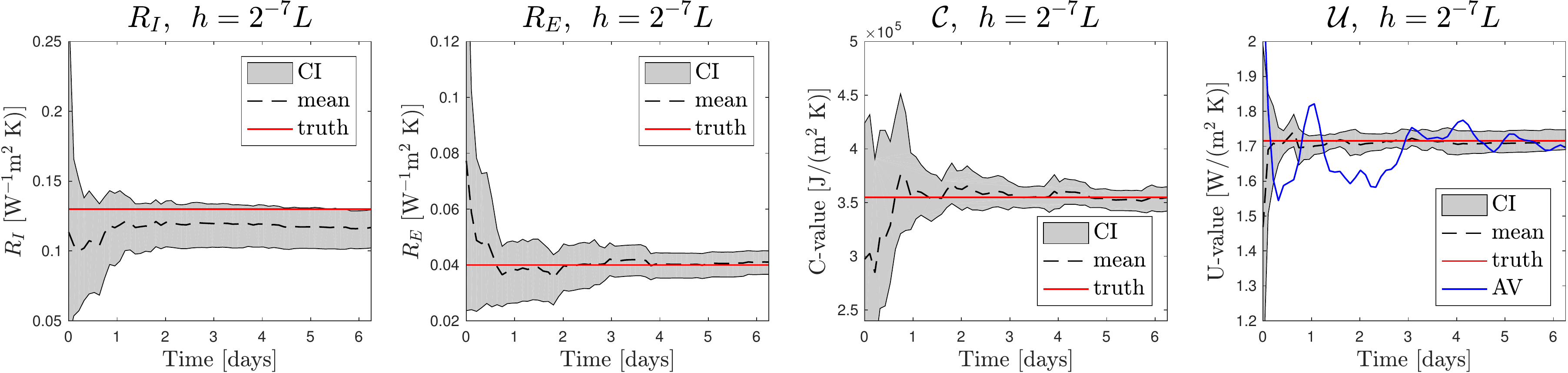}
\includegraphics[scale=0.45]{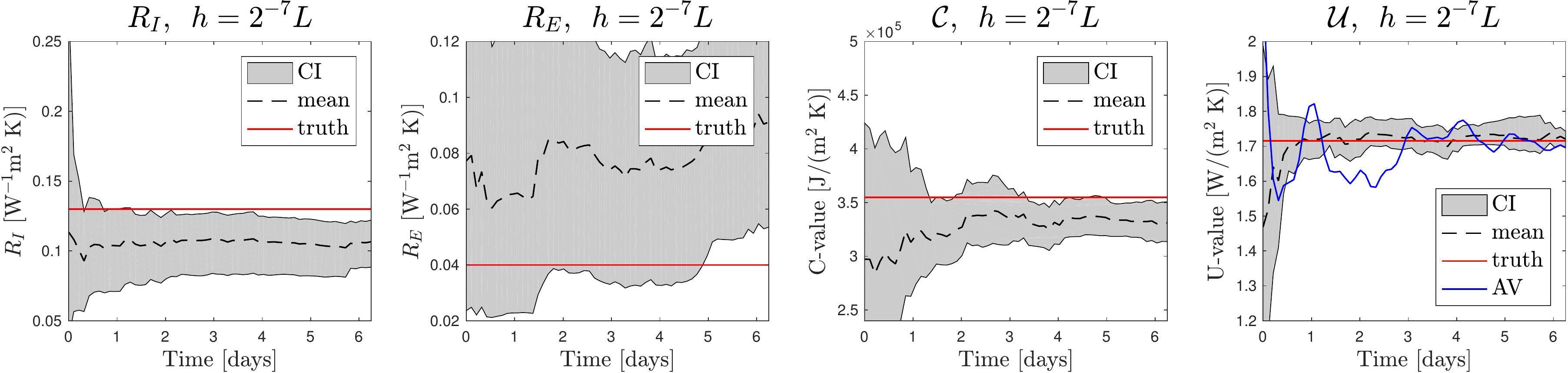}

 \caption{Synthetic experiment. Top row: Sequential posterior mean and posterior $95\%$ credible interval of $\mathbb{P}(R_{I}\vert q_{1:m}) $(left), $\mathbb{P}(R_{E}\vert q_{1:m}) $ (left-middle)$, \mathbb{P}(\mathcal{C}\vert q_{1:m}) $ (middle-right) and $\mathbb{P}(\mathcal{U}\vert q_{1:m}) $ (right), approximated at every assimilation time $\tau_{m}$. The solid red horizontal lines denote the true values $R_{I}^{\dagger}$ (left), $R_{E}^{\dagger}$ (left-middle), $\mathcal{C}^{\dagger}$ (middle-right) and $\mathcal{U}^{\dagger}$ (right). The blue line in the right panel is the U-value computed via the average method. Bottom: Same as top panel but computed using only synthetic measurements of internal heat flux. }\label{Fig8}
\end{center}
\end{figure}

In order to qualitatively monitor the stability of inferred parameters as more measurements are assimilated, in Figure \ref{Fig8} (top) we plot the posterior means and 95\% equal tail credible intervals, at each assimilation time $\tau_{m}$, of the posteriors of $R_{I}$, $R_{E}$,  $\mathcal{C}$ and $\mathcal{U}$. We note that these statistics are fully stabilised after 3 days; the posterior credible intervals are substantially reduced, and the posterior means (dashed line)  of the distributions of $R_{E}$,  $\mathcal{C}$ and $\mathcal{U}$ provide a very good approximation of the true values $R_{E}^{\dagger}$, $\mathcal{C}^{\dagger}$ and $\mathcal{U}^{\dagger}$ (red lines), respectively. In the right panel of Figure \ref{Fig8} (top) we have also included (solid blue line) the running estimate of the U-value ($\mathcal{U}$) obtained via the average method discussed in Section \ref{Intro} (see equation \eqref{average_method}). We can clearly notice that the proposed approach offers a much faster stabilisation and accuracy for the estimation of the U-value. Further evidence of the rapid stability of the sequential posterior uncertainties is provided in Table \ref{Table1}, where, for each of these variables, we display (i) posterior mean (resp. prior mean for $\tau_{0}=0$), (ii) relative error of the posterior mean with respect to the truth and (iii) coefficient of variation (CoV) defined as the ratio between the standard deviation and the mean. From these results we observe that after only 1 day of assimilation, the error with respect to the truth of $\mathcal{U}$ (resp. $\mathcal{C}$) decays from the prior value (at $\tau_{0}=0$) of $14.387\%$ (resp. 16.238\%) to less than 1\% (resp. 2\%). The uncertainty of $\mathcal{U}$ (resp. $\mathcal{C}$), in terms of the CoV, is reduced from 16.889\% (resp 19.627\%) to 0.84\% (resp. 4.01\%). Although the subsequent assimilation of measurements results in a further decrease of the CoV to less than 1\%, it is clear than 1 day of measurements provide enough information to estimate the U-value with high degree of accuracy. In contrast, we observe severe fluctuations in the estimate computed via the average method and errors which are larger than 5\% even after 2 days of the measurement campaign (see last row of Table \ref{Table1}). 

\subsubsection{Predictive capabilities.}

We now demonstrate how our uncertainty estimates of inferred parameters can be used to characterise the thermal performance of our virtual wall. Our aim is to reproduce, within a high-confident uncertainty estimate, (un-assimilated) measurements of heat flux within the validation time window $(6.25~\textrm{days}, 13.50~\textrm{days}]$. To this end, we use our probabilistic estimates of the unknown parameters $R_{I}$, $R_{E}$, $c(x)$, $\kappa(x)$ and $T_{0}(x)$, computed at the final assimilation time $\tau_{60}= 6.25~ \textrm{days}$ (i.e given the assimilation of measurements collected within the interval $(0,\tau_{60}]$) to compute the predictive distributions of internal and external surface heat flux, over the interval $(6.25~\textrm{days}, 13.50~\textrm{days}]$. As discussed in subsection \ref{bayesian_compu}, predictive distributions of heat flux are approximated from the ensemble of model predictions obtained by running the HDM for each ensemble member of inferred parameters. In the middle panels of Figure \ref{Fig3} and Figure \ref{Fig3B} we show the mean and 95\% equal tail credible intervals from these distributions. A good visual agreement to the measurements (red dots) is provided by the mean of the predictive distributions (black line). For comparison purposes, in Figure \ref{Fig3} (middle) we have also included the internal heat flux predictions (solid blue line), at each observation time $t_{m}$, that we compute via a direct calculation with the so-called no-thermal mass  model \cite{biddulph2014inferring} defined by
\begin{eqnarray}\label{nomass}
Q_{I,m}^{AV}=(T_{I,m}-T_{E,m})\mathcal{U}_{AV}
\end{eqnarray}
where, as before, $\mathcal{U}_{AV}$, is the U-value computed via the average method (see equation (\ref{average_method}). Figure \ref{Fig3} clearly shows that a no-thermal mass model is not able to reproduce the thermal performance of the wall. 

In order to qualitatively assess the goodness of fit of the mean of the predictive distributions we consider a chi-squared statistic defined by \cite{Tarantola:2004:IPT:1062404}
\begin{equation}\label{SME}
\chi_{\beta}^2=\frac{1}{M_{T}}\sum_{m=1}^{M_{T}}\frac{(\nu_{\beta,m}-\overline{Q}_{\beta,m})^{2}}{\sigma_{\beta,m}^2},
\end{equation}
where $\overline{Q}_{\beta,m}$ denotes the mean of the predictive distributions of internal ($\beta=I)$ and external ($\beta=E$) heat fluxes computed at each measurement time, $t_{m}$, within the predictive time window $(6.25~\textrm{days}, 13.50~\textrm{days}]$, and $\nu_{\beta,m}$ denotes the corresponding heat flux measurement.  In expression (\ref{SME}), $\sigma_{\beta,m}$ is the standard deviation of the measurement error computed as described in subsection \ref{synthetic_data} and $M_{T}=1800$ is the total number of measurements within the predictive time window. The values of these chi-square statistics, displayed in Table \ref{Table_new} (second column, $h=L2^{-7}$), are close to one thereby suggesting that the mean of the predictive distributions provide a very good fit to the observations. We can also observe from Figure \ref{Fig3} and Figure \ref{Fig3B} (middle panels) that most measurements fall within the credible intervals (grey area), indicating that these intervals capture subsequent (un-assimilated) measurements of surface heat fluxes within the predictive distributions provided by the proposed Bayesian technique.

\begin{table}
\centering                                                                                           
\begin{tabular}{|c|c|c|c|c|c|c|c|c|c|c|}                                                                                                                   
\hline                                                                                                              
time (days) & 0.000 & 0.527 & 1.053 & 1.580 & 2.107 & 2.633 & 3.160 & 3.687 & 4.213 & 4.740 \\                      
\hline                                                                                                              
$R_{I}$ (mean) & 0.113 & 0.106 & 0.119 & 0.118 & 0.121 & 0.120 & 0.120 & 0.120 & 0.119 & 0.117 \\                   
                                                 
$R_{I}$ (rel. err. \%) & 12.712 & 18.508 & 8.689 & 9.344 & 7.084 & 7.535 & 7.822 & 8.071 & 8.583 & 9.840 \\         

$R_{I}$ CoV (\%) & 54.128 & 16.477 & 10.421 & 7.688 & 6.952 & 6.773 & 6.738 & 6.597 & 6.380 & 6.287 \\              
\hline                                                                                                              
$R_{E}$ (mean) & 0.077 & 0.044 & 0.038 & 0.038 & 0.039 & 0.041 & 0.041 & 0.042 & 0.040 & 0.040 \\                   

$R_{E}$ (rel. err. \%) & 93.123 & 10.592 & 4.606 & 5.490 & 1.440 & 3.037 & 3.723 & 3.950 & 0.499 & 0.528 \\         

$R_{E}$ CoV (\%) & 53.844 & 22.993 & 14.466 & 13.747 & 10.487 & 9.993 & 7.026 & 6.677 & 5.915 & 5.884 \\            
\hline                                                                                                              
$\mathcal{C}~\times 10^5$(mean) & 2.973 & 3.287 & 3.608 & 3.540 & 3.654 & 3.600 & 3.552 & 3.548 & 3.598 & 3.539 \\  

$\mathcal{C}$ (rel. err. \%) & 16.238 & 7.396 & 1.635 & 0.271 & 2.926 & 1.426 & 0.069 & 0.047 & 1.346 & 0.313 \\    

$\mathcal{C}$ CoV (\%) & 19.627 & 10.389 & 4.075 & 2.287 & 2.683 & 2.157 & 1.451 & 1.562 & 1.771 & 1.498 \\         
\hline                                                                                                              
$\mathcal{U}$ (mean) & 1.468 & 1.726 & 1.701 & 1.718 & 1.707 & 1.713 & 1.718 & 1.714 & 1.707 & 1.710 \\             

$\mathcal{U}$ (rel.err. \%) & 14.387 & 0.608 & 0.851 & 0.192 & 0.447 & 0.123 & 0.189 & 0.061 & 0.485 & 0.319 \\     

$\mathcal{U}$ CoV (\%) & 16.889 & 1.056 & 0.836 & 0.623 & 0.837 & 0.708 & 0.749 & 1.298 & 1.349 & 1.180 \\          

$\mathcal{U}_{av}$ & 2.191 & 1.596 & 1.821 & 1.628 & 1.618 & 1.634 & 1.744 & 1.721 & 1.762 & 1.683 \\               

$\mathcal{U}_{av}$ rel.err. (\%) & 27.755 & 6.957 & 6.200 & 5.087 & 5.663 & 4.732 & 1.709 & 0.334 & 2.708 & 1.852 \\
\hline                                                                                                              
\end{tabular}  \caption{Synthetic Experiment. Sequential posterior estimates of $R_{I}$, $R_{E}$, $\mathcal{U}$ and $\mathcal{C}$ obtained during the first 4.7 days of the measurement campaign.} 
\label{Table1} 
\end{table}                                                                                    

\begin{table}
\footnotesize               
\centering                                                             
\begin{tabular}{|c|c|c|c|c|c|c|c|c|c|}                                     
\hline                                                                 
&$h=L/2^{-7}$ &$h=L2^{-6}$ &$h=L2^{-5}$ &$h=L2^{-4}$ & $h=L2^{-3}$ &$h=L2^{-2}$&$h=L2^{-1}$\\
\hline              
(syn.) $\chi_{I}^2$  & 1.042 & 1.047 & 1.044 & 1.042 & 1.045 & 1.039 & 1.137 \\   
(syn. )$\chi_{E}^2$  & 0.998 & 1.000 & 1.422 & 1.630 & 1.439 & 1.014 & 3.774 \\
(syn.) $AIS_{I}$  & 5.365 & 5.333 & 5.600 & 5.466 & 13.279 & 24.790 & 32.765 \\
(syn.) $AIS_{E}$  & 6.383 & 6.564 & 6.775 & 6.637 & 15.233 & 26.875 & 34.457 \\                  
  \hline                                                                 
\hline  
(syn.  only internal HF) $\chi_{I}^2$ & 1.046 & 1.082 & 1.101 & 1.067 & 1.085 & 1.113 & 1.050 \\        
(syn.  only internal HF) $\chi_{E}^2$ & 58.259 & 38.726 & 55.544 & 49.166 & 71.208 & 12.340 & 120.711                                                    \\ (syn. only internal HF) $AIS_{I}$ & 5.942 & 6.211 & 6.676 & 6.825 & 7.210 & 7.541 & 11.099 \\                                                      
(syn. only internal HF) $AIS_{E}$ & 9.305 & 8.528 & 8.708 & 9.054 & 9.720 & 10.971 & 45.962 \\      
         
\hline                                                      
\hline                                                      
(BSRIA)  $\chi_{I}^2$ & 1.008 & 0.973 & 0.975 & 1.063 & 1.013 & 1.030 & 1.666 \\ 
(BSRIA) $AIS_{I}$ & 3.683 & 3.976 & 4.102 & 4.157 & 4.047 & 4.218 & 12.963 \\                                 
\hline             
\end{tabular}                                                          
\caption{Chi-squared and average interval score of the predictive distributions of internal and external surface heat fluxes computed for different choices of mesh size $h$ for the discretisation of the HDM.}  
\label{Table_new}  
\end{table}            
                             
\normalsize
As we discussed in subsection \ref{synthetic_data}, synthetic measurements of internal and external heat flux used for the previous synthetic experiments are contaminated with $5\%$ measurement error. Although this selection of error size is informed by the precision of standard heat flux meters, it is important to emphasize that the size of measurement errors has an effect on the quality of these uncertainty estimates. We have conducted further experiments (not shown) that indicate that measurement errors of $1\%$ yields only a marginal improvement over the uncertainty estimates of $\kappa(x)$, $c(x)$ and $T_{0}(x)$ that we discussed in the preceding subsection (with errors of $5\%$). The effect of reducing these errors is more noticeable in the posterior uncertainties of $R_{I}$, $R_{E}$, $\mathcal{C}$ and $\mathcal{U}$ for which smaller measurement errors not only yields smaller posterior variances but also more concentrated around the true values. On the other hand, even when larger measurement errors are introduced (e.g. 10\%), posterior estimates still provide  an accurate identification of the U-value and the C-value. These experiments (not reported) indicate that the proposed method is robust with respect to realistic measurements errors associated with heat flux measurements.

\subsection{The effect of the spatial discretisation of the wall.}\label{mesh_impact}

Let us recall that the results of the previous subsection are obtained by using a fixed (Finite Element) spatial discretisation of the HDM  (\ref{eq:6})-(\ref{eq:9}) with mesh size $h=L/2^7$. In this subsection we study the effect of $h$ on the uncertainty estimates of thermal properties obtained via the Bayesian proposed approach. The selection of mesh size $h$ not only determines the accuracy of the HDM that defines the parameter-to-output maps $\{\cG_{m}\}_{m=1}^{M}$, but also the intrinsic characterisation of the thermal properties. A large mesh size implies a coarse mesh and thus thermal properties characterised only at a few elements. Such a coarse (low-dimensional) characterisation may not be suitable for inferring properties with highly-heterogenous features. In contrast, a small choice of $h$ characterises thermal properties on a very dense computational mesh which could enable us, via the proposed approach, to infer small-scale features of the internal structure of the wall. However, the smaller the mesh size the higher the computational cost of HDM which needs to be solved, at every assimilation time, for each ensemble member of the unknown parameters within the iterative scheme embedded in Algorithm \ref{REnKA_al}. In order to further understand the effect of mesh size $h$ on the accuracy of the posterior estimates, in this subsection we apply Algorithm \ref{IF} with the same synthetic measurements produced in section \ref{synthetic_data}, but with different choices of mesh size $h$. In particular, we are interested in studying the effect of coarse meshes which are analogous to lumped thermal mass models such as those used in existing Bayesian approaches \cite{biddulph2014inferring,gori2017inferring} to characterise the thermal performance of walls. 

\begin{figure}[htbp]
\begin{center}
\includegraphics[scale=0.6]{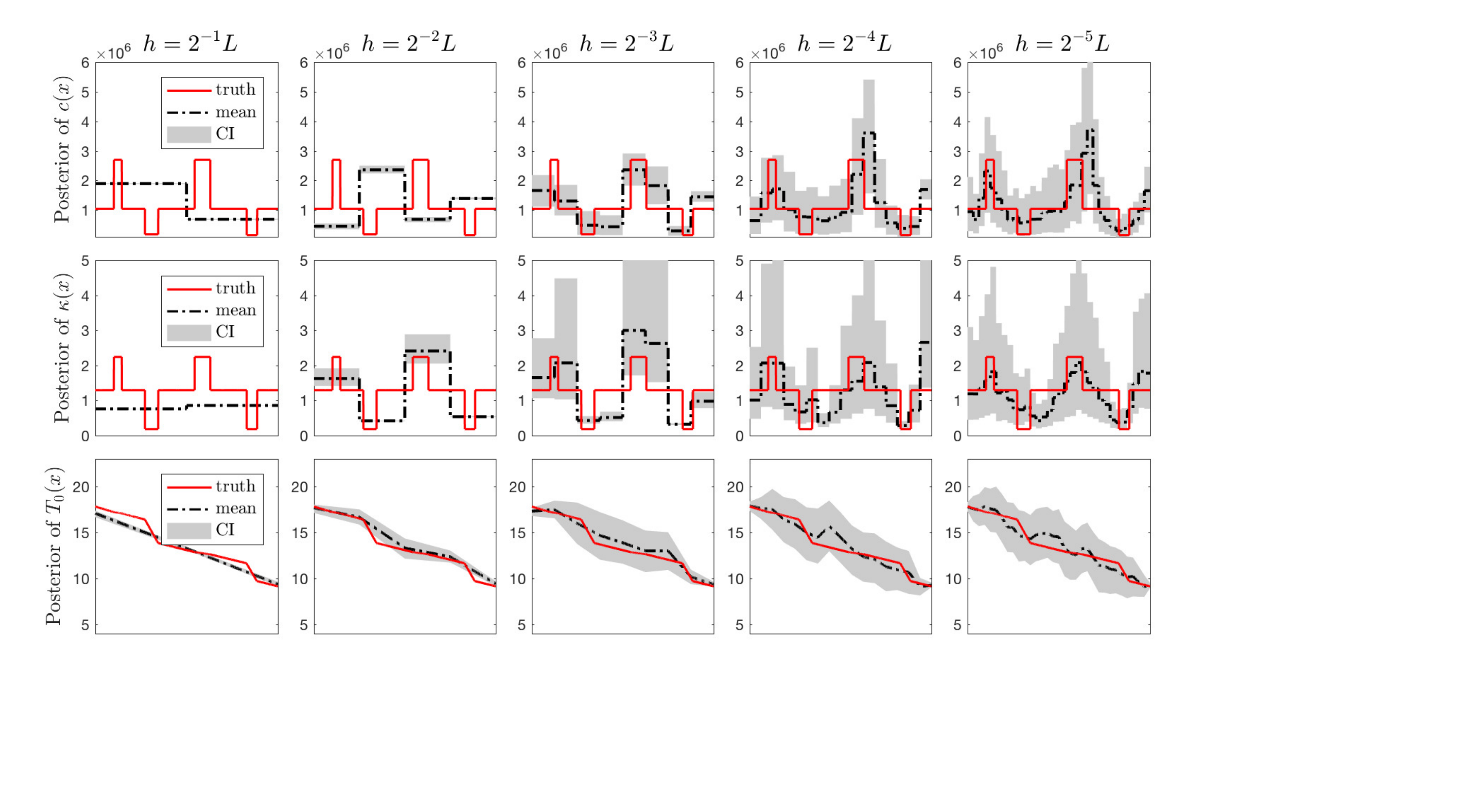}
\vspace{-3cm}
 \caption{Synthetic experiment. Final-time posterior mean and posterior $95\%$ credible intervals of $\mathbb{P}(c\vert q_{1:M})$ (top), $\mathbb{P}(\kappa\vert q_{1:m})$ (middle) and $\mathbb{P}(T_{0}\vert q_{1:M}) $ (bottom), computed using mesh size: (from left to right) $h=L/2^{1},L/2^{2},L/2^3,L/2^4,L/2^5$. Solid red line indicates the true $c^{\dagger}(x)$ (top), $\kappa^{\dagger}(x)$ (middle), and $T_{0}^{\dagger}(x)$ (bottom).}\label{Fig10}

\end{center}
\end{figure}

In Figure \ref{Fig10} we displayed the posterior mean and credible intervals for the variables $c(x)$, $\kappa(x)$ and $T_{0}(x)$, computed at the final assimilation time $\tau_{M}$, obtained via Algorithm \ref{IF} with different mesh sizes ($h=L/2^{i},~~ i=1,\dots, 5$) for the HDM, and with (synthetic) heat flux measurements contaminated with $5\%$ errors.  As we discuss in \ref{discretisation}, we choose a piecewise constant (resp. linear) approximation for the thermal properties $\kappa(x)$ and $c(x)$ (resp. $T_{0}(x)$). From the first and second columns of Figure \ref{Fig10} we note that for $h=L/2,L/2^2, L/2^3$, the posterior means (dashed line) provide a very inaccurate estimation of the true parameters (solid red lines). Moreover, the uncertainty band determined by the posterior intervals clearly fails to capture the true thermal properties of the wall $\kappa(x)$ and $c(x)$.  In particular, the cases $h=L/2$, $h=L/2^2$ and $h=L/2^3$ (i.e. discretisation with only 2, 4 and 8 elements, respectively) reveals the potential detrimental effect on the accuracy of the posterior uncertainty when such a coarse model of the HDM is used for the inference of the unknown properties. As we increase the number of elements (i.e. decrease the mesh size $h$), the uncertainty estimates provided by the posterior means and credible intervals capture the true thermophysical properties of the wall. Note that the measures of the posterior uncertainty for $h=L/2^5$ (Figure \ref{Fig10}) are similar to the ones obtained with the fine mesh $h=L/2^7$ (Figure \ref{Fig5}; middle column). For this particular synthetic experiment, it is clear that decreasing the mesh size below $h=L/2^5$ does not have substantial effect on the uncertainty estimates of the thermal properties. It is worth mentioning that a discretisation of the HDM on $2^5=32$ elements yields a computationally tractable implementation of Algorithm \ref{IF} which, as discussed in subsection \ref{discretisation}, can be performed with a standard high-end computer.

\begin{figure}[htbp]
\begin{center}
\includegraphics[scale=0.475]{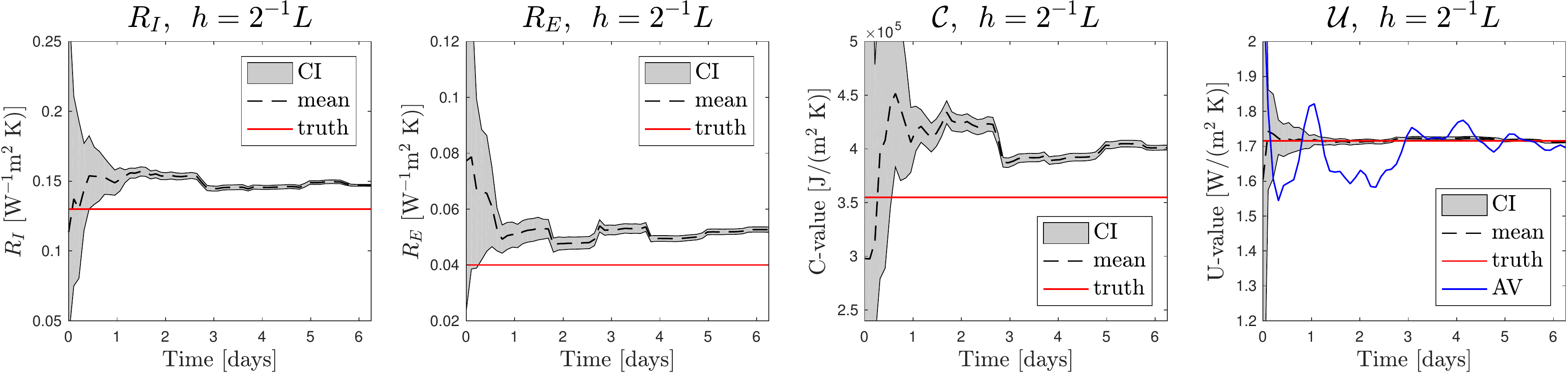}
\includegraphics[scale=0.475]{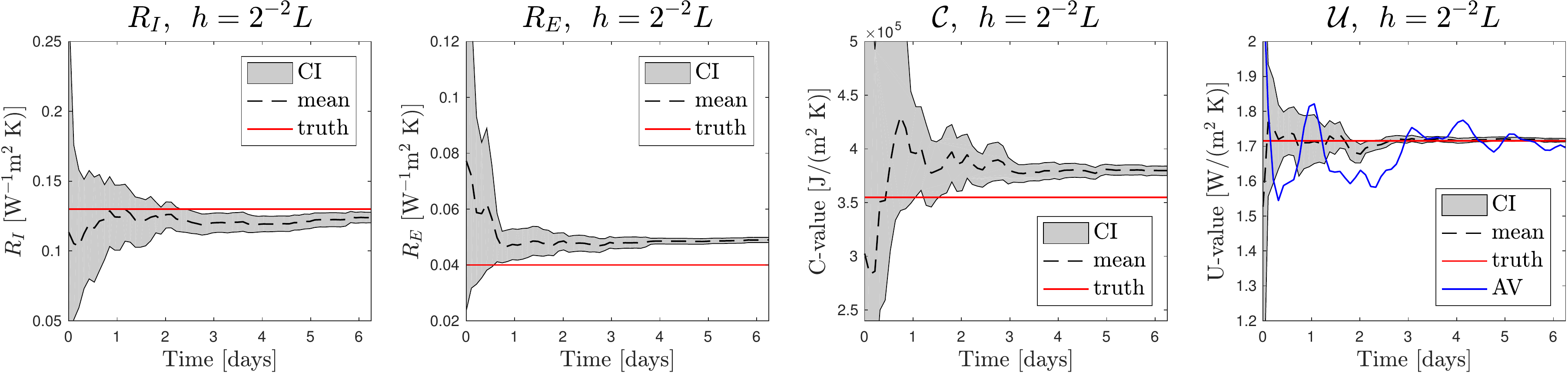}
\includegraphics[scale=0.475]{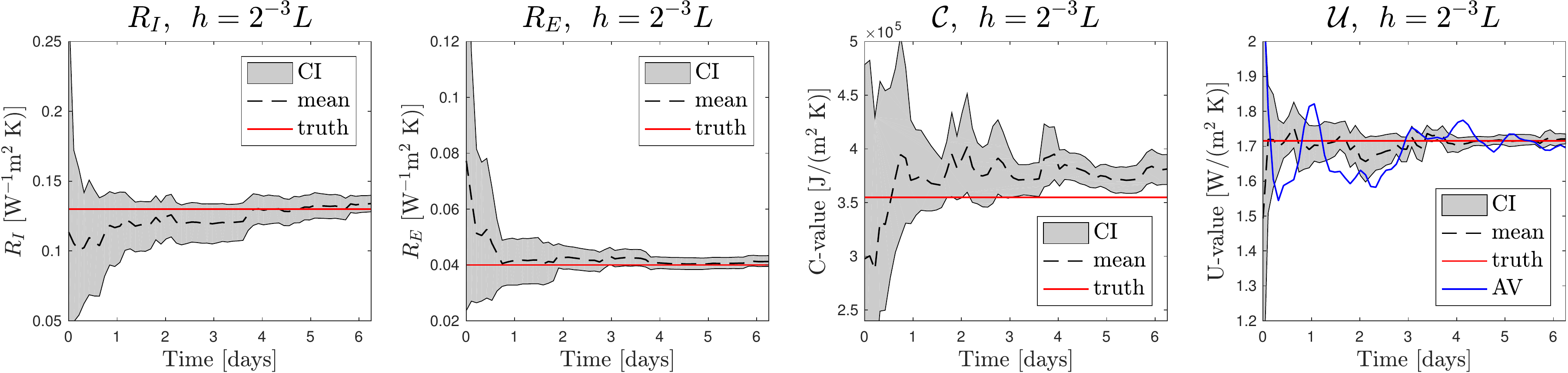}
\includegraphics[scale=0.475]{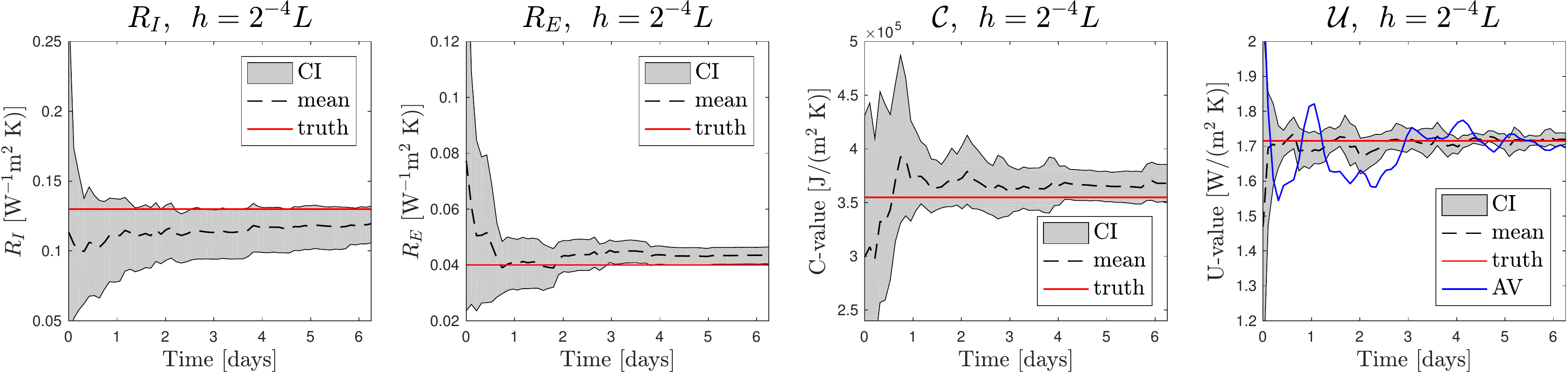}

 \caption{Synthetic experiment. Sequential posterior mean and posterior $95\%$ credible interval of $\mathbb{P}(R_{I}\vert q_{1:m}) $(left), $\mathbb{P}(R_{E}\vert q_{1:m}) $ (left-middle)$, \mathbb{P}(\mathcal{C}\vert q_{1:m}) $ (middle-right) and $\mathbb{P}(\mathcal{U}\vert q_{1:m}) $ (right), computed using mesh size (from top to bottom) $h=L/2$,  $h=L/2^2$, $h=L/2^3$, $h=L/2^4$. The solid red horizontal lines denote the true values $R_{I}^{\dagger}$ (left), $R_{E}^{\dagger}$ (left-middle), $\mathcal{C}^{\dagger}$ (middle-right) and $\mathcal{U}^{\dagger}$ (right). The blue line in the right panel is the U-value computed via the average method.}\label{Fig11}
\end{center}
\end{figure}

\begin{figure}[htbp]
\begin{center}

\includegraphics[scale=0.42]{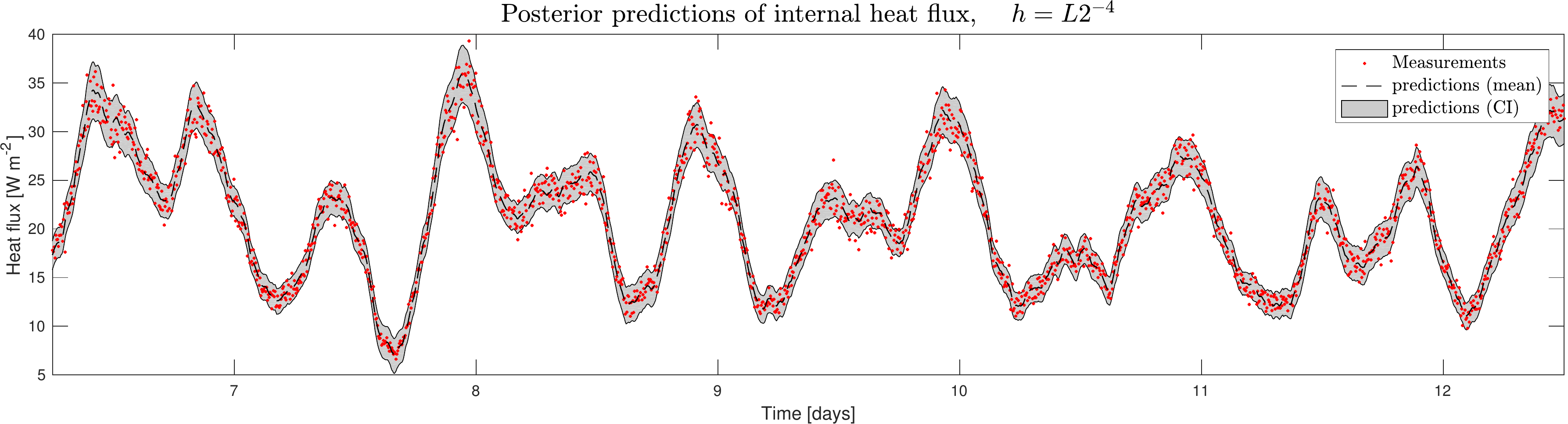}\\
\includegraphics[scale=0.42]{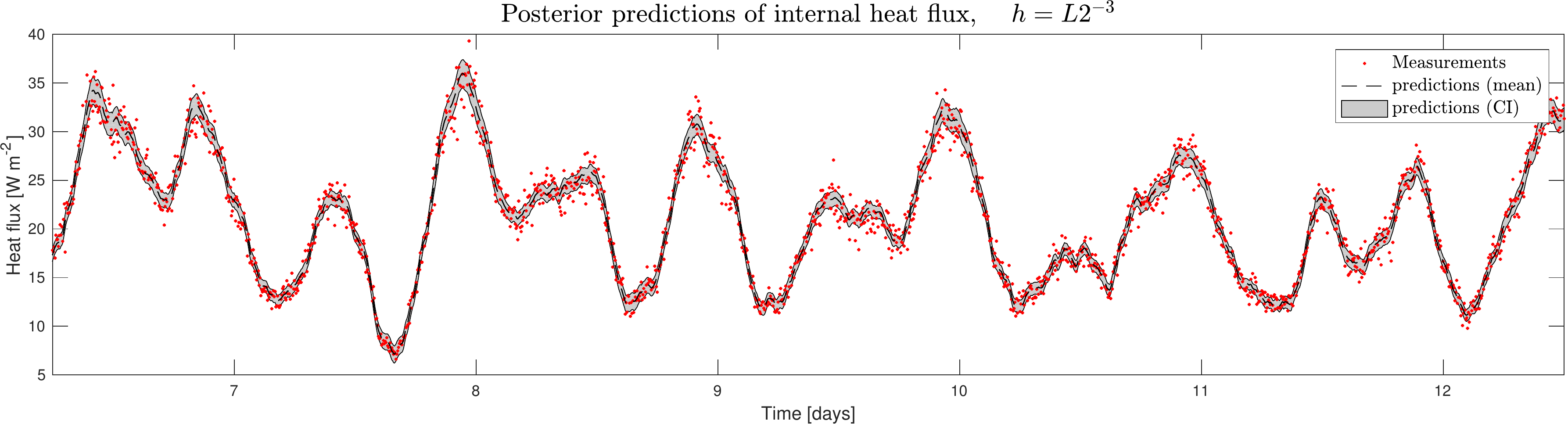}\\
\includegraphics[scale=0.42]{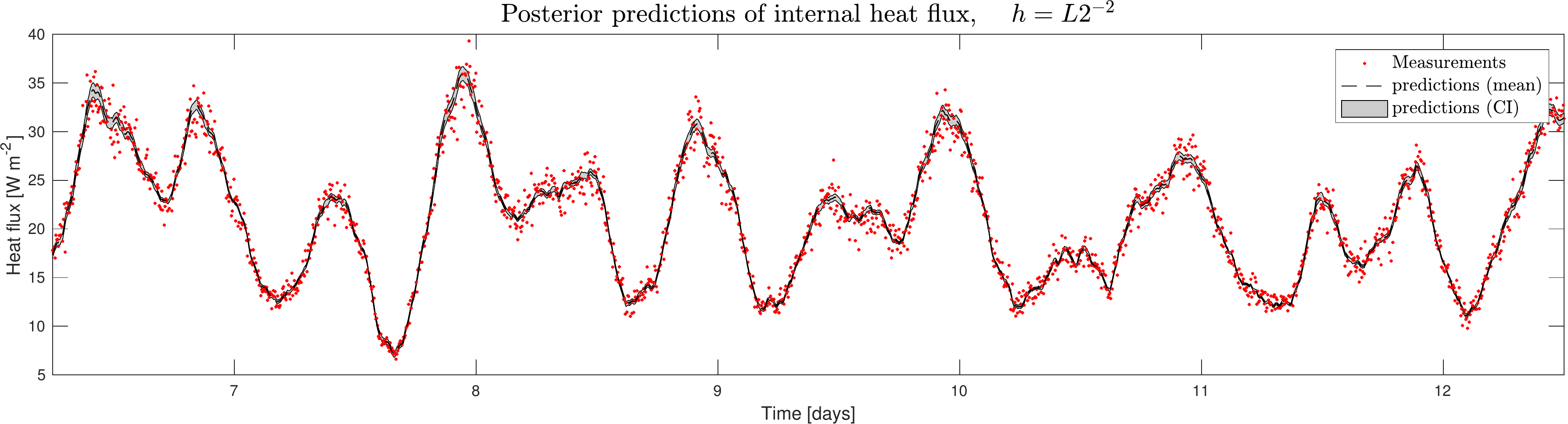}\\
\includegraphics[scale=0.42]{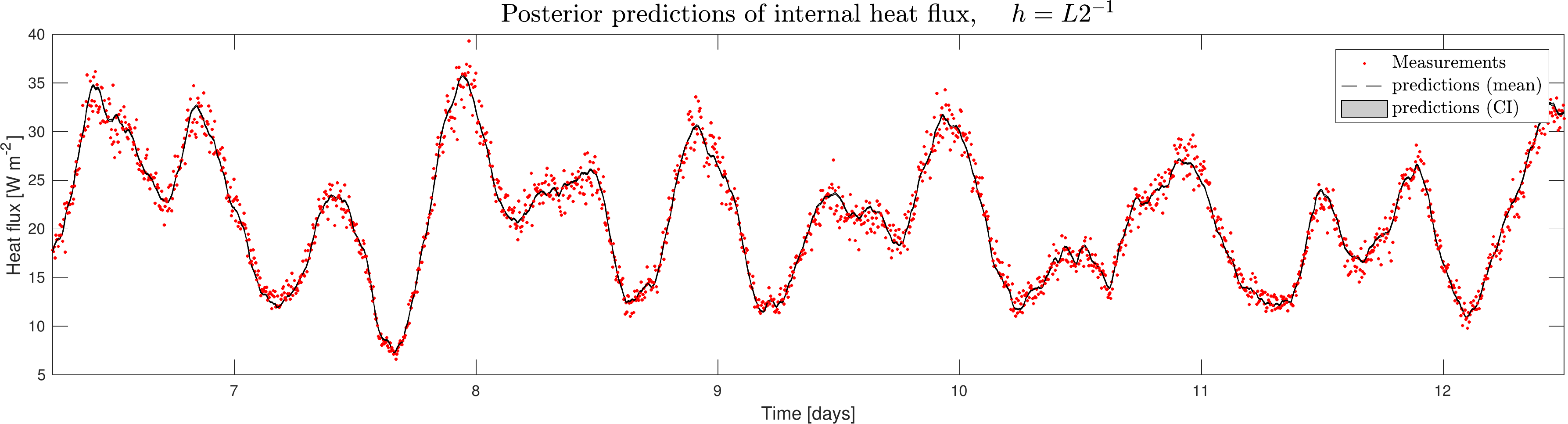} \caption{Synthetic experiment. Posterior mean and $95\%$ credible intervals of internal heat flux predictions generated via a HDM with mesh size: (from top to bottom) $h=L/2^4, h=L/2^3, h=L/2^2, h=L/2$}\label{Fig30}
\end{center}
\end{figure}

\begin{figure}[htbp]
\begin{center}

\includegraphics[scale=0.42]{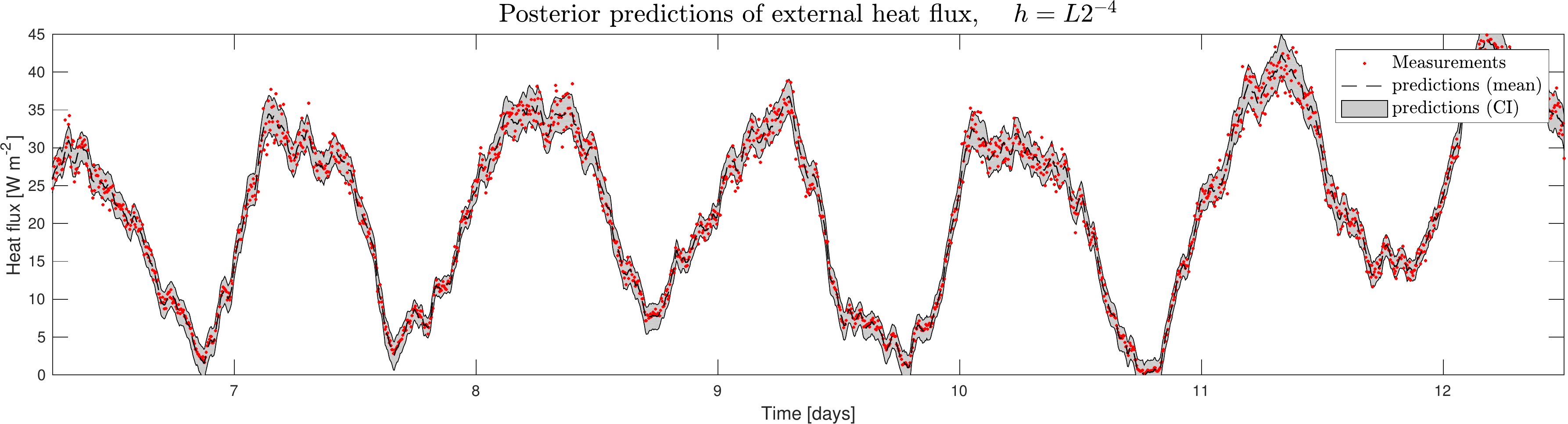}\\
\includegraphics[scale=0.42]{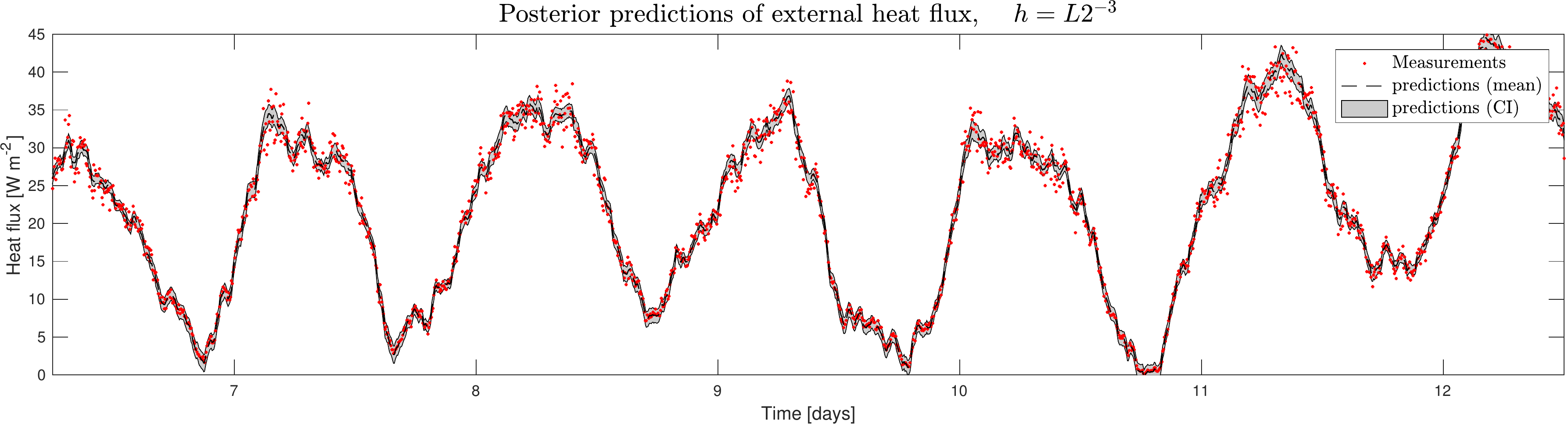}\\
\includegraphics[scale=0.42]{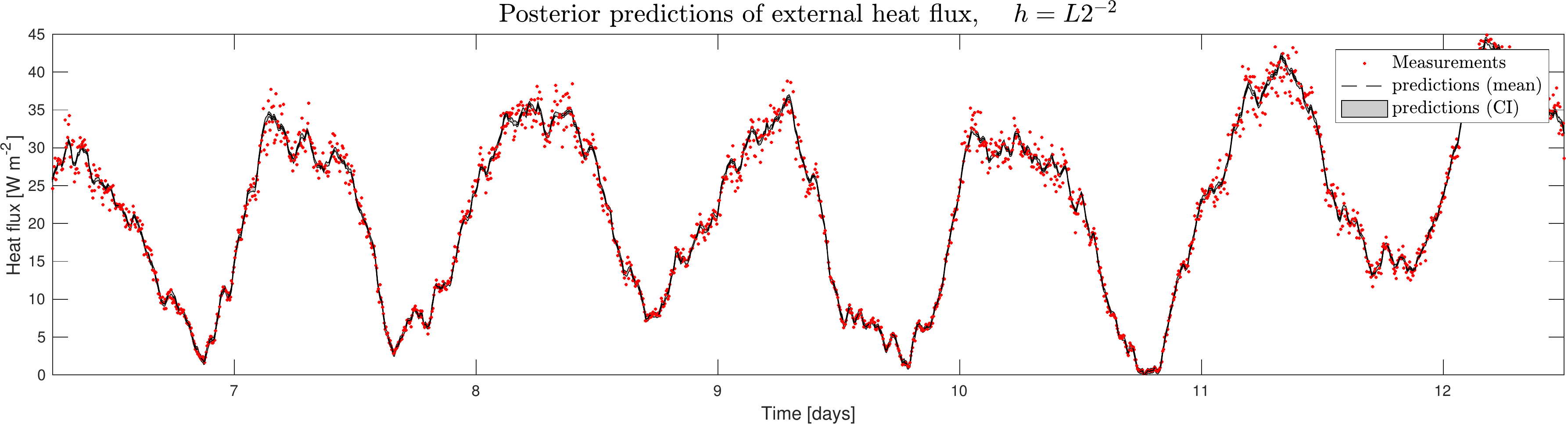}\\
\includegraphics[scale=0.42]{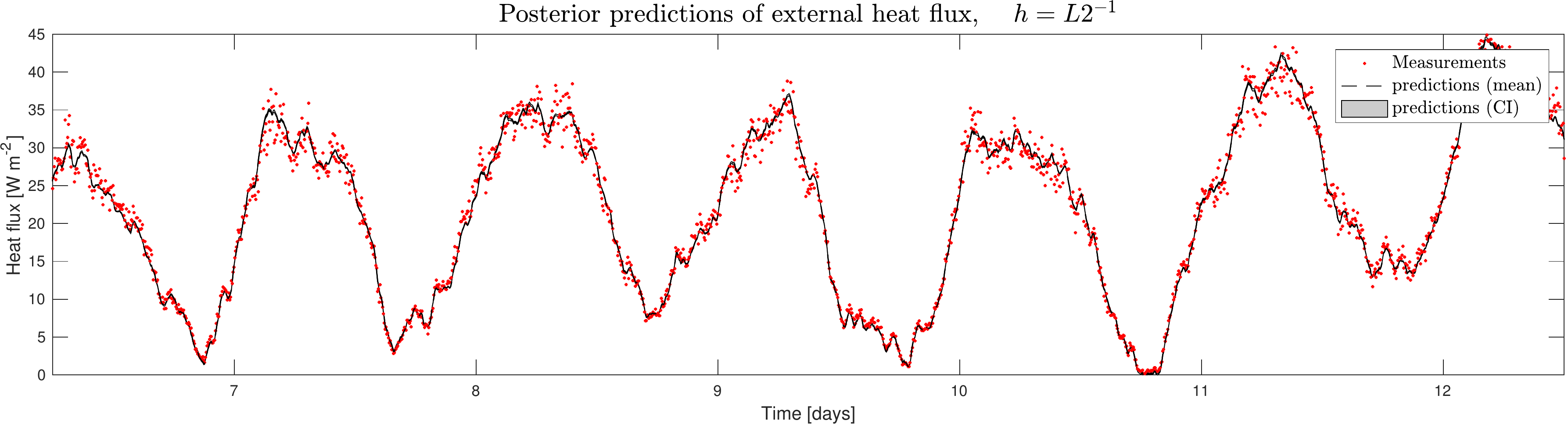}
\caption{Synthetic experiment. Posterior mean and $95\%$ credible intervals of external heat flux predictions generated via a HDM with mesh size (from top to bottom): $h=L/2^4, h=L/2^3, h=L/2^2, h=L/2$}\label{Fig30B}
\end{center}
\end{figure}

In Figure \ref{Fig11} we display the sequential posterior mean and credible intervals of $R_{I}$ and $R_{E}$ as well as the the effective properties $\mathcal{U}$ and $\mathcal{C}$, computed with mesh sizes (from top to bottom) $h=L/2, L/2^2, L/2^3, L/2^4$. From these figures we can note that a significant bias in the posterior means (black dashed line) of $R_{I}$, $R_{E}$ and $\mathcal{C}$ is observed for larger mesh sizes $h=L/2,L/2^2, L/2^3$. Note that for these larger mesh sizes, credible intervals do not capture the true values (red solid line). Finally, in Figure \ref{Fig30} and Figure \ref{Fig30B} we show the posterior predictive distributions of internal and external heat flux obtained with some of our choices of $h$; see Figures \ref{Fig3}-\ref{Fig3B} for the choice $h=L/2^7$. 
For all these choices of $h$, we can visually appreciate that the posterior mean of these predictive distributions of heat flux yields a good fit to the observations. We qualitatively assess the aforementioned predictive capabilities via the chi-square test that we introduced in the previous subsection.  Table \ref{Table_new}  shows the results of this test for some of our choices of $h$. We find that the quality of the fit, that corresponds to values close to one, slightly increases with smaller mesh size. It is clear that even a large selection of mesh size ($h=L/2^2$) results in very good fit to the measurements in the chi-squared sense. However, a closer look at Figures \ref{Fig30}-\ref{Fig30B} reveals that, for some of our choices of large mesh size, the corresponding credible intervals do not capture the measurements that the corresponding calibrated HDM (discretised via a large mesh size) is aim at predicting. This is detrimental to the degree of confidence of these probablistic predictions; our aim is to provide uncertainty estimates (e.g. credible intervals) capable of predicting unobserved measurements within our probabilistic estimates.

In order to assess the degree of confidence of these predictions in a quantitative fashion, we consider the interval score \cite{doi:10.1198/016214506000001437} of the $95\%$ predictive intervals of heat fluxes computed for each selection of $h$. The interval score, applied to a $(1-\alpha)\%$ predictive interval of the form $[l,u]$, around an observation/measurement denoted by $\nu$ can be computed via:
\begin{equation}\label{AIS}
IS=u-l+\frac{2}{\alpha}(l-\nu)\mathbf{1}_{\{\nu<l\}}+\frac{2}{\alpha}(\nu-u)\mathbf{1}_{\{\nu>u\}}
\end{equation}
where $\mathbf{1}_{\{A\}}$ denotes the indicator function defined by
\begin{equation}\label{ind}
\mathbf{1}_{\{A\}}=\left\{\begin{array}{cc}
1 & \textrm{if} ~A~\textrm{holds}\\
0& \textrm{otherwise} \end{array}\right.
\end{equation}
The interval score penalises intervals with small measurement coverage; the higher its value the lowest the confidence of the predictive interval. We apply expression (\ref{AIS}) for each predictive interval of internal and external heat flux, computed at each measurement time $t_{m}$ within the predictive time window $(6.25~\textrm{days}, 13.50~\textrm{days}]$. The average of these interval scores (AIS) over the predictive time window  are displayed in Table \ref{Table_new} which confirms that smaller mesh sizes results in more confident predictions (i.e. larger measurement coverage of the predictive intervals). We further note that even though the choice $h=L/2^2$ gives a good fit to the measurements in the chi-squared sense, it produces a large AIS suggesting that this choice of mesh size results in a predictions with low confidence compared to the ones that we obtain with smaller mesh sizes.

The study from this subsection is particularly relevant in the context of lumped-thermal mass  models \cite{gori2017inferring,biddulph2014inferring,berger2016bayesian} for Bayesian inference of a wall's U-value and heat capacity per unit of area. Lumped-thermal mass  models are effectively coarse-mesh finite difference approximations of the heat equation, and thus involve the solution of a low-dimensional system rather than the more computationally costly (if a small $h$ is used) HDM that we solve within the proposed Bayesian approach. While recent publications \cite{biddulph2014inferring,gori2017inferring,berger2016bayesian} have investigated a wide class of lumped-thermal mass  models of different complexity (i.e. number of resistors/capacitors), these have been primarily concerned with showing that simplified models can provide a good fit to the measurements in a mean-squared sense. The results from this subsection demonstrate that even though a coarse-mesh (simplified) calibrated model of the heat transfer through the wall can successfully fit measurements, the corresponding predictive distributions may not necessarily provide an accurate quantification of the uncertainties in these predictions. As we discussed in Section \ref{Intro}, the ability to accurate quantify uncertanty in predictions of thermal performance is crucial for the computaions of risks under different energy saving measure and, hence, inform decision-making for retrofit interventions.

\subsection{Inferring properties with only internal heat flux measurements}\label{only}

In this section we investigate the application of Algorithm \ref{IF} to infer the unknown HDM parameters $c(x)$, $\kappa(x)$, $T_{0}(x)$, $R_{I}$ and $R_{E}$ given, as before, near-air temperatures at the internal and external surface of the wall, but using heat flux measurements collected only at the internal surface of the wall. This investigation is motivated by standard practices in which only internal heat flux measurements are collected, such as the monitoring procedure described in ISO9869:2014 \cite{ISO9869:2014}. The same formulation from subsection \ref{HDM} and \ref{bayesian_theo} can be applied for the assimilation of only heat flux measurements by simply setting $q_{m}=q_{I,m}$ and $Q_{m}=Q(0,\tau_{m})$ in equations (\ref{eq:1}) and (\ref{eq:9}) respectively. We modify Algorithm \ref{IF} accordingly and use it with the same experimental setting described in subsection \ref{synthetic_data}. This setting is identical to the one that we use to produce the results from subsection  \ref{posterior_num}, except that we now exclude measurements of external heat flux during the inference process. 

In Figure \ref{Fig5} (right column) we show the final-time posterior estimates of $c(x)$, $\kappa(x)$ and $T_{0}(x)$. We observe that these estimates fail to recover the spatial variability of the corresponding true parameters near the external surface of the wall ($x=0$). In fact, the posterior credible intervals in this region, are almost identical to the ones from the priors (see Figure \ref{Fig5} left column). It is only near the internal surface of the wall where we observe that the inferred parameters capture, under uncertainty, the variability in $c(x)$ and $\kappa(x)$. This comes as no surprise since we have not included the external wall heat flux measurements during the inference process. The effect that arises from excluding these measurements can also be noted from the final-time posteriors of $R_{I}$, $R_{E}$, $\mathcal{C}$ and $\mathcal{U}$ shown in Figure \ref{Fig6} (bottom), the corresponding credible intervals reported in Table \ref{Table1}, and the plots of the sequential posteriors shown in Figure \ref{Fig8} (bottom). Indeed, the posterior of the internal surface resistance $R_{I}$ is very similar to the one obtained when both internal and external heat flux measurements are assimilated (see Figure \ref{Fig6} (top)). However, the posterior of the external surface resistance $R_{E}$ is very close to the prior. Again, this indicates that the measurements of internal heat flux are not informative of this variable. Furthermore, from Figure \ref{Fig6} (middle right column) we note that the posterior density for the C-value have a larger variance and captures the true value $\mathcal{C}^{\dagger}$ on the tail of the posterior distribution. When we compare the posterior estimates of the U-value, $\mathcal{U}$, with the one reported in subsection \ref{posterior_num}, we note only a slight increase in the posterior variance (see Figure \ref{Fig6} (left column) and credible intervals in Table \ref{Table1}). This increase is expected from the fact that the uncertainty in the external surface resistance was not reduced as in the previous case. Nevertheless, Figure \ref{Fig8} (left column) reveals rapid and accurate estimates of the of the U-value which clearly outperforms the average method. 

In the bottom panels of Figure \ref{Fig3} and Figure \ref{Fig3B} we show the posterior predictive distributions of surface heat fluxes at the internal and external wall, respectively. Similar to the results from subsection \ref{posterior_num}, the posterior mean of the predictive distributions of the internal heat fluxes shows a good visual agreement with the measurements while credible intervals display good coverage of heat flux measurements. In contrast, credible intervals of the posterior predictions of external heat fluxes (Figure \ref{Fig3B}) show large coverage and the corresponding mean does not seem to adequately fit the measurements of external heat flux. Table \ref{Table_new} (second column $h=L2^{-7}$) further confirms, via the chi-square test, that internal heat flux predictions are properly matched with the predictive mean. These results also indicate that a slightly larger coverage of the credible interval is obtained compared to the ones obtained when both external and internal measurements are assimilated. In addition, predictions of external heat flux provide a very poor fit to the external measurements ($\chi_{E}^2= 58.3$) and a relatively low degree of confidence ($AIS_{E}=9.3$). In Table \ref{Table_new} we also show the chi-square and AIS quantities obtained from assimilating only internal heat flux measurements using different choices of mesh size $h$ in the HDM. Note that the largest mesh size $h=L/2$ yields predictions of internal heat flux that match the measurements of internal heat flux ($\chi_{I}^2= 1.05$). However, for such a coarse mesh, the predictions of external heat flux do not match the corresponding measurements ($\chi_{E}^2= 120.7$).  Table \ref{Table_new} reveals that increasing the mesh size decreases the confidence of the predictions of both internal and external heat flux. 

In summary, accurate estimates of effective thermal properties such as the U-value can be computed from the assimilation of only internal heat flux measurements. However, our results suggest that the Bayesian inversion of both external and internal is essential for an accurate characterisation of the predictions of the thermal performance of the wall. Similar to the results from the preceding subsection, the uncertainty estimates obtained with smaller mesh sizes are capable of providing higher-confidence predictions of the thermal performance compared to those obtained via reduced (coarse-grid) models the HDM. 

\section{Application with real data}\label{BSRIA}

\begin{figure}[htbp]
\begin{center}
\includegraphics[scale=0.4]{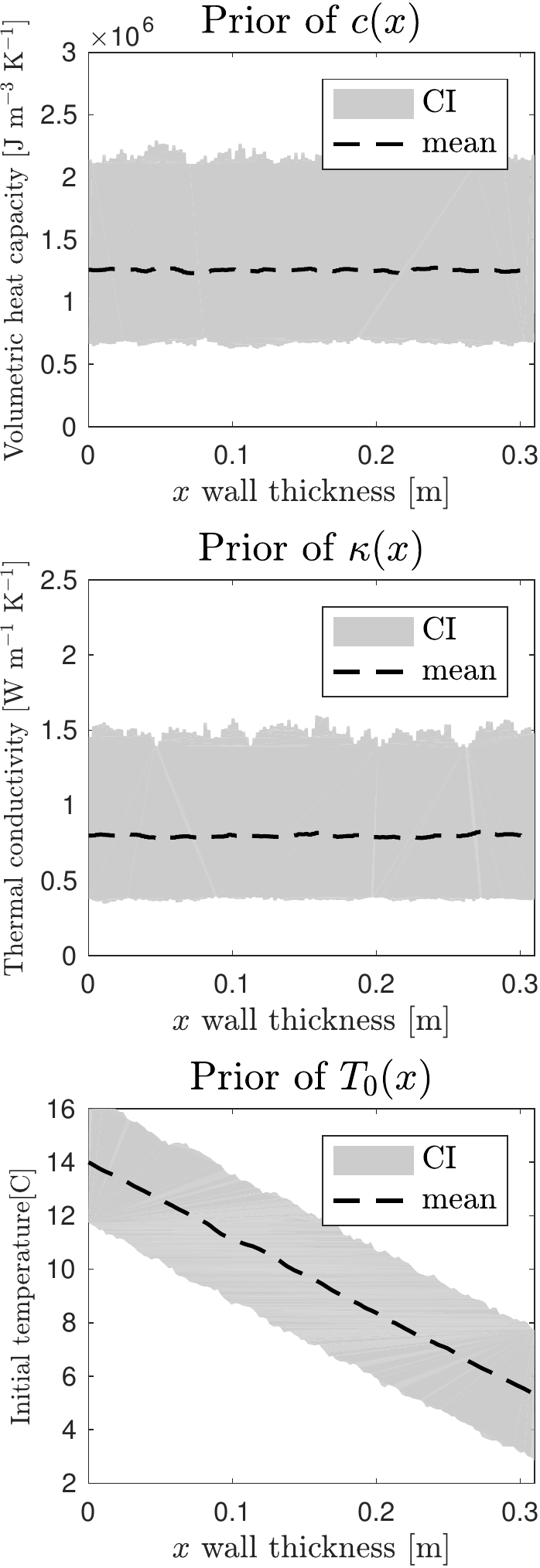}\includegraphics[scale=0.4]{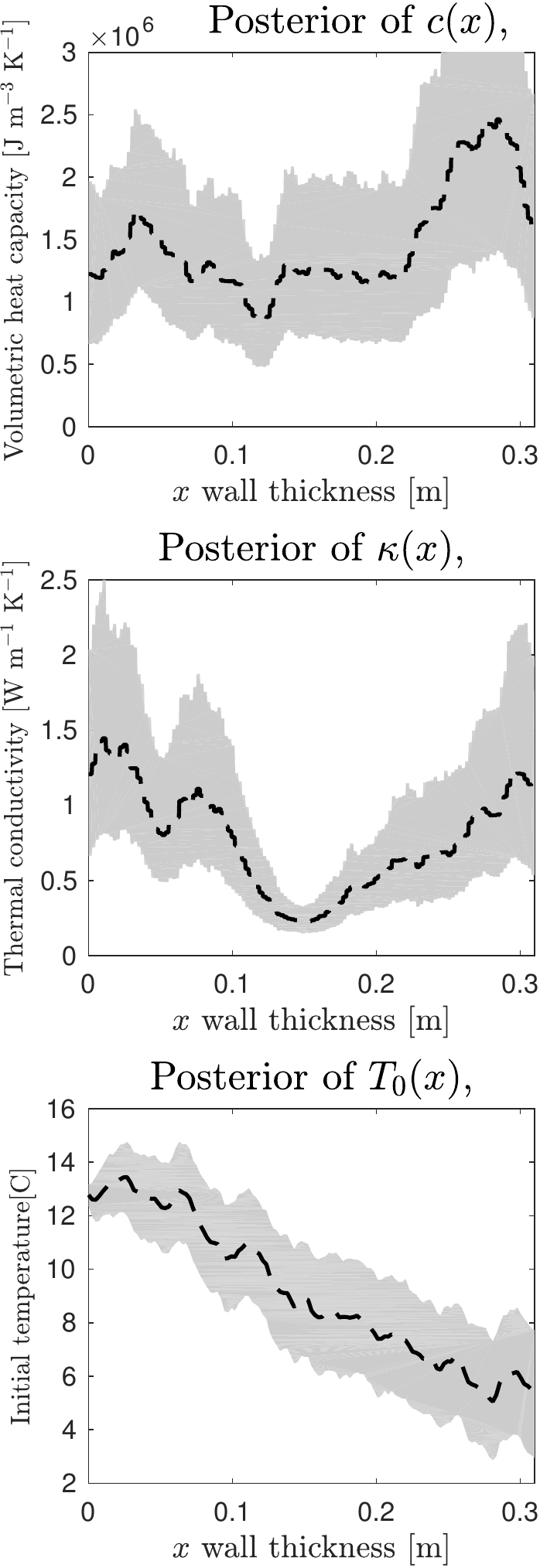}~~~\includegraphics[scale=0.425]{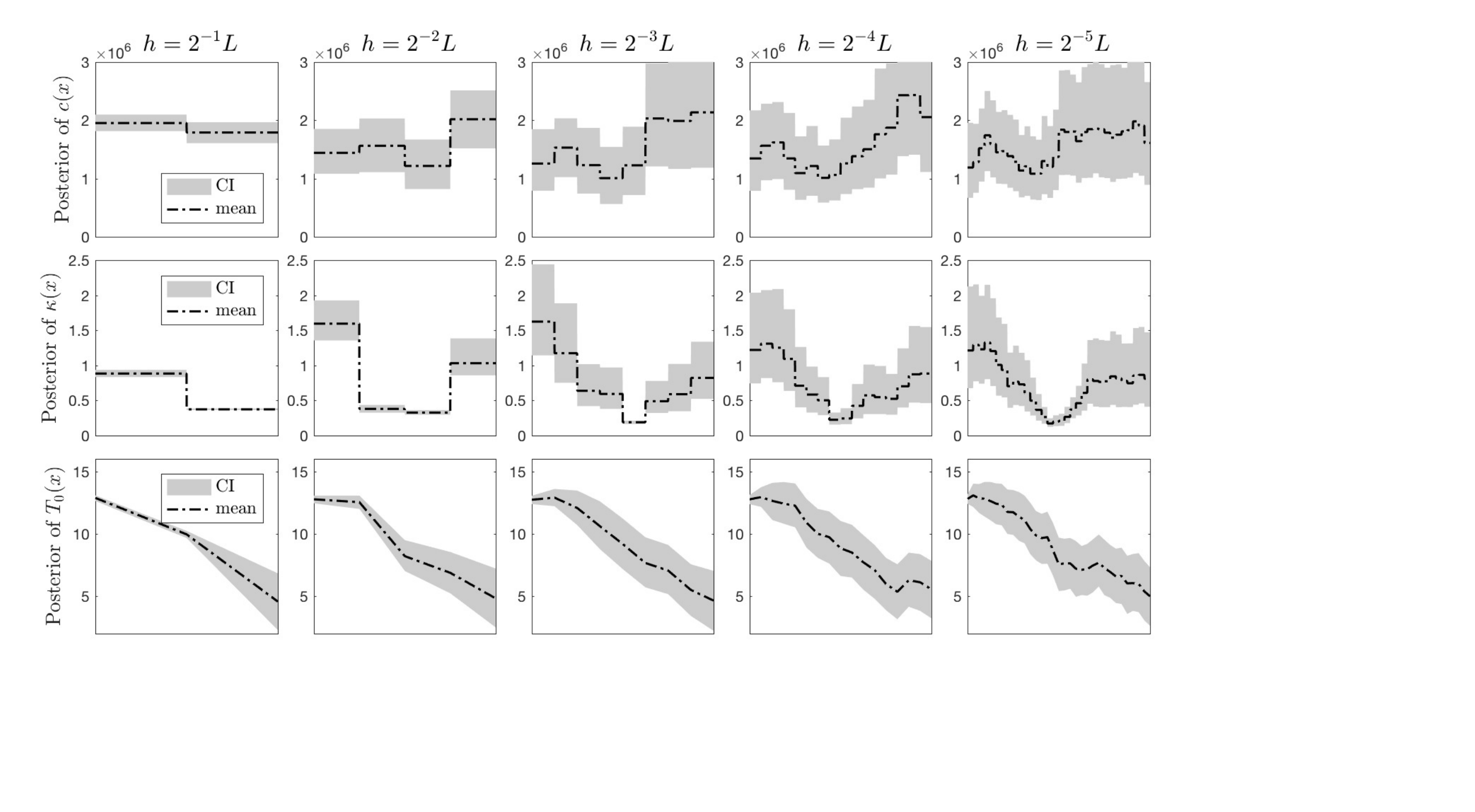}
 \caption{BSRIA experiment. Left columns: Prior and final-time posterior mean and credible intervals of (from top to bottom) $c$, $\kappa$ and $T_{0}$. Right columns:  Final-time posterior mean and posterior $95\%$ credible intervals of $\mathbb{P}(c\vert q_{1:M})$ (top), $\mathbb{P}(\kappa\vert q_{1:m})$ (middle) and $\mathbb{P}(T_{0}\vert q_{1:M}) $ (bottom), computed using mesh size: (from left to right) $h=L/2^{1},L/2^{2},L/2^3,L/2^4,L/2^5$. }\label{Fig14}
\end{center}
\end{figure}

In this section we apply Algorithm \ref{IF} with the real data described in subsection \ref{BSRIA_data}. We emphasise that for this experiment, the underlying true thermal properties and the initial temperature distribution of the wall $\kappa(x)^{\dagger}$, $c^{\dagger}(x)$ and $T^{\dagger}_{0}(x)$ are unknown, and only measurements of internal surface heat flux are available for the inference algorithm. We assimilate/invert measurements collected during the first 8 days of the measurement campaign; subsequent measurements are used for the validation of the predictive distributions of internal heat flux. Values for the surface resistances $R_E^{\dagger}$ and $R_{I}^{\dagger}$ are also unknown; only book values for a generic wall with similar specifications are available. Relevant prior parameters (see Table \ref{parameters}) are selected so that the distributions of $R_{I}$, $R_{E}$, $\mathcal{U}$ and $\mathcal{C}$ are consistent with the range of book values for the brick wall under investigation \cite{cibse2015environmental}. In Figure \ref{Fig14} (left) we show the mean and credible interval of the prior and final time ($\tau_{m}=8$ days) posterior of $c(x)$ (top), $\kappa(x)$ (middle) and $T_{0}(x)$ (bottom). These posterior estimates reveal substantial variability in $c(x)$ and $\kappa(x)$ which may be arguably attributed to internal inhomogeneities such as residual cavities, moisture condensation and/or defects overlooked during visual inspection.

\begin{figure}[htbp]
\begin{center}
\includegraphics[scale=0.475]{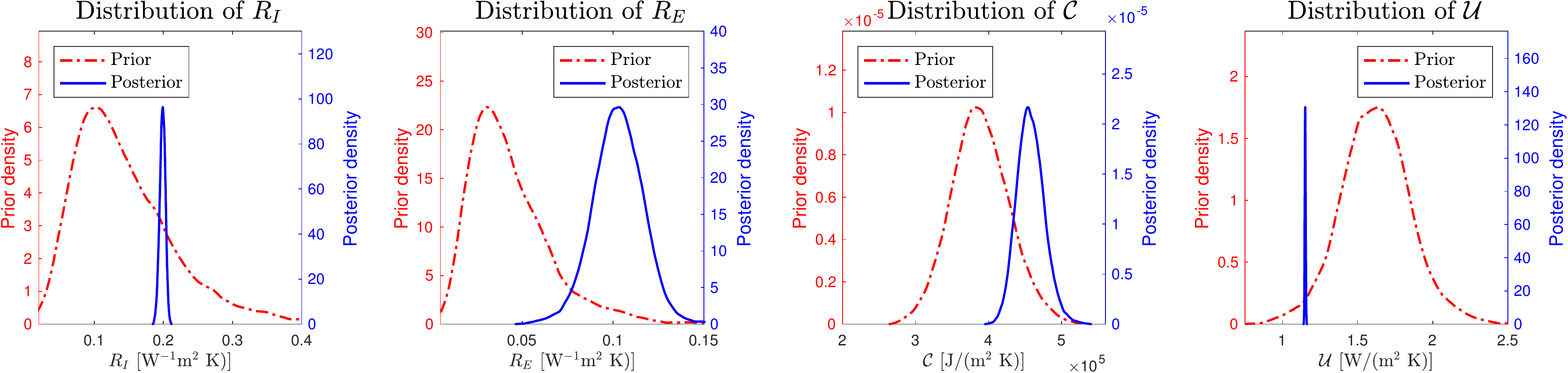}
 \caption{BSRIA experiment.  Prior (dashed-red line) and final-time posterior (solid blue line) of $R_{I}$ (left), $R_{E}$ (left-middle), $\mathcal{C}$ (middle-right) and  $\mathcal{U}$ (right).}\label{Fig15}
\end{center}
\end{figure}

\begin{figure}[htbp]
\begin{center}

\includegraphics[scale=0.45]{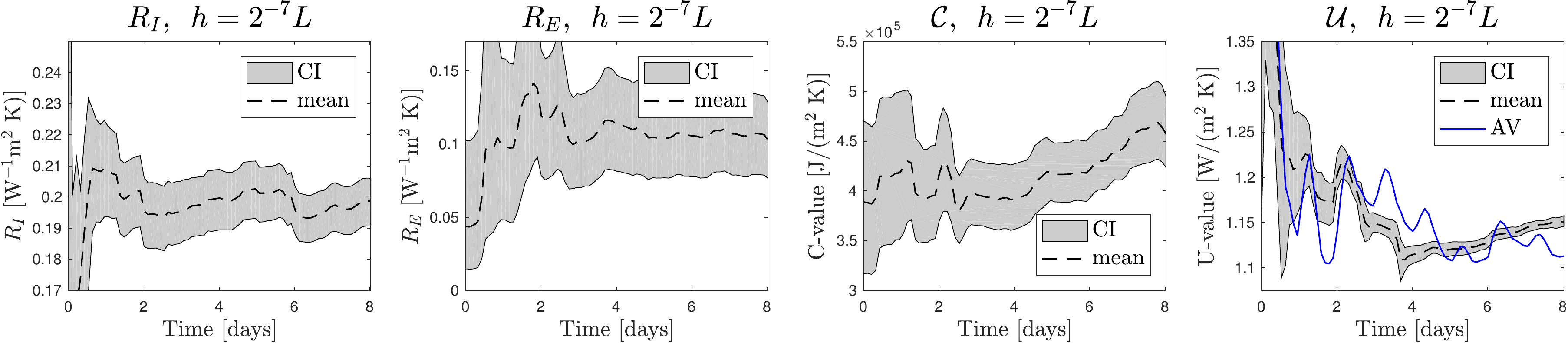}

 \caption{BSRIA experiment.  Sequential posterior mean and posterior $95\%$ credible interval of $\mathbb{P}(R_{I}\vert q_{1:m}) $(left), $\mathbb{P}(R_{E}\vert q_{1:m}) $ (left-middle)$, \mathbb{P}(\mathcal{C}\vert q_{1:m}) $ (middle-right) and $\mathbb{P}(\mathcal{U}\vert q_{1:m}) $ (right). The blue line in the right panel is the U-value computed via the average method.}\label{Fig18}
\end{center}
\end{figure}

The prior and final-time posterior densities of $R_{I}$, $R_{E}$, $\mathcal{C}$ and $\mathcal{U}$ are plotted in Figure \ref{Fig15}. Equal tail 99\% credible intervals for these distributions are shown in Table \ref{literature_data}. Similar to the results obtained in subsection \ref{only}, only the uncertainties of $R_{I}$ and $U$ have been substantially reduced via the assimilation of measurements of internal surface heat flux. The posterior variances of $R_{E}$ and $C$ do not exhibit a significant reduction with respect to their priors although their means have been shifted. Note that our final-time estimates of the posterior mean of the surface resistances $R_{I}$ and $R_{E}$ are substantially different from the book values suggested in CIBSE Guide A \cite{cibse2015environmental}. Sequential posterior means and credible intervals for $R_{I}$, $R_{E}$, $\mathcal{C}$ and $\mathcal{U}$ are shown in Figure \ref{Fig18}, where we also include the running estimates of the U-value computed via the average method (see equation (\ref{average_method})). Values of the mean and CoV for these posteriors are reported in Table \ref{Table3}. We note that after 1.5 days of the measurement campaign, the posterior estimates of the U-value (in terms of CoV) decreases from the prior value of $14\%$ to about 1\%, and seems to stabilise more rapidly than the estimate provided by the average method. 

\begin{figure}[htbp]
\begin{center}

\includegraphics[scale=0.4]{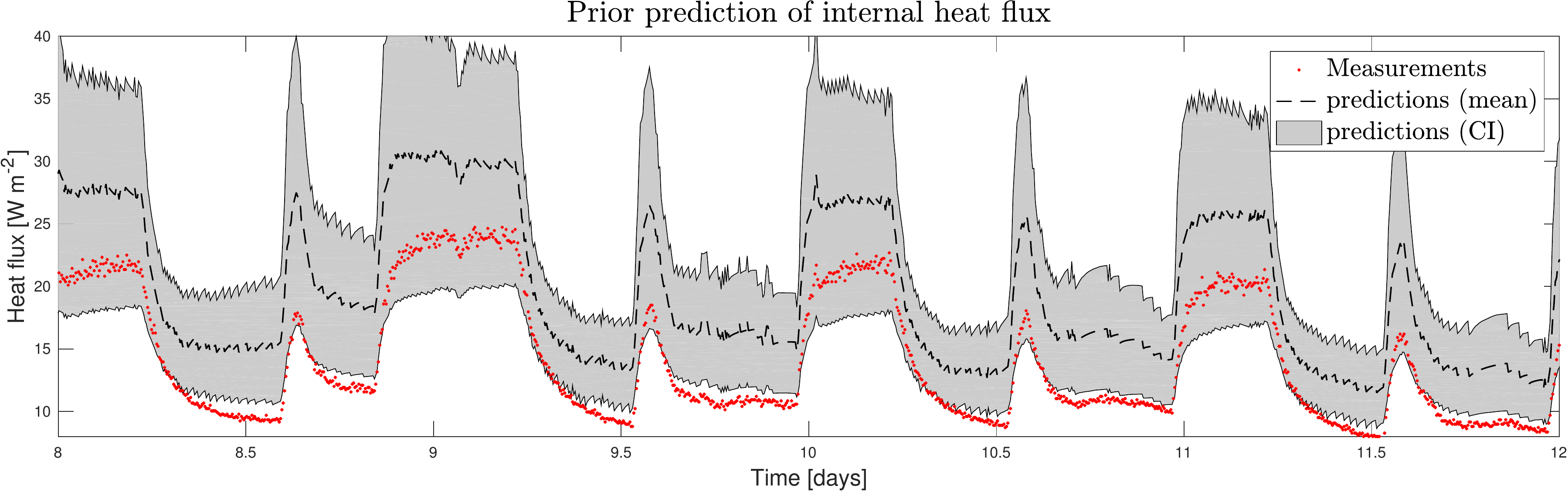}
\includegraphics[scale=0.4]{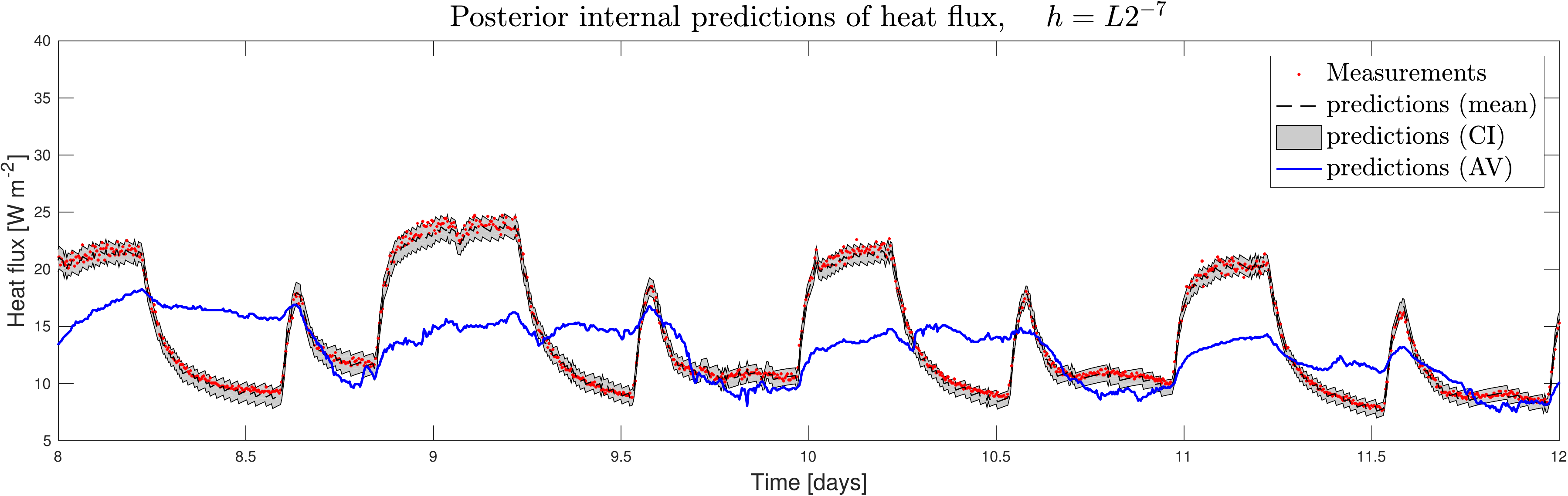}

 \caption{BSRIA experiment. Mean and $95\%$ credible intervals of internal heat flux predictions generated from the prior (top) and the posterior using the HDM with a mesh size of $h=L/2^7$ (bottom) blue line is the plot of heat flux predictions computed via no-thermal-mass model (see (\ref{nomass})). Measurements are displayed in red dots.
}\label{Fig17}
\end{center}
\end{figure}

In Figure \ref{Fig17} we display the prior (top) and posterior (bottom) predictive distribution of internal surface heat flux on the interval [8 days, 12 days]. The posterior predictions are produced with the inferred parameters computed at the final the assimilation time $\tau_{m}=8$ days. While the prior displays large uncertainties that do not fully capture the measurements (red dots), the posterior mean of these predictive distributions (solid black line) provides a good visual fit these (un-assimilated) measurements, confirmed with a chi-square value $\chi_{I}^{2}=1.008$ (see Table \ref{Table_new}, $h=L/2^7$). Moreover, most measurement are enclosed within the predictive interval of interest; high confident credible intervals are confirmed with AIS displayed in Table \ref{Table_new}. In summary, our posterior predictive distributions can accurately capture the uncertainties in the measurements of internal heat flux. In Figure \ref{Fig17} (bottom) we have also included the plot of the internal surface heat flux predictions computed via the no-thermal mass model (\ref{nomass}) with our estimate of the U-value computed with the average method (equation (\refeq{average_method})). It is clear that this model is not able to accurately reproduce the measurements surface heat flux.

\begin{figure}[htbp]
\begin{center}

\includegraphics[scale=0.5]{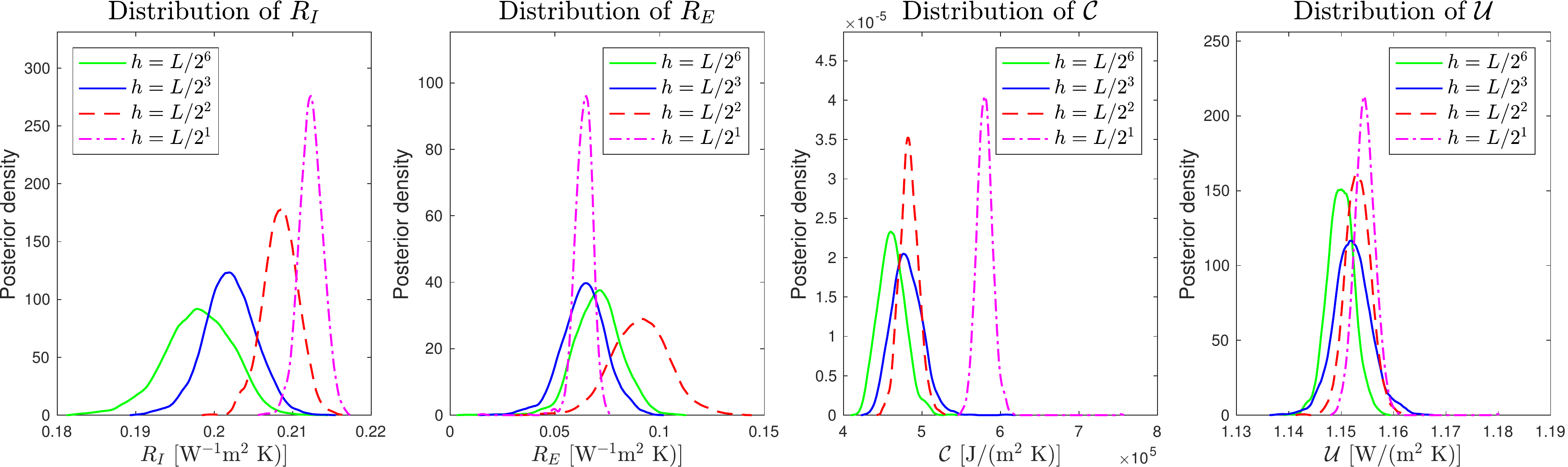}
 \caption{BSRIA experiment. Final time posteriors $\mathbb{P}(R_{I}\vert q_{1:M})$ (left) , $\mathbb{P}(R_{E}\vert q_{1:M})$ (left-middle), $\mathbb{P}(\mathcal{C}\vert q_{1:M})$ (middle-right) and  $\mathbb{P}(\mathcal{C}\vert q_{1:M})$ (right), computed with different selections of mesh size $h=L/2,L/2^{2},L/2^3,L/2^6$. }\label{Fig16BB}
\end{center}
\end{figure}

\begin{figure}[htbp]
\begin{center}

\includegraphics[scale=0.4]{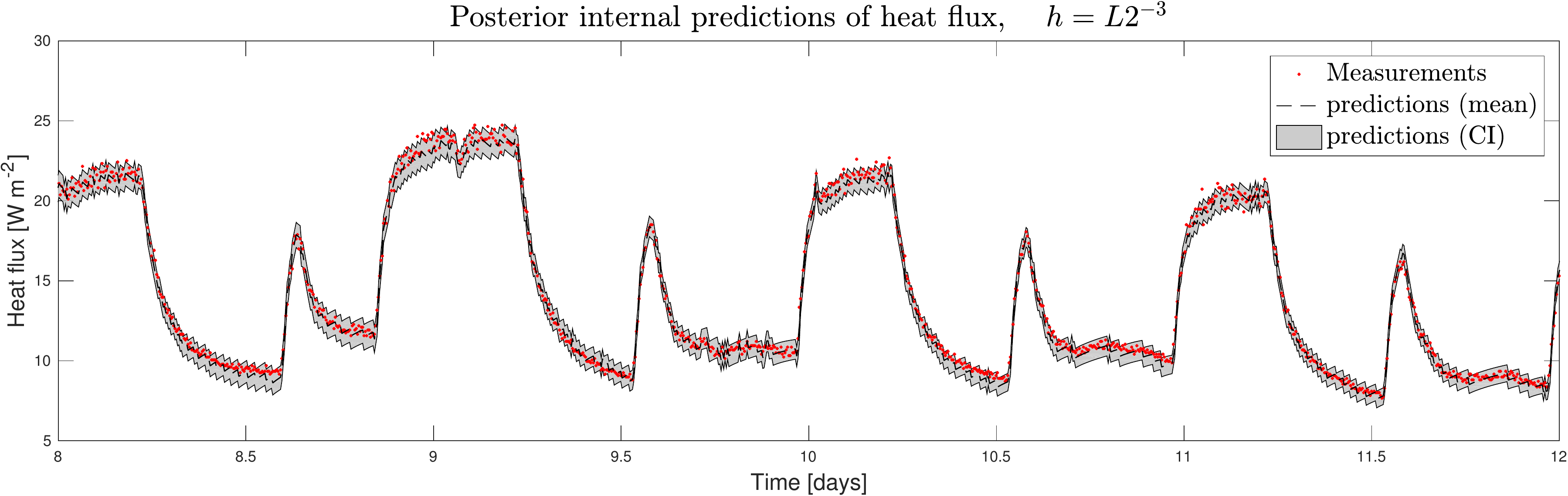}
\includegraphics[scale=0.4]{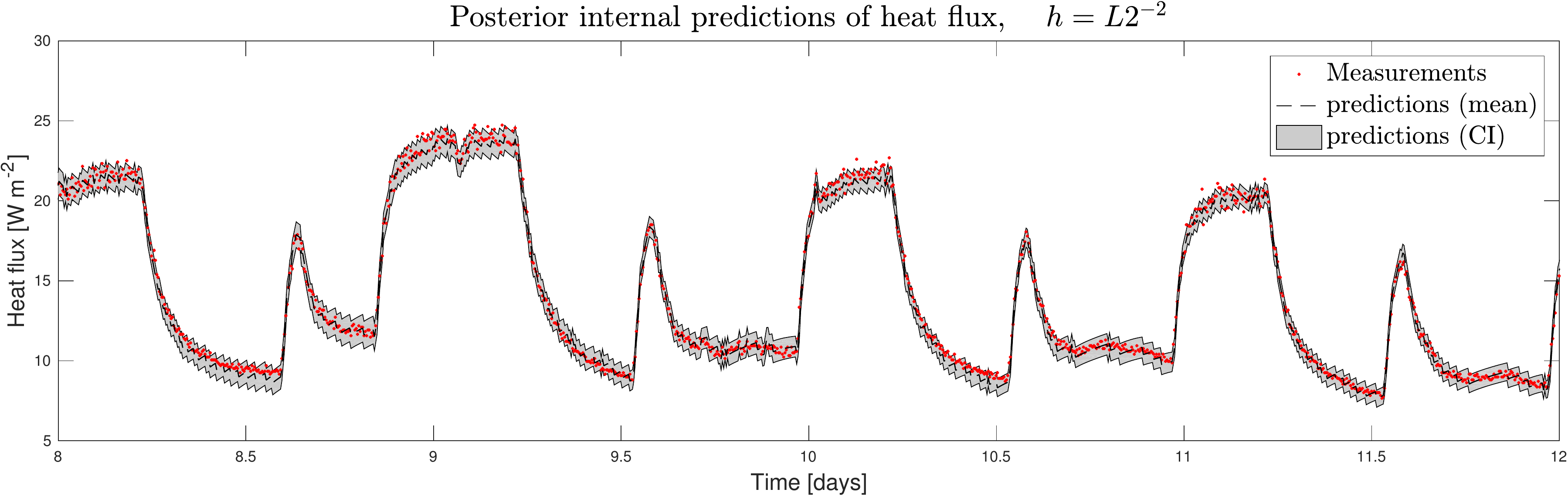}
\includegraphics[scale=0.4]{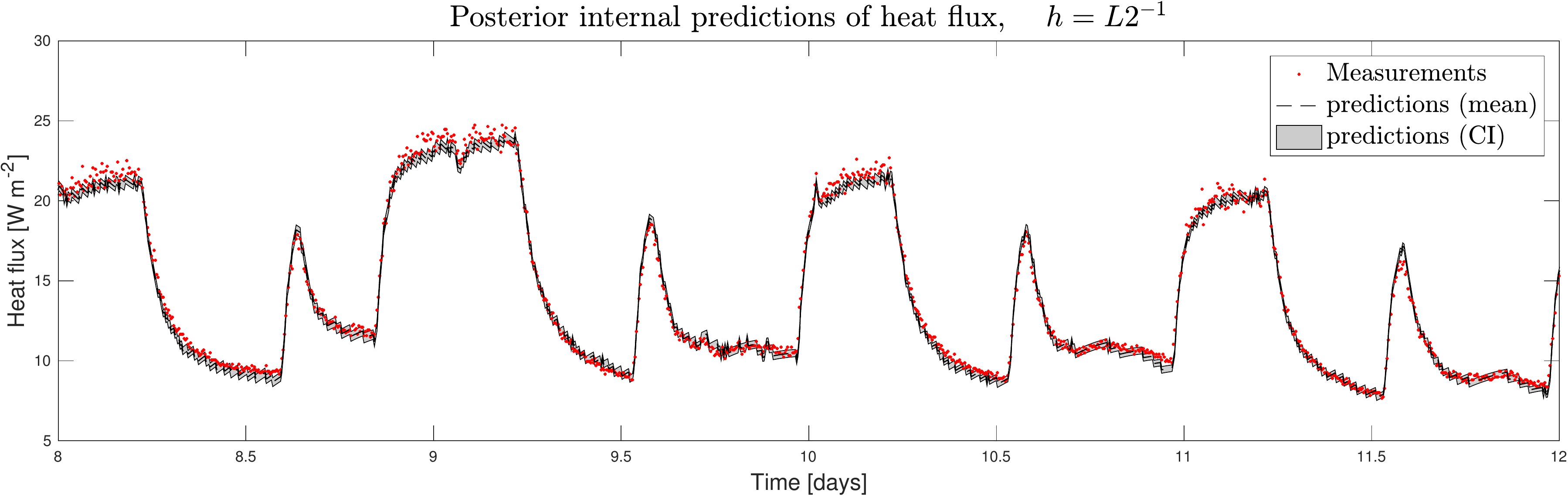}
 \caption{BSRIA experiment. Posterior mean and $95\%$ credible intervals of internal heat flux predictions generated with a mesh size of (from top to bottom) $h=L/2^3$, $h=L/2^2$, $h=L/2$. Measurements are displayed in red dots.}\label{Fig170}
\end{center}
\end{figure}

The effect of the mesh size, $h$, of the HDM on the final-time ($\tau_{m}=8$ days) posterior estimates of thermal properties is shown in Figure \ref{Fig14} (right columns), where we display the posterior mean and credible intervals of the parameters $c(x)$, $\kappa(x)$ and $T_{0}(x)$ for different choices of $h=L/2^{1},L/2^{2},L/2^3,L/2^4,L/2^5$. Final-time posterior densities for the variables $R_{I}$, $R_{E}$, $\mathcal{C}$ and $\mathcal{U}$ are displayed in Figure \ref{Fig16BB}. Assuming that the accuracy of the uncertainty estimate improves with mesh refinement, it is thus clear that larger mesh sizes $h=L/2, h=L/2^2$ (i.e. 2 and 4 elements) seem to introduce biased estimates of $R_{I}$, $R_{E}$ and $\mathcal{C}$ (see Figure \ref{Fig16BB}). Moreover, we notice that the variance of the U-value is slightly underestimated for $h=L/2$. Nevertheless, estimates of the U-value are quite similar regardless of the mesh size used for the HDM within the Bayesian scheme. For some of these choices of $h$, Figure \ref{Fig170} displays plots of the mean and credible intervals of internal heat flux model prediction on the interval [8 days, 12 days]. Note that the predictive mean seems to visually agree with the measurements regardless of the value of $h$. However, similar to the results discussed in the previous section, a very large choice of mesh size ($h=L/2$) yields very low confident prediction due to the collapse of the uncertainty on these predictions. 

Table \ref{Table_new} provides additional evidence that decreasing the mesh size improves the fit to the measurements ($\chi_{I}^{2}$ closer to one) and increases the confidence in our predictions (smaller ASI). These results also indicate that a coarse mesh $h=L/2^2$ (i.e. with 4 elements) can reasonably reproduce the uncertainty in the model predictions of internal surface heat flux. It is important to emphasize that, for these computations, only internal heat flux measurements are assimilated. From our conclusions in subsection \ref{only} we know that internal heat flux can be successfully reproduced via a coarse mesh approximation of the HDM. However, there is no assurance that additional quantities that describe the thermal performance of the wall (e.g. external surface heat flux) can be accurately reproduced via such a coarse model.

\begin{table}                                                                                       
\small
\centering                                                                                                               
\begin{tabular}{|c|c|c|c|c|c|c|c|c|c|c|}                                                                                                                                                        
\hline                                                                                                            
time (days) & 0.000 & 0.733 & 1.467 & 2.200 & 2.933 & 3.667 & 4.400 & 5.133 & 5.866 & 6.600 \\                    
\hline                                                                                                            
$R_{I}$ (mean) & 0.147 & 0.209 & 0.199 & 0.195 & 0.195 & 0.199 & 0.200 & 0.201 & 0.200 & 0.195 \\                                                                                                              
$R_{I}$ CoV (\%) & 54.001 & 4.139 & 2.985 & 2.755 & 2.628 & 2.517 & 2.416 & 2.230 & 2.136 & 2.064 \\              
\hline                                                                                                            
$R_{E}$ (mean) & 0.044 & 0.097 & 0.126 & 0.117 & 0.101 & 0.116 & 0.110 & 0.104 & 0.106 & 0.109 \\                                                                                                                      
$R_{E}$ CoV (\%) & 53.667 & 36.112 & 27.276 & 17.663 & 15.733 & 15.205 & 14.778 & 13.880 & 13.744 & 13.502 \\     
\hline                                                                                                            
$\mathcal{C}~\times 10^5$(mean) & 3.888 & 4.147 & 3.944 & 4.251 & 3.968 & 3.928 & 4.003 & 4.165 & 4.180 & 4.404 \\
$\mathcal{C}$ CoV (\%) & 9.943 & 8.852 & 6.465 & 5.909 & 4.299 & 4.344 & 4.283 & 3.670 & 3.709 & 3.644 \\         
\hline                                                                                                            
$\mathcal{U}$ (mean) & 1.619 & 1.221 & 1.179 & 1.213 & 1.150 & 1.110 & 1.120 & 1.121 & 1.126 & 1.140 \\           
$\mathcal{U}$ CoV (\%) & 14.038 & 3.167 & 1.102 & 0.740 & 0.471 & 1.034 & 0.338 & 0.336 & 0.279 & 0.245 \\        
$\mathcal{U}_{av}$ & 1.969 & 1.173 & 1.154 & 1.212 & 1.168 & 1.162 & 1.160 & 1.116 & 1.109 & 1.133 \\             
\hline                                                                                                                                                                                               
\end{tabular}                                                                                       
\caption{BSRIA Experiment. Sequential posterior estimates of $R_{I}$, $R_{E}$, $\mathcal{U}$ and $\mathcal{C}$ obtained during the first 4.4 days of the measurement campaign.}
\label{Table3}  
\end{table}              

\clearpage
\newpage 
\section{Summary, conclusions and future directions.}\label{conclusions}

We proposed a Bayesian approach to sequentially infer thermophysical properties of any wall, given in-situ measurements from the walls's internal and external near-air temperatures and surface heat fluxes. The proposed approach was encoded in a computational methodology (Algorithm \ref{IF}) that uses these measurements to sequentially characterise the posterior distribution of the unknown parameters of a HDM of the wall based on the 1D heat equation. These parameters included the thermal conductivity and volumetric heat capacity that we characterised with spatially varying functions defined at every location inside the wall. Scalar values of the wall's surface resistance were also inferred within the proposed framework. Posteriors of the unknown parameters were transformed into probability distributions of the U-value and C-value, as well as the predictive distributions of surface heat flux. 

We use both synthetic and real measurements to test and validate the proposed computational approach, and demonstrated that it enables a fast and accurate estimation of the effective thermophysical properties of the wall (i.e. U-value and C-value) together with a measure of their uncertainties. We  additionally showed that our technique can be used, on-the-fly, to determine the duration of the measurement campaign required to achieve a specified level of uncertainty in the posterior estimates of the thermophysical properties. For example, our numerical results showed that 1\% coefficient of variation in the estimates of the U-value can be achieved within the first day of the measurement campaign. Furthermore, these estimates were more accurate, and achieved through shorter measurement campaigns, compared to the ones we computed via the average method suggested by the ISO9869:2014  \cite{ISO9869:2014}. The proposed technique can thus be applied to conduct cost-effective measurement campaigns that accurately estimate the U-value of walls in existing dwellings. These estimates can be used to inform some normative/certification models, such as the Reduced data Standard Procedure Assessment used in the UK, to assess the energy efficiency of an existing dwelling, and ultimately, aid in the development of low-carbon policies.

The capability of the proposed technique to infer spatially-variable thermophysical properties of the wall via the Bayesian calibration of the HDM (with a relatively fine discretisation) allowed us to demonstrate that the proposed technique can (i) statistically detect unknown inhomogeneities within the wall; and (ii) produce model predictions that achieve accurate and high-confident predictions of the thermal performance of the wall (i.e. surface heat flux). This capability also enabled us to show that using coarse-grid discretisations of the HDM for the Bayesian inversion of in-situ measurements can lead to inaccurate statistical predictions of internal and external surface heat flux, even though accurate estimates of the U-value were obtained. These results suggest that  an accurate in-situ estimation of the U-value obtained via the calibration of a heat transfer model does not necessarily ensure that the underlying model produces an accurate probabilistic description of the thermal performance of the wall. In practice, a sufficiently-resolved HDM calibrated via the proposed approach can be useful within decision-making workflows since it enables the simulation of the wall's thermal performance under different retrofit interventions which, in turn, facilitates the computation of probabilities of the financial outcomes associated to those renovation measures.

Although the focus of the present work is to infer the thermal properties of the wall characterised via the input parameters in the 1D-based HDM of the wall, we recognise that additional sources of uncertainty in the prediction of the wall's thermal performance may exist. Failing to account for these uncertainties can be detrimental to the accuracy of the estimates of inferred parameters. Uncertainties of those kind could arise from unaccounted sources of heat (e.g. solar radiation) in the proposed HDM within the Bayesian calibration framework. In addition, the one-dimensional assumption of the heat transfer trough the wall used in the present work, albeit used in most existing work for the characterisation of a wall's thermal performance, has been recently identified as a source of modeling errors during the thermal characterisation walls under the presence of thermal bridge effects that arise, for example, from material defects and/or moisture penetration \cite{en9030126}. Further work should then incorporate 3D models of the thermal performance of the wall within the proposed Bayesian approach, and the use of additional measurement technologies (e.g. time-lapse thermography \cite{cite-key}) in order to capture thermal bridge effects within the posterior estimates of the thermophysical properties.

\section*{Acknowledgements}

The authors are grateful to Phillip Biddulph and Tadj Oreszczyn for sharing the BSRIA data. The authors would also like to acknowledge the financial support obtained for Lia De Simon through the FP7 framework sponsorship of the Holistic Energy-efficient Retrofit of Buildings (HERB) project.

\appendix

\section{The Prior}\label{prior}

As we discussed in subsection \ref{bayesian_theo}, the Bayesian approach enables us to incorporate prior knowledge of the unknown parameters $\kappa(x)$, $c(x)$, $T_{0}(x)$, $R_{I}$ and $R_{E}$ that wish to infer. In this subsection we discuss our selection of the prior distribution for these variables and how samples can be drawn from these distributions. We recall that samples from these priors are needed in order to generate the initial ensemble that we must specify in order to initialise Algorithm \ref{IF}.

For simplicity, let us first assume that, under the prior, the unknown parameters $\kappa(x)$, $c(x)$, $T_{0}(x)$ $R_{I}$ and $R_{E}$ are independent random functions/variables and so the joint prior can be written as 
 \begin{eqnarray}\label{eq:18}
\mathbb{P}(u)=\mathbb{P}(\kappa,c,T_{0},R_{I},R_{E})=\mathbb{P}(\kappa)\mathbb{P}(c)\mathbb{P}(T_{0})\mathbb{P}(R_{I})\mathbb{P}(R_{E}),
\end{eqnarray}
where $\mathbb{P}(\kappa)$, $\mathbb{P}(c)$, $\mathbb{P}(T_{0})$, $\mathbb{P}(T_{0})$, $\mathbb{P}(R_{I})$ and $\mathbb{P}(R_{E})$ are the priors of $\kappa(x)$, $c(x)$, $T_{0}(x)$, $R_{I}$ and $R_{E}$, respectively. In a practical context, our prior knowledge (e.g. from previous experiments) of these thermophysical properties could suggest correlations between these parameters; these correlations can be incorporated within the proposed framework. 

We know proceed to specify each of the priors in the right hand side of (\ref{eq:18}). For the scalars $R_{I}$ and $R_{E}$ we consider log-normal priors. More specifically, we assume $R_{I}=\omega_{I}\exp{\Psi_{I}}$ and $R_{E}=\omega_{E}\exp{\Psi_{E}}$, where $\Psi_{I}$ and $\Psi_{E}$ are normal random variables with zero mean and variances $\sigma_{I}$ and $\sigma_{E}$, respectively. Samples from these distributions can be easily generated via standard statistical software. We have selected these log-normal priors simply to ensure that samples from these distributions are positive quantities. However, other choices of priors (e.g. uniform) could be considered within the proposed Bayesian approach. In practice, these priors should be constructed from empirical distributions computed from historical data over a number of dwellings of the same type. 

In order to define priors for the functions $\kappa(x)$, $c(x)$, $T_{0}(x)$ we propose the use of Gaussian random fields \cite{lord2014introduction}. Our aim is to use Gaussian priors to characterise, via a wide class of functions, the spatial variability in the wall's thermal properties. In particular, for the prior of the function $\kappa(x)$ we consider lognormal Gaussian distributions for which we assume that (under the prior), $\kappa(x)$ can be written as 
 \begin{eqnarray}\label{eq:19}
\kappa(x) =\omega_{\kappa}\exp(\Psi_{\kappa}(x)) 
\end{eqnarray}
where $\omega_{\kappa}$ is a positive constants and $\Psi_{\kappa}$ is a stationary Gaussian random function/field (GRF) with zero mean and covariance operator $C_{\kappa}$. The generation of a random functions from such a distribution can be easily achieved by means of the Karhunen-Loeve (KL) expansion \cite{lord2014introduction}. The choice of $C_{\kappa}$ determines the regularity of the family of functions which are samples from this distribution. For the present work we consider covariance operators $C_{\kappa}$, induced by the Whittle-Matern covariance function given by \cite{whittlematern}
 \begin{equation}\label{eq:20}
f_{\sigma, \nu, l}(x,y) =\sigma_{\kappa}^{2}\frac{2^{\nu-1}}{\Gamma(\nu)}\Bigg(\frac{\vert x-y\vert }{l} \Bigg)^{\nu}K_{\nu}\Bigg(\frac{\vert x-y\vert }{l}\Bigg),
\end{equation}
where $\nu>0$ is a parameter that controls the regularity/smoothness of the samples, $l$ is the characteristic length scale, $\sigma_{\kappa}^2$ is the variance, $\Gamma$ is the gamma function, and $K_{\nu}$ is the modified Bessel function of the second kind of order $
\nu$. 

\begin{figure}[htbp]
\begin{center}

\includegraphics[scale=0.55]{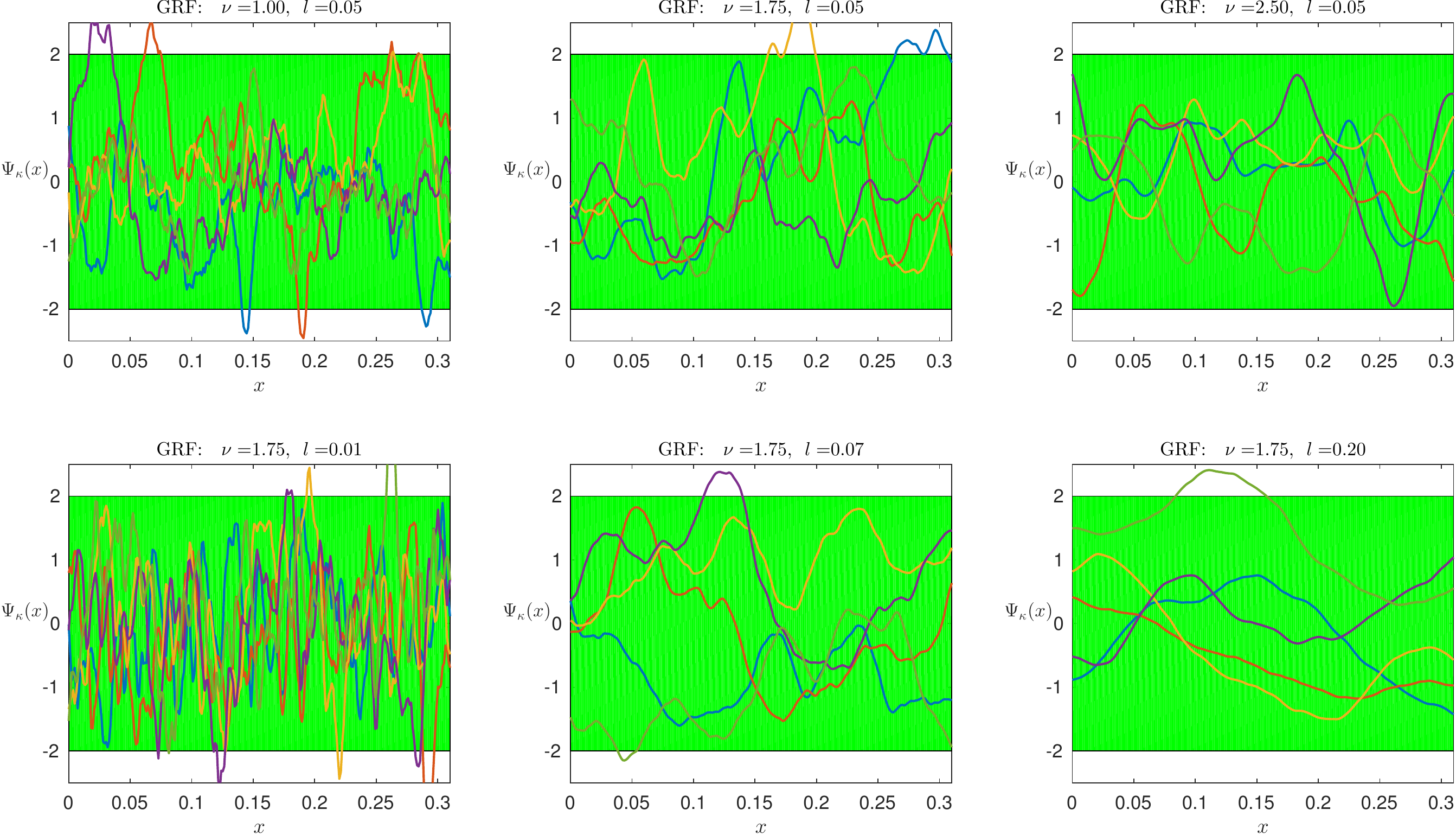}

 \caption{Samples of Gaussian random functions with Whittle-Matern (WM) correlations with different intrinsic length scales $l$ and smoothness parameter $\nu$. } \label{Fig0}
\end{center}
\end{figure}

By means of the Karhunen-Loeve expansion we generate, and display in Figure \ref{Fig0}, five (discretised) GRF with different choices of $\nu$ and $l$.  Note that larger $\nu$'s results in samples which are more regular functions of $x$. Similarly, smaller $l$ yields samples which are less spatially correlated. We can then exploit the variability provided by these random family of functions to characterise, a priori, the spatial variability of the thermal properties of the wall. Furthermore, the selection of the prior above, in terms of the parameterisation in (\ref{eq:19}) ensures that the (unknown) function $\kappa(x)$ is always positive. In addition, the prior mean and variance of $\kappa(x)$ are constants (with respect to $x$) and given by
\begin{eqnarray}\label{eq:21}
\mathbb{E}(\kappa)=\omega_{\kappa}\exp\Bigg(\frac{1}{2}\sigma_{\kappa}^2\Bigg)   ,\qquad \mathbb{V}(\kappa)=\omega_{\kappa}^{2}\exp(\sigma_{\kappa}^2)(\exp(\sigma_{\kappa}^2)-1)
\end{eqnarray}
Note that we can then select $\omega_{\kappa}$ and $\sigma_{\kappa}^{2}$ so that the prior mean, $\mathbb{E}(\kappa)$ reflects our prior knowledge of the thermal conductivity given, for instance,  by visual inspection or book values.

The prior $\mathbb{P}(c)$ for the unknown $c(x)$ is defined in a similar fashion to the one for $\kappa$ above. For the prior of the initial temperature $T_{0}(x)$ we first note from (\ref{eq:7})-(\ref{eq:8}), evaluated at $t=0$, that 
\begin{eqnarray*}
T(0,0)=& T_{I}(0)-R_{I}Q(0,0)\\
T(L,0)=&R_{E}Q(L,0)+T_{E}(0)
\end{eqnarray*}
provides an expression for the surface temperature at time $t=0$. We recall that $T_{I}(0)$ and $T_{E}(0)$ are values of internal and external near-air temperature at the initial time.  These values are recorded in-situ via the variables $T_{I,0}^{\dagger}$ and $T_{E,0}^{\dagger}$. Similarly, $Q(0,0)$ and $Q(L,0)$ are the initial-time internal and external surface heat fluxes which are observed in-situ; the corresponding observations of these quantities are $q_{I,0}$ and $q_{E,0}$, respectively. We therefore propose to construct a Gaussian prior for $T_{0}(x)$ with a mean given by
\begin{eqnarray} \label{ape_last1}
\omega_{T_{0}}(x)\equiv \overline{R}_{E}q_{E,0}+T_{E,0}^{\dagger}+\frac{(L-x)}{L}(T_{I,0}^{\dagger}-\overline{R}_{I}q_{I,0}-\overline{R}_{E}q_{E,0}-T_{E,0}^{\dagger})
\end{eqnarray}
which reflects our prior knowledge that the initial temperature is a random fluctuation around the linear interpolation between the internal and external surface temperatures at $t=0$, approximated via in-situ measurements. In expression \eqref{ape_last1}, the values of $\overline{R}_{E}$ and $\overline{R}_{I}$ correspond to the means of the prior (log-normal) distributions of $R_{I}$ and $R_E$ introduced above. A covariance function of the form (\ref{eq:20}) is also used for the prior covariance of $T_{0}(x)$. 

We refer the reader to \cite{lord2014introduction} for further technical details of the KL approach to generate five (discretised) GRF.  We have included this approach in Algorithm \ref{prior_al} which provides the steps for the generation of an initial ensemble of $J$ particles of the aforementioned prior for the variables $\kappa(x)$, $c(x)$, $T_{0}(x)$, $R_{E}$ and $R_I$. It is important to mention that our selection of priors defined in this subsection induces the following parameterisation
\begin{eqnarray} \label{ape_last2}
u=(\kappa,c,T_{0},R_{I},R_{E})=(\omega_{\kappa}\exp\Psi_{\kappa},\omega_{c}\exp\Psi_{c},\omega_{T_{0}}+\Psi_{T_{0}}, \omega_{I}\exp\Psi_I,\omega_{I}\exp\Psi_E)=F(\Psi)
\end{eqnarray}
in terms of the random function/variables:
$$\Psi\equiv (\Psi_{\kappa},\Psi_{c},\Psi_{T_{0}},\Psi_I,\Psi_E).$$
This parameterisation will be used within the application of the REnKA algorithm that we review in \ref{REnKA}.

\begin{figure}[htbp]
\begin{center}

\includegraphics[scale=0.75]{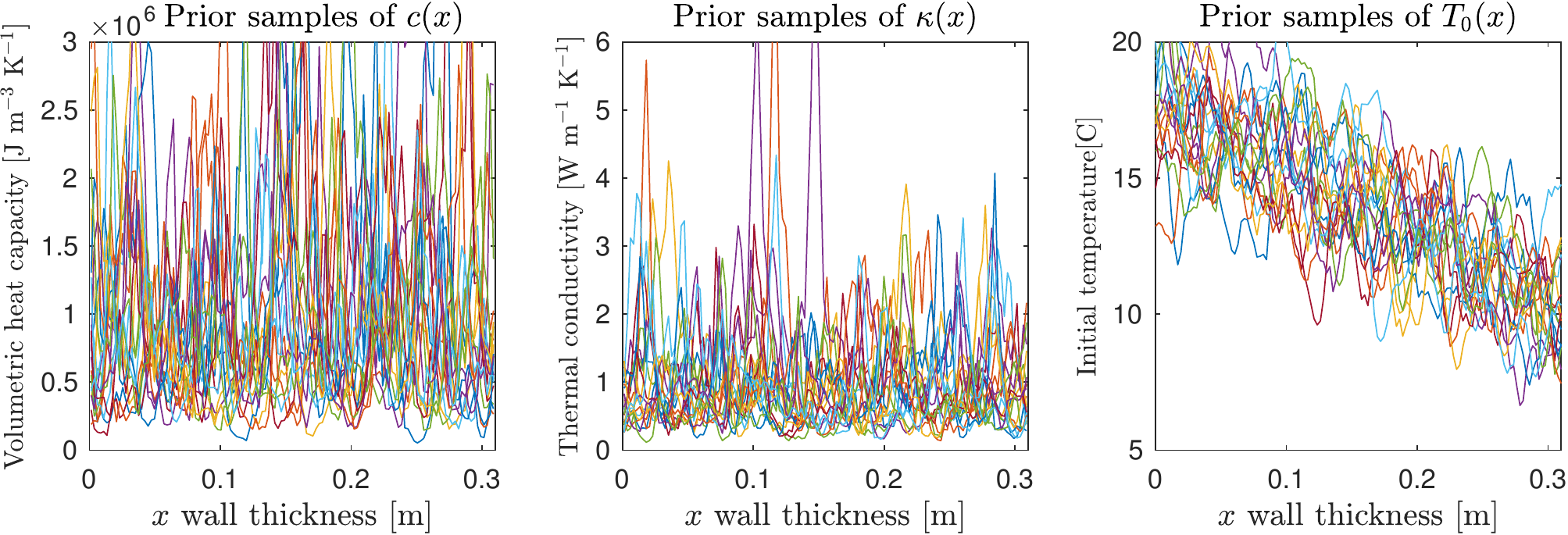}\\
 \caption{Samples from the priors $\mathbb{P}(c)$ (left), $\mathbb{P}(\kappa)$ (middle) and $\mathbb{P}(T_{0}) $ (right)}\label{Fig2}

\end{center}
\end{figure}

For the experiments of Section \ref{synthetic} and Section \ref{BSRIA} we use parameters in the priors (or hyperparameters) as displayed in Table \ref{parameters}. These hyperparameters are used in Algorithm \ref{prior_al} to generate the set of prior samples that we display in Figure \ref{Fig2}. As stated earlier, our choice of hyperparameters is aimed at capturing rapid changes in the thermophysical properties within the wall. However, it is possible to extend the proposed technique via hierarchical parameterisations \cite{0266-5611-34-5-055009} to enable the estimation of these hyperparameters from in-situ measurements within the computational framework.

\begin{algorithm}
\caption{Generation of the prior ensemble $\{(\kappa_{0}^{(j)}(x),c_{0}^{(j)}(x),T_{0,0}^{(j)}(x),R_{I,0}^{(j)},R_{E,0}^{(j)})\}_{j=1}^{J}$}\label{prior_al}
Input parameters: $J$ (number of ensemble members), $\omega_{\beta},\sigma_{\beta}$ ( for $\beta\in \{\kappa,c,T_{0},I,E\})$ and  $\nu_{\beta},l_{\beta}$, (for $\beta\in \{\kappa,c,T_{0}\})$.
\begin{algorithmic}
\For{$\beta\in \{\kappa,c,T_{0}\}$}
\State(1)  Construct the discretised covariance ($N_x\times N_x)$ matrix $C_{\beta}$,  by means of 
$$[\mathcal{C}_{\beta}]_{i,j}=f_{\sigma_{\beta}, \nu_{\beta}, l_{\beta}}(x_{i},x_{j})$$
\State where $\{x_{i}\}_{i=1}^{N_{x}}$ are the nodal points of the discretisation of the domain $[0,L]$.
\State(2) Compute the eigenvalue-eigenvector pair $\{\lambda_{\beta,i},w_{\beta,i}\}_{i=1}^{N_{x}}$ of $C_{\beta}$
\State(3) Construct KL expansion:\For{$j=1\dots J$}
\State Sample $\xi_{\beta,i}\sim N(0,1)$ ($i=1,\dots,N_{x}$) and construct
$$\Psi_{\beta}^{(j)}\equiv \sum_{i=1}^{N_{x}}\lambda_{\beta,i}^{1/2}w_{\beta,i}\xi_{\beta,i}$$
\EndFor
\EndFor \State 
\For{$j=1\dots J$}

Sample $\Phi_{I}^{(j)}\sim N(0,\sigma_{I})$ and  $\Phi_{E}^{(j)}\sim N(0,\sigma_{E})$
\EndFor

Output:
$$\kappa_{0}^{(j)}=\omega_{\kappa}\exp\Psi_{\kappa}^{(j)}, \quad c_{0}^{(j)}=\omega_{c}\exp\Psi_{c}^{(j)},\quad T_{0,0}^{(j)}=\omega_{T_{0}}+\Psi_{T_{0}}^{(j)}, \quad R_{I,0}^{(j)}=\omega_{I}\exp\Psi_I^{(j)}, \quad R_{E,0}^{(j)}=\omega_{I}\exp\Psi_E^{(j)}$$
for $j=1,\dots,J$

\end{algorithmic}
\end{algorithm}

\begin{table}                                                                                       
\small
\centering                                                                                                               
\begin{tabular}{|c|c|c|c|c|c|c|c|c|c|}                                                                             
\hline                                                                                                                  
& & $\beta=\kappa$ &$\beta=c$&$\beta=T_{0}$&$\beta=I$&$\beta=E$\\
\hline            
&$\nu_{\beta}$  &1.05 &      1.05 &  1.05&-&- \\
&$l_{\beta}$  &$0.62\times 10^{-2}$& $0.62\times 10^{-2}$&  $10^{-2}$&-& -\\
Synthetic Experiment &$\sigma_{\beta}$  &  0.65  &  0.7 & 1.87&0.5&0.5\\
&$\omega_{\beta}$  &$0.75$ & $7.5\times 10^5$& See (\ref{ape_last1})&$\ln{0.1}$&$\ln{0.07}$\\
\hline                                                                                                                 
&$\nu_{\beta}$  &1.05 &      1.05 &  1.05 &-&- \\
&$l_{\beta}$  &0.0103&  0.0103&  0.0103&-&- \\
BSRIA Experiment &$\sigma_{\beta}$  &  0.35  &  0.3 & 1.22&0.5&0.5\\
&$\omega_{\beta}$  &$0.75$ & $1.2\times 10^6$& See (\ref{ape_last1})&$\ln{0.13}$&$\ln{0.04}$\\
\hline                                                                                                                 
\end{tabular}           
\caption{Parameters for the priors generated via Algorithm \ref{prior_al}.}
\label{parameters}  
\end{table}       

\section {Regularising ensemble Kalman algorithm}\label{REnKA}

In this subsection we briefly discuss the regularising ensemble Kalman algorithm (REnKA) that we use for the Bayesian updating step of Algorithm \ref{IF}. Suppose that, at the assimilation time $\tau_{m}$, we have an ensemble $\{u_{m-1}^{(j)}\}_{j=1}^{J}=\{(\kappa_{m-1}^{(j)}(x),c_{m-1}^{(j)}(x),T_{m-1,0}^{(j)}(x),R_{I,m-1}^{(j)},R_{E,m-1}^{(j)})\}_{j=1}^{J}$ that provides a particle/sample approximation of the posterior ${\mathbb P}(u|q_{1:m-1})$; this distribution takes into account all measurements collected in the interval of time $[0,\tau_{m-1}]$. The aim of REnKA is to use new measurements of surface heat flux $q_{m}$ and near-air temperatures ($T_{I,m}^{\dagger}$,$T_{E,m}^{\dagger}$), collected in the interval $(\tau_{m-1},\tau_{m}]$, in order to produce a new ensemble of particles $\{u_{m}^{(j)}\}_{j=1}^{J}$ that approximates the distribution ${\mathbb P}(u|q_{1:m})$, which now includes all measurements in the interval $[0,\tau_{m}]$. 

Crucial to the REnKA scheme is the idea of tempering which introduces a sequence of $p+1$ distributions $\{\mu_{r}(u)\}_{r=0}^{p}$ defined by 
\begin{equation}\label{eq:16Ape}
\mu_{r}(u)= c_{r} {\mathbb P}_{\eta}(q_{m}-\cG_{m}(u, T_{I,1:m}^{\dagger}, T_{1:mE}^{\dagger}))^{\phi_{r}} {\mathbb P}(u |q_{1:m-1}).
\end{equation}
where $c_{r}$ is a normalisation constant and $\phi_{r}$  is a tempering parameter that satisfy $0=\phi_{0}\leq \dots<\phi_{r}<\dots\phi_{p}=1$. From (\ref{eq:16Ape}) and (\ref{eq:16}) we observe that, for $r=0$, $\mu_{r}(u)={\mathbb P}(u |q_{1:m-1})$, while for $r=p$, we have $\mu_{r}(u)= {\mathbb P}(u|q_{1:m})$. Expression (\ref{eq:16Ape}) thus defines a sequence of intermediate distributions between the available posterior at time $t=\tau_{m-1}$ and the one that wish to approximate at time $t=\tau_{m}$. Note that expression (\ref{eq:16Ape}) can be written in the following recursive fashion:
\begin{eqnarray}\label{eq:17Ape}
\mu_{r}(u)= \frac{c_{r}}{c_{r-1}}\mu_{r-1}(u){\mathbb P}_{\eta}(q_{m}-\cG_{m}(u, T_{I,1:m}^{\dagger}, T_{1:mE}^{\dagger}))^{(\phi_{r}-\phi_{r-1})} \nonumber \\
\propto\frac{c_{r}}{c_{r-1}}\mu_{r-1}(u)  \exp\Big[- (\phi_{r}-\phi_{r-1}) \vert\vert \Gamma_{m}^{-1/2}(q_{m}-\cG_{m}(u, T_{I,1:m}^{\dagger}, T_{1:mE}^{\dagger}))\vert\vert^2\Big],
\end{eqnarray}
where the last expression follows from our Gaussian assumption on the measurement error of heat flux measurements. 

Starting with the ensemble $\{u_{m-1,0}^{(j)}\}_{j=1}^{J}=\{u_{m-1}^{(j)}\}_{j=1}^{J}$ that approximates $\mu_{0}={\mathbb P}(u|q_{1:m-1})$, the central idea of REnKA is to construct ensemble-based Gaussian approximations of each distribution $\mu_{r}$ defined via (\ref{eq:16Ape}). Each ensemble member from these Gaussian approximations is updated by an expression derived from Kalman-like formulations. The selection of the tempering parameter $\phi_{r}$ is carried out with the adaptive approach of \cite{Kantas} based on the Effective Sample Size defined by
\begin{eqnarray}\label{eq:18Ape}
ESS_{r}(\phi)\equiv  \Bigg[\sum_{j=1}^{J}(\mathcal{W}_{r-1}^{(j)}[\phi])^2\Bigg]^{-1}
\end{eqnarray}
with
\begin{eqnarray}\label{eq:19Ape}
\mathcal{W}_{r-1}^{(j)}[\phi] = \frac{\exp\Big[- (\phi-\phi_{r-1}) \vert\vert \Gamma_{m}^{-1/2}(q_{m}-\cG_{m}(u^{(j)}, T_{I,1:m}^{\dagger}, T_{1:mE}^{\dagger}))\vert\vert^2\Big]}{\sum_{s=1}^{J}\exp\Big[- (\phi-\phi_{r-1}) \vert\vert \Gamma_{m}^{-1/2}(q_{m}-\cG_{m}(u^{(s)}, T_{I,1:m}^{\dagger}, T_{1:mE}^{\dagger}))\vert\vert^2\Big]}.
\end{eqnarray}
More specifically, at an iteration level $r$, the tempering parameter $\phi_{r}$ is selected so that 
\begin{eqnarray}\label{eq:18Ape}
ESS_{r}(\phi_{r})=J_{thresh}
\end{eqnarray}
where $J_{thresh}$ is a used-defined tunable parameter. Intuitively, this selection of the tempering parameters ensure a smooth/regularised transition between the distributions $\mu_{r}$, thus avoiding a collapse of the particles which may be, in turn, detrimental to the accuracy of the inference scheme. The full algorithm is display in Algorithm \ref{REnKA_al}. Note that the implementation is  carried out in therms of the parameterisation refined via (\ref{ape_last2}). Further details on the derivation of the scheme can be found in \cite{SMC_REnKA}.

\begin{algorithm}
\caption{Regularising ensemble Kalman algorithm (REnKA)}\label{REnKA_al}
Inputs: $\Psi_{m-1}^{(j)}$ (recall $u_{m-1}^{(j)}=F(\Psi_{m-1}^{(j)})$), $J_{thresh}$, $q_{m}$, $\Gamma_{m}$, $T_{I,m}^{\dagger}$, $T_{E,m}^{\dagger}$
\begin{algorithmic}

\State(1) Set $r=0$ and $\phi_{0}=0$, and $\{\Psi_{m-1,0}^{(j)}\}_{j=1}^{J}=\{\Psi_{m-1}^{(j)}\}_{j=1}^{J}$
\While {$\phi_{r}<1$}
\State(a) $r\gets r+1$
\State(b) For $j=1\dots, J$, compute
$$\cG_{m,r-1}^{(j)}\equiv \cG_{m}(F(\Psi_{m-1,r-1}^{(j)}),T_{I,1,\dots,m}^{\dagger},T_{E,1\dots,m}^{\dagger})$$
\State This involves solving \eqref{eq:6}-\eqref{eq:9} from $t=0$ to $t=\tau_{m}$.
\State(c)  \textbf{Compute tempering parameter $\phi_{r}$}: \\
\If{ $\min_{\phi\in (\phi_{r-1},1)}\textrm{ESS}_{r}(\phi)>J_{thresh}$}
\State set $\phi_{r}=1$.
\Else
\State Compute $\phi_{r}$ such that $\textrm{ESS}_{r}(\phi)\approx J_{tresh}$ using a bisection algorithm on $(\phi_{r-1},1]$.
\EndIf
\State(d)  Construct the following empirical covariances:
\State
\begin{eqnarray*}
A_{r-1}=&\frac{1}{J-1}\sum_{j=1}^{J}(\cG_{m,r-1}^{(j)}-\overline{\cG}_{m,r-1})(\cG_{m,r-1}^{(j)}-\overline{\cG}_{m,r-1})^{T},\\
B_{r-1}=&\frac{1}{J-1}\sum_{j=1}^{J}(\Psi_{m-1,r-1}^{(j)}-\overline{\Psi}_{m-1,r-1})(\cG_{m,r-1}^{(j)}-\overline{\cG}_{m,r-1})^{T}
\end{eqnarray*}
\State where $\overline{\cG}_{m,r-1}\equiv \frac{1}{J}\sum_{j=1}^{J}\cG_{m,r-1}^{(j)}$ and $\overline{\Psi}_{m-1,r-1}\equiv \frac{1}{J}\sum_{j=1}^{J}\Psi_{m-1,r-1}^{(j)}$. 
\State(e) Update each ensemble member:
 \begin{eqnarray}\label{eq:m16}
\Psi_{m-1,r}^{(j)} =\Psi_{m-1,r-1}^{(j)}+B_{r-1}(A_{r-1}+\alpha_{m-1,r}\Gamma   )^{-1}(q_{m,r}^{(j)}-\cG_{m,r-1}^{(j)})
\end{eqnarray}
where
$$\alpha_{m-1,r}=(\phi_{r}-\phi_{r-1})^{-1}\qquad q_{m,r}^{(j)}=q_{m}+\eta_{m,r}^{(j)},\qquad \eta_{m,r}^{(j)}\sim N(0,\alpha_{m-1,r}\Gamma_{n});$$
\EndWhile

Output: $\Psi_{m}^{(j)}\equiv \Psi_{m-1,r}^{(j)}$. The particle approximation of $\mathbb{P}(u\vert q_{1,\dots,m})$ is provided by $\{u_{m}^{(j)}\}_{j=1}^{J}=\{F(\Psi_{m}^{(j)}\}_{j=1}^{J}$
\end{algorithmic}
\end{algorithm}

\section*{\refname}
\bibliographystyle{elsarticle-num} 
\bibliography{Thermal_mass,Reducing_carbon_emissions,Numerical_modelling,RCNetworks,Ensemble_bib,Instrumentation,Bayesian_inverse}



\end{document}